\documentclass[12pt]{amsart}
\usepackage[utf8]{inputenc}
\usepackage{amsfonts}
\usepackage{natbib}
\usepackage{bm}
\usepackage[usenames,dvipsnames,svgnames,table]{xcolor}
\usepackage[font=footnotesize,labelfont=bf]{caption}
\usepackage{colortbl}
\usepackage{booktabs}
\usepackage{fixltx2e}
\usepackage{graphicx}
\usepackage{float}
\usepackage{bbm}
\usepackage[hidelinks]{hyperref}
\usepackage{standalone}
\usepackage{mwe}
\usepackage[inline]{enumitem}
\usepackage{subcaption}
\usepackage{indentfirst}
\usepackage[margin=1.25in]{geometry}

\usepackage[right]{lineno}
\usepackage{amsmath}  
\usepackage{etoolbox} 

\newcommand*\linenomathpatch[1]{%
  \cspreto{#1}{\linenomath}%
  \cspreto{#1*}{\linenomath}%
  \csappto{end#1}{\endlinenomath}%
  \csappto{end#1*}{\endlinenomath}%
}

\linenomathpatch{equation}
\linenomathpatch{gather}
\linenomathpatch{multline}
\linenomathpatch{align}
\linenomathpatch{alignat}
\linenomathpatch{flalign}

\RequirePackage{calc}

\newcommand{\pr}{\text{Pr}}
\newcommand{\phat}{\widehat{\pr}}
\newcommand{\btau}{\boldsymbol{\tau}}

\newcommand{\bX}{\boldsymbol{X}}
\newcommand{\bY}{\boldsymbol{Y}}

\newcommand{\E}{\mathbb{E}}
\newcommand{\var}{\text{Var}}
\newcommand{\sd}{\text{SD}}
\newcommand{\se}{\text{SE}}
\newcommand{\semcmc}{\widehat{\se}_{\text{MCMC}}}
\newcommand{\seess}{\widehat{\se}_{\text{MCESS}}}
\newcommand{\limvar}{\sigma^2_{\text{lim}}}
\newcommand{\postvar}{\sigma^2_{\pi}}
\newcommand{\cov}{\text{Cov}}
\newcommand{\cor}{\text{Cor}}

\newcommand{\summary}{\hat{\mathcal{S}}_n}

\newcommand{\ESS}{\text{ESS}}
\newcommand{\frechetCorrelationESS}{\textsf{frechetCorrelationESS}}
\newcommand{\splitFrequencyESS}{\textsf{splitFrequencyESS}}
\newcommand{\minPseudoESS}{\textsf{minPseudoESS}}
\newcommand{\medianPseudoESS}{\textsf{medianPseudoESS}}
\newcommand{\approximateESS}{\textsf{approximateESS}}
\newcommand{\foldedRankmedoidESS}{\textsf{foldedRankmedoidESS}}
\newcommand{\CMDSESS}{\textsf{CMDSESS}}
\newcommand{\totalDistanceESS}{\textsf{totalDistanceESS}}
\newcommand{\jumpDistanceBootstrapESS}{\textsf{jumpDistanceBootstrapESS}}
\newcommand{\jumpDistanceBootstrapUnsmoothedESS}{\textsf{jumpDistanceBootstrapUnsmoothedESS}}

\newcommand{\beginsupplement}{%
        \setcounter{table}{0}
        \renewcommand{\thetable}{S\arabic{table}}%
        \setcounter{figure}{0}
        \renewcommand{\thefigure}{S\arabic{figure}}%
     }

\definecolor{lightergray}{rgb}{0.9, 0.9, 0.9}

\begin{document}

\title[Phylogenetic Monte Carlo Error]{How trustworthy is your tree? Bayesian phylogenetic effective sample size through the lens of Monte Carlo error}

\begin{center}

\author[A.F. Magee, M. D. Karcher, F. A. Matsen IV, V.N. Minin]{Andrew F. Magee$^1$, Michael D. Karcher$^2$, Frederick A. Matsen IV$^3$, and Volodymyr M. Minin$^4$}

\maketitle

\noindent {\small \it
$^1$Department of Biology, University of Washington, Seattle, WA, 98195, USA\\
$^2$Department of Mathematics and Computer Science, Muhlenberg College, Allentown, PA, 18104, USA\\
$^3$Howard Hughes Medical Institute, Fred Hutchison Cancer Research Center, Departments of Genome Sciences and Statistics, University of Washington, Seattle, WA, 98109, USA\\
$^4$Department of Statistics, University of California, Irvine, CA, 92697, USA}
\end{center}

\medskip
\noindent{\bf Corresponding authors:} Frederick A. Matsen IV (matsen@fredhutch.org) and Volodymyr M. Minin (vminin$@$uci.edu)\\

\newpage

\section*{Abstract}
Bayesian inference is a popular and widely-used approach to infer phylogenies (evolutionary trees).
However, despite decades of widespread application, it remains difficult to judge how well a given Bayesian Markov chain Monte Carlo (MCMC) run explores the space of phylogenetic trees.
In this paper, we investigate the Monte Carlo error of phylogenies, focusing on high-dimensional summaries of the posterior distribution, including variability in estimated edge/branch (known in phylogenetics as ``split'') probabilities and tree probabilities, and variability in the estimated summary tree.
Specifically, we ask if there is any measure of effective sample size (ESS) applicable to phylogenetic trees which is capable of capturing the Monte Carlo error of these three summary measures.
We find that there are some ESS measures capable of capturing the error inherent in using MCMC samples to approximate the posterior distributions on phylogenies.
We term these tree ESS measures, and identify a set of three which are useful in practice for assessing the Monte Carlo error.
Lastly, we present visualization tools that can improve comparisons between multiple independent MCMC runs by accounting for the Monte Carlo error present in each chain.
Our results indicate that common post-MCMC workflows are insufficient to capture the inherent Monte Carlo error of the tree, and highlight the need for both within-chain mixing and between-chain convergence assessments.

\section*{Introduction}

Bayesian inference via Markov chain Monte Carlo (MCMC) is widely used in phylogenetic estimation \citep{nascimento2017biologist}.
MCMC enables the generation of samples according to arbitrary distributions, such as the posterior distribution of a phylogenetic model, though it must draw autocorrelated samples \citep{geyer2011introduction}.
In fact, it is only in the limit of running the analysis infinitely that we are guaranteed the summaries of our MCMC samples will converge to the true values of the corresponding posterior summaries.
Users are thus left with the task of determining whether inference from their MCMC samples are trustworthy.
In practice, this entails assessing whether any given chain appears to be stationary and assessing how well it is mixing \citep{kass1998markov}.
In this paper, we focus on the issue of mixing as embodied in the notion of the effective sample size (ESS).
The effective sample size is closely related to the notion of Monte Carlo error, which describes the error in parameter estimation due to using sampling-based approaches \citep{geyer2011introduction,vats2021revisiting,vehtari2021rank}.

A phylogenetic model describes the evolutionary history of a set of samples for which we have character data (such as DNA sequence data) \citep{felsenstein2004inferring,lemey2009phylogenetic,chen2014bayesian}.
The core of a phylogenetic model is a discrete tree topology, which specifies how the samples are related to each other via a series of nested relationships \citep{thompson1975human,felsenstein1981evolutionary}.
Describing the process of evolution along the tree also requires branch lengths, which specify the amount of evolutionary change, and a substitution model, which specifies the relative rates of change between character states \citep{felsenstein1981evolutionary}.

Phylogenetic posterior distributions are complex objects, and it is known that sampling from them via MCMC can be quite difficult \citep{lakner2008efficiency,hohna2012guided,whidden2015quantifying,zhang2020using,meyer2019adaptive}.
Overall performance of the MCMC chain is often diagnosed by looking at the trace (collection of samples through time) of the log-likelihood or the log-posterior density, though this should only be the first step in a more rigorous pipeline \citep{lemey2009phylogenetic}.
\@\texttt{Tracer} is a popular tool for summarizing and visualizing MCMC samples from phylogenetic software, which will automatically compute the ESS for all values a program records, including continuous model parameters and the densities of the log-likelihood and log-posterior \citep{rambaut2018posterior}.
\texttt{Tracer} flags parameters if the ESS is below 200, commonly taken as a rule-of-thumb minimum \citep{lanfear2016estimating}.
This can be useful for determining whether continuous model parameters have been sampled appropriately, though \citet{fabreti2021convergence} argue for a more stringent cutoff of 625.

These tools and guidelines do not address a central question: how well did a given MCMC run sample from the posterior distribution of tree topologies, defined as unweighted tree graphs or, more informally, as branching structures of phylogenetic trees?
Tree sampling is challenging, and previous theoretical \citep{mossel2005phylogenetic} and empirical \citep{harrington2020properties} work has demonstrated decoupling between the mixing of the log-likelihood and the sampling of the tree.
For assessing MCMC convergence of trees, standard practice involves running multiple chains and comparing the estimated ``split'' probabilities \citep{lemey2009phylogenetic}.
Splits are bipartitions of taxa (tips of the tree), correspond to edges in an unrooted phylogeny, and can be useful for comparing tree distributions even when there are many sampled tree topologies \citep{lemey2009phylogenetic}.
Each split corresponds to a particular evolutionary grouping of taxa, and so the posterior probability of a particular split measures the strength of support for a hypothesized evolutionary relationship.
Comparing split probabilities between runs addresses whether multiple chains are sampling from similar distributions, but it does not address how well those chains are mixing.

Ideally, a phylogenetic workflow should rigorously assess the quality of MCMC samples by examining both between-chain and within-chain diagnostics for all model parameters.
Here and in the rest of this paper we set aside the related issue of determining the ``burn-in'' (the point at which the chain has reached the stationary distribution); for work on this see \citep{kelly2021lagged} and \citep{fabreti2021convergence}.
For continuous model parameters, between-chain convergence can be assessed with the potential scale reduction factor (PSRF).
This is reported for each variable by the Bayesian phylogenetic inference package MrBayes \citep{ronquist2012mrbayes}, and both univariate and multivariate versions are available in \texttt{R} packages like \texttt{coda} \citep{rcoreteam,plummer2006coda}.
Mixing of continuous model parameters is easily addressed by examining the ESS of all variables in \texttt{Tracer}, or by computing a multivariate ESS \citep{vats2019multivariate} in the \texttt{R} package \texttt{mcmcse}.
For the tree, split-based comparisons such as the average (or maximum) standard deviation of split frequencies (ASDSF) are a common approach to examine between-chain convergence.
However, approaches for examining the mixing of the tree remain under-studied, and form the focus of this paper.

One promising approach to understanding MCMC mixing performance for the tree is to extend the notion of effective sample size to trees.
\citet{lanfear2016estimating} present two approaches for computing a single ESS for tree topologies which they term the approximate ESS and the pseudo-ESS (both available in the \texttt{RWTY} package \citep{warren2017rwty}).
They additionally present several simulation-based validations of their ESS measures, but do not address how those ESS measures capture Monte Carlo error.
\citet{gaya2011align} and \citet{fabreti2021convergence} alternatively consider taking the ESS individually for each split by representing it as a 0/1 random variable, which allows for standard ESS computations.
However, this approach does not account for correlation in the presence or absence of splits due to the shared tree topology.
Further, samples of trees can be summarized in many ways, and there is no obvious way to link a vector of per-split-probability ESS values to the Monte Carlo error of other key summary measures, like the probabilities of different tree topologies or the summary tree (a single tree taken to be representative of a sample of trees).

In this paper, we seek to understand the relationship between an ESS for phylogenies and phylogenetic Monte Carlo error.
This requires special consideration because trees are complex and high-dimensional objects.
Thus, classical means of linking ESS to Monte Carlo error are not directly applicable, and a considerable amount of the paper will be dedicated to making this link.
We note that we will often use ``tree'' as synonymous with the topology of the phylogenetic tree, as we focus on the challenges posed by the discrete tree structure,
and we will refer to ESS measures for phylogenies interchangeably as either tree or topological ESS measures.
The goal of such measures is to adequately describe the mixing and autocorrelation of the MCMC samples of phylogenies such that the Monte Carlo error of phylogenetic quantities can be addressed.

Classically, the ESS is defined using the variance of the sample mean.
Imagine we wish to estimate the mean, $\mu$, of some distribution with known (true) variance $\sigma^2$ given $n$ samples from this distribution.
If the $n$ samples were independent, we would have a direct link between the variance of our sample mean and the number of samples, given by $\text{Var}(\hat{\mu}) = \sigma^2/n$, where $\hat{\mu}$ is the sample mean.
However, when samples are dependent this will underestimate the true variance of the sample mean.
The ESS is a hypothetical number of independent samples which corrects for this and yields the correct variance (or standard error) of our estimator of the mean \citep{vats2021revisiting}, which can be defined by $\text{Var}(\hat{\mu}) = \sigma^2/\text{ESS}$.
The ESS is important to MCMC-based Bayesian inference because we must use correlated samples to estimate quantiles of the posterior distribution.
Run infinitely long, it is guaranteed that the posterior mean of a parameter calculated from MCMC samples will be infinitesimally close to the true posterior mean.
Given finite run lengths we must account for the resultant error to understand how precise our estimates are \citep{neal1993probabilistic,kass1998markov}.
The Markov chain central limit theorem establishes that the sampling distribution of a mean converges asymptotically to a normal distribution \citep{jones2004markov}.
Thus, if we know the ESS of a (real-valued) model parameter, we can construct confidence intervals for it, making the ESS a key quantity for Bayesian inference.

In this paper, we focus on two questions: for what aspects of estimated phylogenies might a tree ESS capture Monte Carlo error, and how might we compute an ESS of phylogenies?
Specifically, we consider whether an ESS of phylogenies is informative about the sampling variability of three summary measures (Figure~\ref{fig:conceptual_figure}):
\begin{enumerate*}[label=(\arabic*)]
  \item the probabilities of splits in the tree,
  \item the probabilities of tree topologies, and
  \item a summary tree.
\end{enumerate*}
Then, we describe several different ways to compute putative measures of tree ESS, including approaches of \citet{lanfear2016estimating} and newly-derived approaches.
These methods can be broken down into three categories:
\begin{enumerate*}[label=(\Alph*)]
  \item generalizing continuous variable ESS identities to trees,
  \item computing the ESS on a reduced-dimensional representation of the trees, and
  \item \textit{ad-hoc} methods.
\end{enumerate*}
We test these putative definitions and approaches to a tree ESS via simulations.
In these simulations, we take probability distributions on phylogenetic trees inferred from real data and run an MCMC sampler targeting this known posterior.
This allows us to compute brute-force estimates of sampling variability to which to compare the estimates based on our ESS measures.

We conclude with a case study, applying all ESS measures to six datasets of frogs and geckos from Madagascar \citep{scantlebury2013diversification}.
Time-calibrated phylogenies of these groups were originally used to test hypotheses about adaptive radiations, and the datasets were revisited by \citet{lanfear2016estimating} as benchmarks for their ESS measures.
Using these datasets, we demonstrate how tree ESS can be used to construct confidence intervals on split probabilities.
When combined with a common visual multi-chain convergence diagnostic, the split probability plot, this allows us to decompose between-chain disagreement into disagreements that can be attributed to low sample size, and disagreements that cannot.
This highlights both the importance of within-chain measures of Monte Carlo error, but also how they form only a part of the larger picture of MCMC convergence.
Indeed, our results show that multiple chains are always necessary to be confident in phylogenetic MCMC convergence.
Furthermore, these multiple-chain runs show that even with confidence intervals derived using our best ESS estimates, pairs of runs can have distinctly different estimates of split probabilities.
Taken together, our empirical and simulated results make clear the importance of directly assessing how well the tree topology mixes using a tree-specific ESS.

\section*{Methods}
We first offer a brief overview of this section before proceeding into the methods.
In the first subsection, we present a brief summary of Bayesian phylogenetic inference as a point of reference.
Next, we review background information on the ESS of one-dimensional Euclidean random variables, including how it is linked to the estimator variance of the sample mean (defined above) and how it is computed.
Then, we detail three Markov chain Monte Carlo standard error (MCMCSE) measures which we will use to assess the performance of tree ESS measures.
Specifically, we investigate Monte Carlo variability in estimated split probabilities, estimated tree probabilities, and the estimated summary tree.
Lastly, we consider four tree ESS measures in several different families of approaches, including two methods from \citet{lanfear2016estimating}.

\subsection*{Phylogenetic inference}
\begin{figure}[ht]
  \centering
  \includegraphics[width=0.825\textwidth]{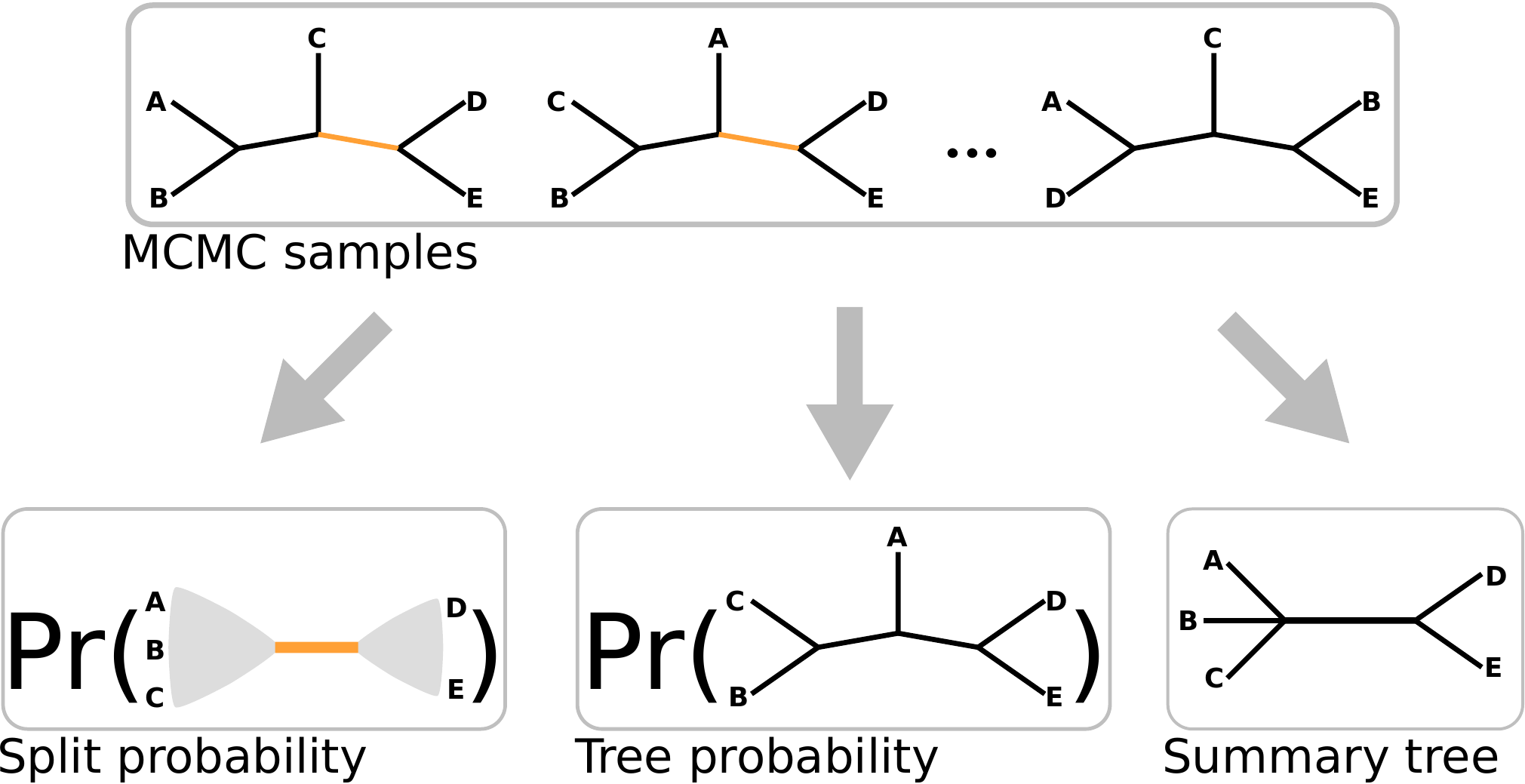}
  \caption{
    A cartoon depiction of a set of MCMC samples of a phylogenetic tree topology on five tips and the three summaries of the posterior distribution on trees we consider in this study.
    Split probabilities (left) are estimated by the fraction of MCMC samples which have a given bipartition of taxa, marginalizing out the rest of the tree topology.
    The depicted split partitions the tips into the tip-sets $\{A,B,C\}$ and $\{D,E\}$ (the edge corresponding to this split is colored orange in the MCMC samples).
    Tree probabilities (center) are estimated by the fraction of MCMC samples which have a given tree topology.
    Summary trees (right) are single trees which serve as a point estimate for all trees.
    Some, like the MRC tree, may not be fully resolved.
    The depicted MRC tree is unresolved (does not establish the particular topology with respect to) for the tip-set $\{A,B,C\}$.
  }%
  \label{fig:conceptual_figure}
\end{figure}

Phylogenetic inference is centrally concerned with the estimation of a phylogenetic tree topology, $\tau$, from character data $y$ such as DNA sequences \citep{lemey2009phylogenetic}.
Standard inference approaches require a number of continuous parameters, $\boldsymbol{\xi}$, and compute the likelihood, $\pr(y \mid \tau, \boldsymbol{\xi})$ using the Felsenstein pruning algorithm \citep{felsenstein1981evolutionary}.
The continuous parameters, $\boldsymbol{\xi}$, define the branch lengths (the time parameter of a continuous-time Markov chain process of evolution), the relative rates of change between different character states (the transition rate matrix), and possibly heterogeneity among sites in the relative rate of change (the among-site rate variation model).
In Bayesian phylogenetic inference, we use MCMC to sample from the $\pr(\tau, \boldsymbol{\xi} \mid y)$ and marginalize out $\boldsymbol{\xi}$ to obtain the distribution of interest, $\pr(\tau \mid y)$.
In this article, we consider unrooted tree topologies, where the topology depicts a set of nested evolutionary relationships without directionality (Figure~\ref{fig:conceptual_figure}).
Unrooted trees can be defined by their collection of edges, which are often referred to as splits or bipartitions of taxa.
Every fully resolved tree with $n_{\text{taxa}}$ tips contains $n_{\text{taxa}} - 3$ non-trivial splits (internal edges) and $n_{\text{taxa}}$ trivial splits (pendant edges which exist in all trees).
One common summary of the posterior distributions on trees is the majority-rule consensus (MRC) tree.
The (strict) MRC tree includes every split estimated to posterior probability greater than 0.5, and only those splits \citep{margush1981consensus,lemey2009phylogenetic}, and is widely implemented in software for both Bayesian and maximum likelihood inference (where it summarizes bootstrap support).
The set of all these splits correctly defines a tree \citep{semple2003phylogenetics}, though it may not be completely resolved (see Figure~\ref{fig:conceptual_figure} for an example of an unresolved MRC tree).
The posterior probabilities of splits (henceforth simply split probabilities) themselves are useful as measures of posterior support for particular evolutionary relationships.

\subsection*{Classical definitions of ESS}
Before we dive into the classical definition of ESS, we briefly review the Markov chain central limit theorem.
Assume we have a set of MCMC samples (with any burn-in period previously removed as needed) $X_1,\dots,X_n$, a function $g$, and use $\bar{g}_n$ to denote the sample mean using $n$ samples:
\begin{equation}
  \bar{g}_n = \frac{1}{n} \sum_{i=1}^n g(X_i).
\end{equation}
Let $\pi$ denote the posterior distribution and $\mathbb{E}_{\pi}[g] = \int_X g(x) \pi(dx)$ the true posterior expectation.
Using the notation of \citet{flegal2008markov}, the Markov chain central limit theorem tells us that the asymptotic distribution of $\bar{g}_n$ is,
\begin{equation}
  \bar{g}_n \sim \text{Normal}(\mathbb{E}_{\pi}[g],\sigma^2_g/n).
  \label{eqn:mcmc_clt}
\end{equation}
Following \citet{flegal2008markov}, we can write $\sigma^2_g$ as the sum of a series of autocovariances at increasing time lags,
\begin{equation}
  \sigma^2_g = \var_{\pi}(g(X_1)) + 2 \sum_{i=2}^{\infty}\cov_{\pi}(g(X_1),g(X_i)),
  \label{eqn:limiting_variance_definition}
\end{equation}
where we assume $X_1 \sim \pi$ and use $\pi$ as a subscript to denote this.
For the rest of this section, we will assume that $g$ is the identity function, such that $g(x) := x$, and $\bar{g}_n$ becomes the average of the MCMC samples, $\bar{g}_n = \bar{X}$.
For clarity of terminology, we will distinguish between the two variance quantities of interest.
We will use $\postvar$ to represent the variance of the posterior distribution,
\[
  \postvar = \var_{\pi}(X_1).
\]
We will refer to $\sigma^2_g$ as the limiting variance and represent it as $\limvar$.
In this simpler case, where we only care about the sample mean, Equation~\ref{eqn:mcmc_clt} tells us that
\begin{align}
  \frac{\limvar}{n} &= \var(\bar{X}). \label{eqn:mcmc_clt_simple}
\end{align}
This reveals why $\limvar$ is an important value: it is tied directly to the variance of our estimate of the mean.
Thus, it allows us to construct confidence intervals for the posterior mean computed from MCMC samples.

The effective sample size (ESS) of a set of MCMC samples is the (hypothetical) number of independent samples which would have the same variance of the sample mean \citep{liu2008monte}.
Written as an equation, this yields $\postvar / \ESS = \limvar / n$, which we can rearrange as
\begin{equation}
  \ESS = n \frac{\postvar}{\limvar}.
  \label{eqn:ess_grand_definition}
\end{equation}
While in practice we only know $n$, we can easily estimate $\postvar$ from the MCMC samples as $(1/n) \sum_{i=1}^n (X_i - \bar{X})^2$.
The difficulty in estimating the ESS lies in estimating either $\limvar$ or $\postvar / \limvar$.
In practice, there are a number of different approaches to estimating these quantities, which do not always yield the same answers \citep{fabreti2021convergence}.

\subsection*{Assessing performance of candidate ESS measures}

In this paper, we deal with situations where there is no clear theoretical basis for deriving an ESS measure.
To determine whether a putative ESS measure works, we turn to the definition of the effective sample size as the number of independent samples with the same standard deviation of the empirical mean as the standard deviation of the empirical mean obtained using MCMC samples.
Briefly, the idea of our testing setup is that we draw two sets of samples from a known target distribution.
First this is done with MCMC, producing many sets of autocorrelated samples.
Second, we estimate the ESS values from the MCMC chains using a given ESS measure, then draw that number of samples iid from the target to obtain a Monte Carlo error for that ESS measure.
We then estimate the standard error of both sets of samples.
(There will be Monte Carlo error in our estimated Monte Carlo error, but for a sufficiently large number of sets of samples it will be negligible.)
If a putative ESS measure works, then, by the definition of the effective sample size, the standard errors computed based on both sets of samples should be nearly the same.
A graphical depiction of this setup is shown in Figure~\ref{fig:graphical_methods}.
We now explain this idea in more detail, starting with the mean of a Euclidean random variable and then moving on to summaries of phylogenetic posterior distributions.

\subsubsection{Assessing the ESS for continuous parameters}
Classical ESS measures ``work'' in the sense that they allow us to correctly compute the standard error of our estimate of the posterior mean $\hat{\theta}$.
Let us review the Monte Carlo standard error of a Euclidean variable in the context of Bayesian inference via MCMC\@.
We want to estimate the true posterior mean with the mean of our $n$ MCMC samples, $\hat{\theta}$.
The standard deviation of our estimator $\hat{\theta}$ is $\sd(\hat{\theta}) = \sqrt{\E[ {(\hat{\theta} - \E[\hat{\theta}])}^2 ]}$.
This is the standard deviation of the sampling distribution, also called the standard error (SE).
In general, there is not a known closed form solution for the actual sampling distribution of $\hat{\theta}$, and thus for $\se(\hat{\theta})$.
But if we were to run $m$ independent replicate analyses, we could use a brute-force approach to estimate the \textit{true} Markov chain Monte Carlo SE (MCMCSE) as,
\[
  \semcmc(\hat{\theta}) =  \sqrt{ \frac{1}{m} \sum_{i=1}^m {( \hat{\theta}_i - \widehat{\E}[\hat{\theta}] )}^2 },
\]
where $\hat{\theta}_i$ is the estimate of $\theta$ using the $i$th set of MCMC samples and $\widehat{\E}[\hat{\theta}]$ is estimated using all $m \times n$ MCMC samples drawn.
This is an estimate of the true standard error of the mean given $n$ MCMC samples.
Imagine that we had the ability to draw samples in an identically and independently distributed (iid) manner from the true posterior.
Then, as a comparison, for each run $i = 1,\dots,m$, we could compute the effective sample size, $\ESS_i$, and draw $\ESS_i$ samples iid according to the posterior distribution $\pr(\theta \mid y)$.
This gives us a second set of samples, for which we can repeat our above brute-force calculation of the Monte Carlo error.
These $m$ sets of ``ESS-equivalent'' iid samples allow us to estimate $\seess(\hat{\theta})$, the Monte Carlo estimate of the standard error when drawing samples iid from the true posterior distribution according to the computed ESS.
We should expect to find that $\seess(\hat{\theta}) \approx \semcmc(\hat{\theta})$, because the classical ESS is derived to give us the hypothetical number of independent samples such that we have the same variance of the posterior mean, and thus the same SD and same SE.
We will use the discrepancy between $\seess(\hat{\theta})$ and $\semcmc(\hat{\theta})$ to judge the quality of various ESS measures.

\subsubsection{Assessing the ESS for trees}
When running MCMC targeting multivariate distributions, the usual approach in high dimensions is to compute the ESS separately for each univariate variable.
However, posterior distributions on phylogenetic trees are more complex than typical high-dimensional vector-valued distributions, which precludes this approach.
Instead, we will consider several different summaries of posterior distributions of trees, and the Monte Carlo SE in the estimates of those summaries.
We now write out more concretely our testing approach, which is a generalization of the above idea of how ESS measures should work.

\begin{figure}[htp]
  \centering
  \includegraphics[angle=270,width=\textwidth]{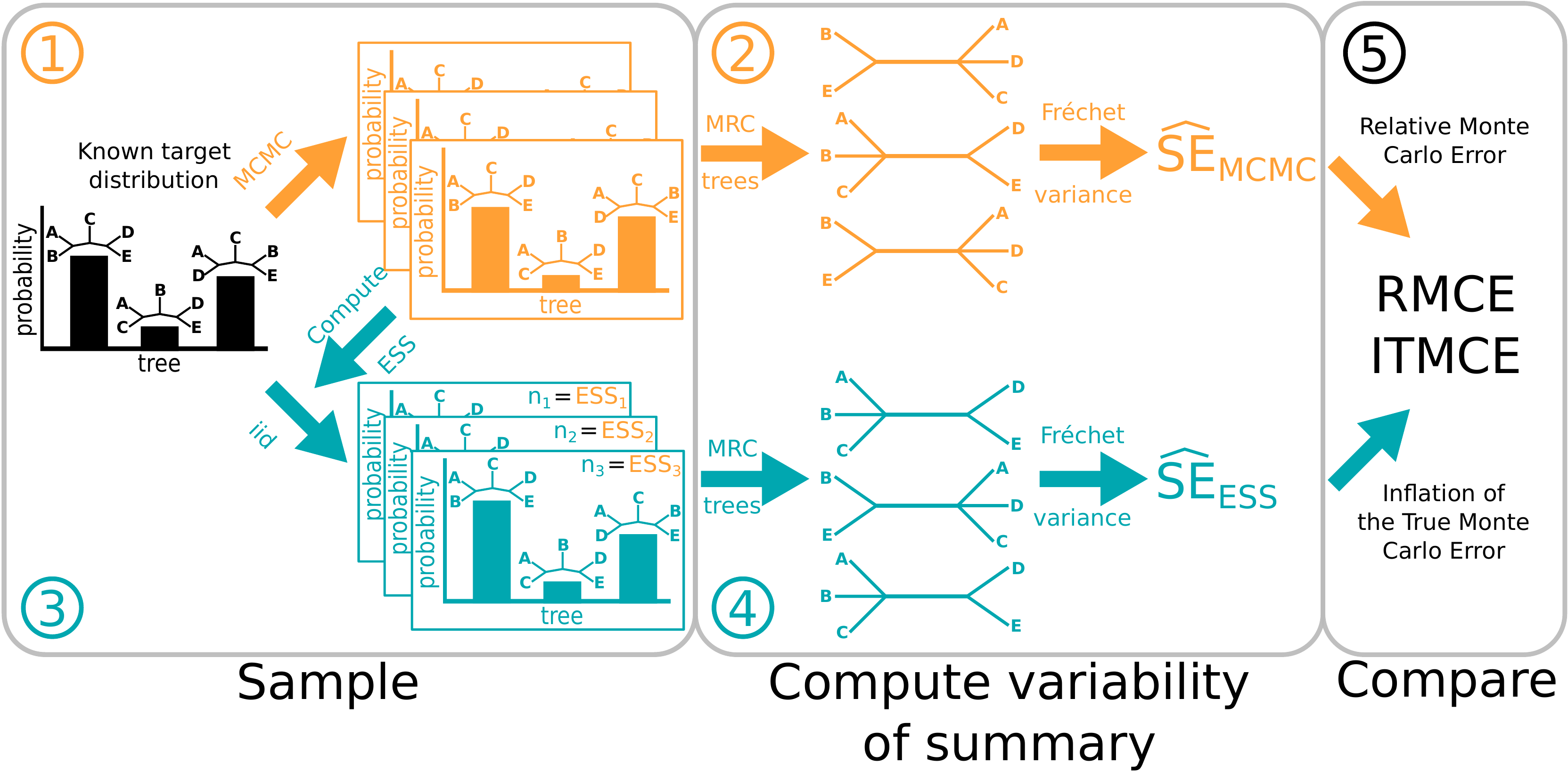}
  \caption{
    A graphical depiction of steps 1-5 of our approach to testing tree ESS measures, highlighting the case where we examine the Fr{\'e}chet standard error of the MRC tree.
    The general workflow is the same for the other cases, which calculate the standard error of split and tree probabilities.
    Steps involving MCMC are shown on top in orange, while steps involving iid samples are shown on bottom in teal.
    We begin with a (known) target distribution.
    Before we can consider Monte Carlo errors, we must sample from this distribution.
    We do this first using an MCMC algorithm (Step 1), and then by simply drawing iid samples (Step 3).
    The key is that we draw iid samples according to the effective sample size of each of the MCMC chains from Step 1.
    Here we have $m=3$ MCMC chains, so we draw $m=3$ sets of ESS-equivalent samples.
    We must then summarize each of these sets of samples.
    In this paper, we consider three distinct summaries, as depicted in Figure~\ref{fig:conceptual_figure}.
    Here we show the MRC tree as the summary, yielding $m$ MRC trees from MCMC chains (Step 2, left) and $m$ MRC trees from the ESS-equivalent samples (Step 4, left).
    The variability of these summaries is a brute-force estimate of the Monte Carlo error.
    The Fr{\'e}chet standard deviation of the $m$ MRC trees from MCMC gives us the Markov chain Monte Carlo standard error, $\semcmc$ (Step 2, right).
    The Fr{\'e}chet standard deviation of the $m$ MRC trees from the ESS-equivalent iid samples gives us the $\seess$ (Step 4, right).
    (For split and tree probabilities, the $\semcmc$ and $\seess$ are the usual standard error rather than a Fr{\'e}chet standard error.)
    If an ESS measure works, then it is informative about the number of independent samples which would have the same Monte Carlo error as a given set of MCMC samples.
    Thus, we should expect the $\semcmc$ and $\seess$ to be similar, and the last step is to compare them.
    We use the RMCE and ITMCE, which are measures of relative difference.
  }%
  \label{fig:graphical_methods}
\end{figure}

We will assume that we have $m$ phylogenetic MCMC runs, and that each has $n$ samples.
We will assume for the moment that we have both a topological ESS measure and a way to draw tree topologies iid from the posterior distribution, $\pr(\tau \mid y)$.
Let $\summary$ be a summary of $n$ tree samples (possibly obtained via MCMC) from posterior distribution $\pr(\tau \mid y)$.
For example, $\summary$ could be the probability of a particular split or tree topology, or a summary tree such as the MRC tree.
Our approach to testing topological effective sample sizes can then be written out as follows (Figure~\ref{fig:graphical_methods}):
\begin{enumerate}[label=\arabic*.]
  \item Run $m$ independent MCMC chains.
  \item Using the chains from step 1, compute the brute-force estimate of the \emph{true} Markov chain Monte Carlo SE, $\semcmc(\summary)$.
  \item For each chain $i = 1, \dots, m$, compute that chain's ESS, $\ESS_i$, and draw $\ESS_i$ trees independently from the true posterior distribution $\pr(\tau \mid y)$. Round $\ESS_i$ to the nearest integer such that we draw an integer number of trees.
  \item Using the set of ESS-equivalent trees drawn in step 3, compute the ESS-equivalent estimate of the MCMCSE, $\seess(\summary)$.
  \item Compare $\seess(\summary)$ and $\semcmc(\summary)$.
\end{enumerate}

Generally speaking, if a putative measure of the topological ESS works with respect to an MCMCSE measure, then $\seess(\summary) \approx \semcmc(\summary)$, and the closer the ESS-based estimate of the SE is, the better that ESS estimator works.
If we have overestimated the ESS, we will find $\seess < \semcmc$ (as we have drawn too many independent samples and thus the SE of the estimate from them will be too small).
On the other hand, if we have underestimated the ESS, we will find $\seess > \semcmc$ (as we have drawn too few independent samples and thus the SE of the estimate from them will be too large).
As the scales of the Monte Carlo SEs are not inherently meaningful, we instead measure the relative error in the estimated Monte Carlo error (the relative Monte Carlo error, RMCE),
\begin{equation}
  \text{RMCE} = \frac{\semcmc - \seess}{\semcmc}.
\end{equation}
This quantity is negative when we have underestimated the ESS, positive when we have overestimated the ESS, and tells us the relative proportion by which we have mis-estimated the MCMCSE using our ESS measure.
In some cases, however, a more useful measure is given by the inflation (or deflation) of the true Monte Carlo error (ITMCE), which is defined by the relation,
\begin{equation}
  \text{ITMCE} = \frac{\semcmc}{\seess}.
\end{equation}
This value measures how inflated the \textit{true} Monte Carlo error is relative to the estimated Monte Carlo error, which can be seen by rewriting the above as $\semcmc = \seess \times \text{ITMCE}$.
In essence, the ITMCE measures how much wider the true confidence intervals should be relative to our estimated intervals.
The ITMCE is greater than one when we have overestimated the ESS, and less than one when we have underestimated it.

The RMCE and ITMCE are complementary; the RMCE is useful to quantify when ESS is underestimated, while the ITMCE clearly shows when ESS is overestimated.
To see this, start by noting that by definition $\semcmc$ and $\seess$ are non-negative.
Thus, $-\infty < \text{RMCE} \leq 1$ and $0 \leq \text{ITMCE} < \infty$.
Underestimation of the ESS is represented by $\text{RMCE} < 0$ and $\text{ITMCE} < 1$.
In this underestimation case the RMCE may take any value in $(-\infty,0]$, while the ITMCE compresses this regime into the range $[0,1]$.
Overestimation of the ESS is represented by $\text{RMCE} > 0$ and $\text{ITMCE} > 1$.
In the overestimation case the RMCE is compressed, stuck in $[0,1]$, while the ITMCE may take any value in $[0,\infty)$.
Thus, the RMCE has more effective resolution when the ESS is underestimated, while the ITMCE has more resolution when the ESS is overestimated.

In the introduction, we outlined three quantities a tree ESS might reflect; now we will more rigorously define three MCMCSE measures based on these, noting that they are all measures of the Monte Carlo standard error (MCMCSE).
Broadly, we contemplate whether a tree ESS measure might reflect the standard deviations of our estimates of split probabilities, tree probabilities, or a summary tree.

\subsubsection*{ESS and split probabilities}
We may wish for ESS of trees to reflect the quality of our estimates of the split probabilities.
Let us denote the probability of this a particular split $p$, the estimate of the split probability from the $i$th MCMC run as $\hat{p}_{\text{MCMC}}^i$ (Step 1).
Then, we can estimate the true MCMCSE of the split probability as (Step 2),
\[
  \semcmc(\hat{p}^i) = \sqrt{ \frac{1}{m} \sum_{i=1}^m {(\hat{p}_{\text{MCMC}}^i - \widehat{\E}[\hat{p}_{\text{MCMC}}])}^2 },
\]
where $\widehat{\E}[\hat{p}_{\text{MCMC}}]$ is the average probability of the split across all $m$ chains.
Let us denote the estimate of the split probability computed from $\ESS_i$ tree topologies drawn iid from the posterior distribution as $\hat{p}_{\text{MCESS}}^i$ (Step 3, we discuss how to do this in the section ``Faking phylogenetic MCMC'').
Then, using the ESS-equivalent tree samples, we can compute $\seess$ (Step 4),
\[
  \seess(\hat{p}^i) = \sqrt{\frac{1}{m} \sum_{i=1}^m {(\hat{p}_{\text{MCESS}}^i - \widehat{\E}[\hat{p}_{\text{MCESS}}])}^2 },
\]
and then we can compare $\semcmc(\hat{p}^i)$ and $\seess(\hat{p}^i)$ (Step 5).

We note two places that $\semcmc(\hat{p}^i)$ appears in existing phylogenetic practice.
The standard error of the probability of a split is better known as the standard deviation of the split probability (SDSF).
The average across all splits is called the ASDSF and is commonly used to diagnose the convergence of multiple MCMC chains.
MrBayes also uses the $\semcmc(\hat{p}^i)$ from the (usually 2-4) independent chains run to quantify the Monte Carlo error in the consensus trees.

\subsubsection*{ESS and tree probabilities}
We may wish for ESS of trees to reflect the quality of our estimates of the tree topology probabilities.
As these are probabilities, we can compute $\semcmc(\hat{p}^i)$ and $\seess(\hat{p}^i)$ exactly as for split probabilities.

\subsubsection*{ESS and the summary tree}
We may wish for the ESS of trees to reflect the quality of our estimates of the summary tree.
This is less straightforward than with the probabilities of splits or tree probabilities, as trees are non-Euclidean objects.
Hence, we turn to a Fr{\'e}chet generalization of MCMCSE (we discuss Fr{\'e}chet generalizations more thoroughly in the section Calculating the ESS by generalizing previous definitions).
In this paper, we focus on the majority rule consensus (MRC) tree as the summary tree.
As our measure of variability between topologies, we consider the squared Robinson-Foulds (RF) distance \citep{robinson1981comparison} (which we describe in more detail below).
Thus, our MCMCSE measure here is a given by,
\[
  \semcmc(\widehat{\tau}) = \sqrt{ \frac{1}{m} \sum_{i=1}^m {d(\widehat{\tau}_{\text{MCMC}}^i,\widehat{\E}[\widehat{\tau}_{\text{MCMC}}])}^2 },
\]
where $d(\cdot,\cdot)$ is the RF distance.
Here, we define $\widehat{\tau}_{\text{MCMC}}^i$ to be the MRC tree for the $i$th MCMC run, and $\widehat{\E}[\widehat{\tau}_{\text{MCMC}}]$ to be the MRC tree obtained by pooling all $m$ runs to estimate split frequencies.
Then, using the ESS-equivalent tree samples (Step 3), we can compute $\seess$ (Step 4) analogously,
\[
  \seess(\widehat{\tau}) = \sqrt{ \frac{1}{m} \sum_{i=1}^m {d(\widehat{\tau}_{\text{ESS}}^i,\widehat{\E}[\widehat{\tau}_{\text{ESS}}])}^2 }.
\]

\subsubsection*{Faking phylogenetic MCMC}
Our goal is to use the testing procedure outlined in the section ``Assessing performance of candidate ESS measures'' to test whether tree ESS measures work appropriately.
This requires that we have a distribution from which we can both draw MCMC samples (Step 1), to compute $\semcmc$ (Step 2), and draw iid samples (Step 3), to compute $\seess$ (Step 4).
Because it is impossible to draw iid samples from a Bayesian phylogenetic posterior distribution---else we would not need MCMC in the first place---we set up a simulated (or ``fake'') phylogenetic MCMC\@.
In brief, our setup first uses MCMC samples from real-data phylogenetic analyses to define a categorical distribution on phylogenetic tree topologies.
We can then use phylogenetic MCMC moves to draw MCMC samples from this distribution (Step 1), and using those we can estimate $\semcmc$ (Step 2).
Because our target distribution is based on real-data phylogenetic posterior distributions, the resulting Monte Carlo error should be representative of the Monte Carlo error of real-data phylogenetic posterior distributions.
Importantly, because we have a known, categorical, target distribution, we can draw iid samples from the categorical distribution (Step 3), and using those can estimate $\seess$ (Step 4).
By comparing $\semcmc$ and $\seess$ (Step 5), we can assess whether the tree ESS measures work on target distributions that are known exactly yet which share key features of real phylogenetic posterior distributions.
We now outline in more detail how we obtain this categorical distribution and run phylogenetic MCMC to approximate it.

We start with an estimate of tree probabilities based on MCMC samples from a real phylogenetic posterior distribution.
In this paper we re-use posterior distributions inferred by \citet{whidden2020systematic} on a set of standard benchmark datasets, though any phylogenetic MCMC run could be used.
This gives us a vector of trees $\btau$ and an associated estimate of the probability mass function $\phat(\btau)$.
As the number of trees in $\btau$ may be prohibitively large, one may wish to take only the top 95\% highest posterior density (HPD) set of trees, or other such subset.
Depending on the real-data MCMC algorithm which produced $\btau$, and depending on the truncation, these trees may not be connected with respect to the MCMC proposal used for the fake phylogenetic MCMC.
Thus, we keep only the largest connected subset of $\btau$.
We then take this set of trees and their estimated probabilities to be the \textit{true} probabilities for our fake MCMC (we re-normalize the probabilities if trees are removed, though this is not strictly necessary).
We define any tree $\Psi \not \in \btau$ to have zero probability.

To run a fake phylogenetic MCMC sampler on this known target distribution, we use nearest neighbor interchange (NNI, the interchanging of two subtrees across an edge in the tree) proposals.
We first construct a list of all NNI neighbors for each tree, such that $N(\Psi)$ is the set of all neighbors of $\Psi$.
Then, to initialize the MCMC run, we randomly draw a starting state according to $\phat(\btau)$.
At each step thereafter, if the current state for the MCMC is $\Psi$, we propose a tree $\Psi^*$ uniformly at random from $N(\Psi)$ (excluding $\Psi$), and the acceptance probability for the move is $\min(1,\phat(\Psi^*)/\phat(\Psi))$.
This move is symmetric and has a Hastings ratio of 1 \citep{lakner2008efficiency}.

Although this is by necessity an artificial setup, we believe that it is an improvement over previous benchmarking exercises.
The simulation approach of \citet{lanfear2016estimating} is based on accepting all proposed MCMC moves and assuming all trees were uniformly probable.
They consider bimodal distributions by mixing sets of trees based on a small number of MCMC moves from distant starting points.
In contrast, our approach is based on real-data posterior distributions of trees.
This allows for multimodality, as well as uneven connectivity of trees, such that some trees have many neighbors (with notable posterior probability) and others have few, which can make exploration difficult as some trees may be hard to reach.
Additionally, using real posterior probabilities replicates unevenness, such that some datasets may be particularly rugged and others relatively flat \citep{hohna2012guided}.
Importantly, our approach includes rejected proposals, an important feature given that acceptance rates for phylogenetic MCMC proposals are notoriously low.
Our setup allows for complete exploration of the target distribution with chains of sufficient length, whereas with a uniform distribution on all trees this is a practical impossibility.
As we can initialize the chain from the target distribution, we can ignore burn-in and focus on mixing while accommodating these realistic features.
We can improve the speed and memory requirements by tweaking the proposal step and neighbor tracking, which we discuss in the Supplemental Materials.
We implement the functions needed to run fake phylogenetic MCMC on arbitrary phylogenetic posterior distributions in the R package \texttt{treess}.

\subsection*{Computing a tree ESS}
In this paper, we consider four different tree ESS measures, which fall into the following three categories:
\begin{itemize}
  \item ESS measures based on Fr{\'e}chet generalizations of Equation~\ref{eqn:ess_grand_definition} to trees (we discuss Fr{\'e}chet generalizations in the next section)
  \begin{itemize}
    \item[] 1. The Fr{\'e}chet Correlation ESS ($\frechetCorrelationESS$)
  \end{itemize}
  \item ESS measures based on projecting the tree to a single dimension and computing the ESS of that using standard univariate approaches
  \begin{itemize}
    \item[] 2. The median pseudo-ESS ($\medianPseudoESS$)
    \item[] 3. The minimum pseudo-ESS ($\minPseudoESS$)
  \end{itemize}
  \item \textit{Ad-hoc} ESS measures
  \begin{itemize}
    \item[] 4. The approximate ESS ($\approximateESS$)
  \end{itemize}
\end{itemize}
In the next sections, we present short sketches of our new approaches and the\\ $\medianPseudoESS$ of \citet{lanfear2016estimating}.
We refer readers to \citet{lanfear2016estimating} for explanation of the $\approximateESS$.
In the supplement, we describe six additional measures, as well as a more detailed derivation of the $\frechetCorrelationESS$.
We will use the notation $\btau$ for a vector of phylogenies, $d(\tau_i,\tau_j)$ to denote the distance between trees $i$ and $j$, and $\boldsymbol{D} = \{D_{ij}\} = \{d(\tau_i,\tau_j)\}$ for the distance matrix of all trees in $\btau$.

All tree ESS measures presented depend on the distance matrix between all samples in the posterior.
While in general any tree distance can be used, in this paper we focus on the Robinson-Foulds (RF) distance~\citep{robinson1981comparison}.
The RF distance considers an unrooted tree as a collection of splits, and measures the number of splits by which two trees differ.
For trees $\tau_A$ and $\tau_B$, let $\mathcal{A}$ be the set of splits in $\tau_A$ and $\mathcal{B}$ the set of splits in $\tau_B$, and further denote by $\mathcal{A}^c$ the set of splits not in $\tau_A$ and by $\mathcal{B}^c$ the set of splits not in $\tau_B$.
The RF distance is $d_{\text{RF}} = |\mathcal{A} \cap \mathcal{B}^c| + |\mathcal{B} \cap \mathcal{A}^c|$.
That is, the RF distance is the number of splits in $\tau_A$ but not $\tau_B$ plus the number of splits in $\tau_B$ but not $\tau_A$.
If both trees are fully resolved, then any split which is in $\tau_A$ but not in $\tau_B$ must correspond to a split which is in $\tau_B$ but not in $\tau_A$, so $|\mathcal{A} \cap \mathcal{B}^c| = |\mathcal{B} \cap \mathcal{A}^c|$, and $d_{\text{RF}} = 2 \times |\mathcal{A} \cap \mathcal{B}^c|$.
If one or both trees are unresolved, which may be the case with consensus trees, a split present in $\tau_A$ may simply be absent in $\tau_B$ without implying the presence of a split which is in $\tau_B$ but not in $\tau_A$.
For two resolved trees on $n_{\text{taxa}}$ taxa, the maximum RF distance is twice the number of internal edges, or $2 n_{\text{taxa}} - 6$.
When one or both trees are unresolved, the maximum distance is lowered.
While there are some limitations of RF distance, it has many benefits, primarily that it is interpretable and easily computable in a wide variety of software.
Additionally, RF distance is related to tree space exploration via NNI moves, as two (fully resolved) trees separated by one NNI are separated by an RF distance of 2.

\subsubsection*{Calculating the ESS by generalizing previous definitions}
In this section, we will attempt to generalize the definition of ESS in Equation~\ref{eqn:ess_grand_definition} using concepts borrowed from the notions of Fr{\'e}chet mean and Fr{\'e}chet variance \citep[see for example][and references therein]{dubey2019frechet}.
For this and all other measures presented in this paper, we provide detailed derivations in the supplement.

The Fr{\'e}chet mean and variance generalize the concepts of means and variances to other complete metric spaces.
Where the variance is the average squared deviation from the mean, the Fr{\'e}chet variance is the average squared distance from the Fr{\'e}chet mean.
In the case where $X$ is continuous and one-dimensional and the Euclidean distance is chosen, the Fr{\'e}chet mean is the classical mean and the Fr{\'e}chet variance is the classical variance.
We note that the Fr{\'e}chet mean and variance of unrooted phylogenies with branch lengths have been studied previously \citep[see][and references therein]{brown2020mean} using Billera-Holmes-Vogtmann (BHV) space \citep{billera2001geometry}.
However, in this paper we are purely interested in topological features and so use the purely topological RF distance metric.
RF distances reflect the most challenging part of phylogenetic MCMC, the tree topology itself, and are thus linked directly to topological quantities of interest, like summary trees and split probabilities.
BHV distances are somewhat less interpretable (they summarize both branch length and topological differences, thus small distances can reflect large topological changes) and are computationally burdensome.

The Fr{\'e}chet correlation ESS, or $\frechetCorrelationESS$, is a generalization of what we will call the sum-of-correlations ESS, which we will now review in the case of a single continuous random variable $X$.
The autocorrelation function defines the correlation between samples at times separated by lag $s$, and can be related to the autocovariance \citep{vonstorch2001statistical},
\[
  \rho_s = \cor(X_t,X_{t+s}) = \frac{\cov(X_t,X_{t+s})}{\postvar}.
\]
This means we can re-write Equation~\ref{eqn:limiting_variance_definition} as,
\[
  \limvar = \postvar \times \Bigg( 1 + 2 \sum_{s=1}^{\infty}\cor(X_0,X_s) \Bigg) = \postvar \times \Bigg( 1 + 2 \sum_{s=1}^{\infty}\rho_s \Bigg).
\]
Re-arranged, this can be combined with Equation~\ref{eqn:ess_grand_definition} to get,
\[
  \ESS = \frac{n}{1 + 2\sum_{s=1}^{\infty} \rho_s}.
\]
In practice, we run into difficulties if we apply this definition naively.
As there are fewer MCMC samples separated by large time lags, as $s$ gets larger our estimates $\hat{\rho}_s$ get noisier, and for odd time lags it is possible to have $\rho_s < 0$, meaning we cannot smooth the estimates without care.
Following \citet{vehtari2021rank}, we can overcome these limitations in several steps.
First, they suggest summing adjacent correlations, defining $\hat{P}_{s'} = \hat{\rho}_{2s'} + \hat{\rho}_{2s'+1}$, which guarantees that $\hat{P}_{s'} > 0$.
The estimated $\hat{P}_{s'}$ may not be monotonically decreasing, but the actual curve $P_{s'}$ should be, so \citet{vehtari2021rank} smooth the curve by setting $\hat{P}_{s'}$ to the minimum of all preceding values.
We then find the largest time lag $k$ such that $\hat{P}_{s'} > 0$ for all $s' = 1,\dots,k$, and sum the series only up to this time, yielding the final estimator (note the $-1$ appears because we start indexing at $0$ and $\hat{\rho}_0 = 1$),
\begin{equation}
  \widehat{\ESS} = \frac{n}{ -1 + 2\sum_{s'=0}^{k} \hat{P}_{s'}}.
  \label{eqn:ess_correlation_practical}
\end{equation}
We will refer to this as the sum-of-correlations ESS\@.

Employing Equation~\ref{eqn:ess_correlation_practical} to estimate an effective sample size for trees requires us to define a correlation between vectors of trees.
It can be shown that, for Euclidean variables $X$ and $Y$, the covariance is related to Euclidean distances.
If we denote the Euclidean distance between $X$ and $Y$ with $\Delta = d(X,Y)$, one can derive $\cov(X,Y) = 1/2 \times (\var(X) + \var(Y) - \E[\Delta^2]  + (\E[X] - \E[Y])^2)$ (see supplement).
In a setting where $X$ and $Y$ represent samples from the same MCMC run at some time lag $t$, we might expect $(\E[X] - \E[Y])^2$ to be negligible so long as sufficient burn-in has been discarded.
In this case, we can instead approximate the covariance as $\cov(X,Y) \approx \frac{1}{2}(\var(X) + \var(Y) - \E[\Delta^2])$.
By replacing Euclidean distances with arbitrary distance metrics and means/variances with Fr{\'e}chet means/variances, we can use these relationships to define a generalization of the covariance and thus of the (auto)correlation.
We propose to estimate the autocorrelation between MCMC samples of trees at time lag $s$ as:
\begin{align}
  \hat{\rho}_s = \frac{ \frac{1}{2} (\widehat{\var}(\btau_{t}) + \widehat{\var}(\btau_{t+s}) - \widehat{\E}[\Delta^2]) }{ \sqrt{\widehat{\var}(\btau_{t}) \widehat{\var}(\btau_{t+s})} }.
  \label{eqn:frechet_autocor}
\end{align}
Here, $\btau_{t}$ and $\btau_{t+s}$ represent vectors of MCMC samples of trees separated by a time lag of $s$, $\widehat{\var}$ is the estimated Fr{\'e}chet variance based on RF distance, and $\widehat{\E}[\Delta^2]$ is the average squared RF distance between tree vectors $\btau_{t}$ and $\btau_{t+s}$.
Once we obtain our estimates $\hat{\rho}_s$, we use the sum-of-correlations approach to estimate the ESS (Equation~\ref{eqn:ess_correlation_practical}).
A more thorough derivation of Equation~\ref{eqn:frechet_autocor}, including how to compute $\widehat{\var}(\btau)$, is available in the Supplementary Materials.

\subsubsection*{Approaches to calculating the ESS by projecting the tree to a single dimension}
The pseudo-ESS of \citet{lanfear2016estimating} projects trees to a single dimension and computes the ESS of this transformed variable.
It is computed in several steps.
First, $r$ ``reference'' trees are selected uniformly from among all $n$ posterior trees.
For the $i$th reference tree, we turn the sequence of trees sampled by the MCMC into a real-valued sequence by computing the distance between each such tree and the reference.
We can then compute the ESS of this real-valued sequence to obtain an $\ESS_i$.
Lastly, we must summarize the vector of estimates, $\ESS_1,\dots,\ESS_r$.
\citet{lanfear2016estimating} use the median as a point estimate and suggest reporting the 95 percentile range as well.
We depart slightly from the original approach by using all $n$ samples in the posterior as reference trees rather than choosing a subset.
As the ESS is a single number (the hypothetical size of an equivalent set of iid samples), we do not examine ranges but rather focus on point estimates.
What \citet{lanfear2016estimating} refer to as the pseudo-ESS (the median of the $n$ distinct ESS values), we will refer as the $\medianPseudoESS$\@.
In addition to the median, as a potentially conservative estimator we also investigate the performance of the minimum of the ESS over reference trees, which we call the $\minPseudoESS$\@.
For both of these approaches, we first compute the dimension-reduced representation and then use the \texttt{R} package \texttt{coda} to compute the ESS\@.
In the Supplemental Materials, we present three additional approaches to computing the tree ESS using projection, and a discussion of the \texttt{coda} approch to computing the ESS\@.

\section*{Results}

\subsection*{Tree ESS measures reflect Monte Carlo variability in split probabilities, tree probabilities, and the summary tree}

\begin{figure}
  \centering
  \begin{subfigure}[ht]{\textwidth}
    \centering
    \includegraphics[width=\textwidth]{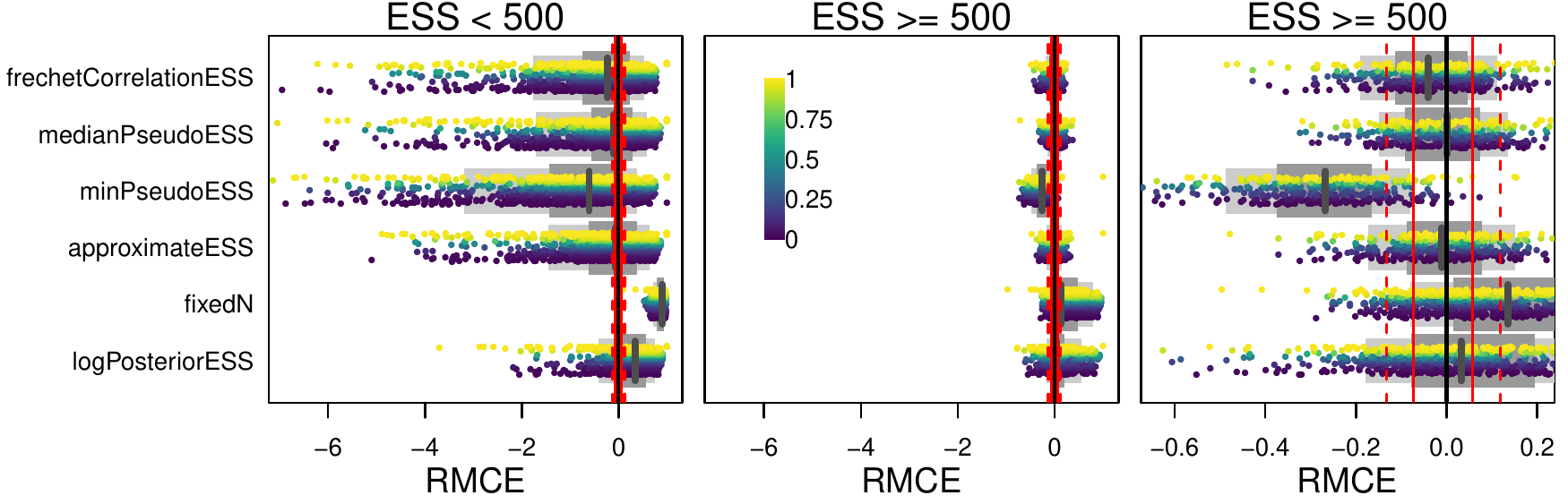}
  \end{subfigure}
  \begin{subfigure}[ht]{\textwidth}
    \centering
    \includegraphics[width=\textwidth]{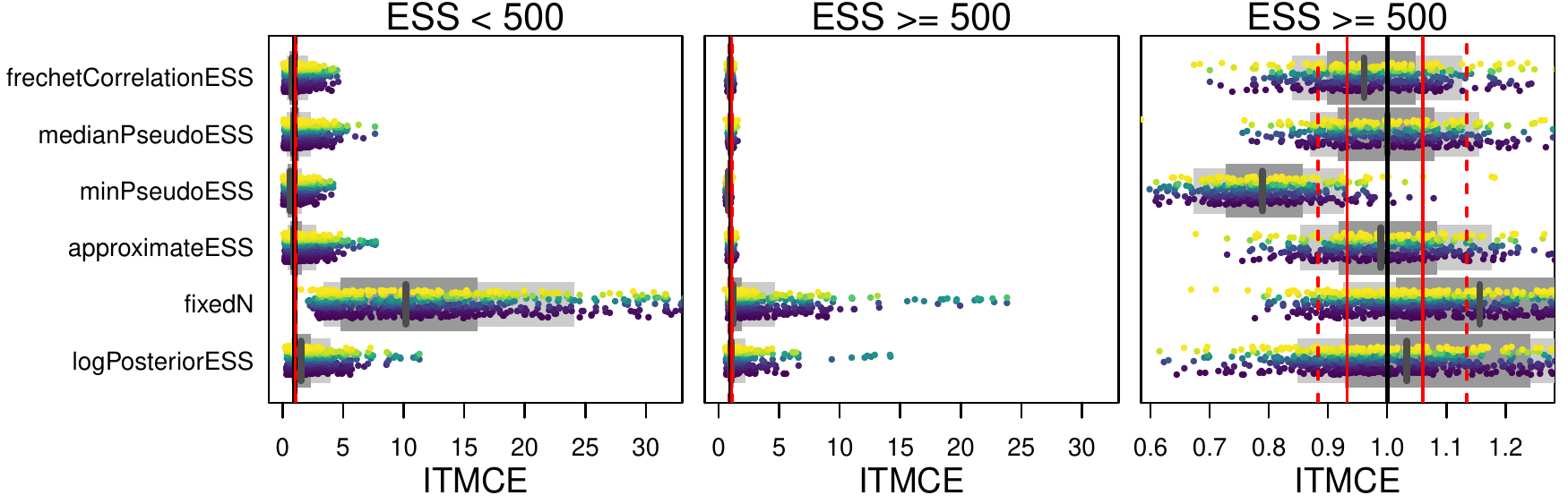}
  \end{subfigure}%
  \caption{
    The RMCE ($(\semcmc - \seess) / \semcmc$) and ITMCE ($\semcmc / \seess$) for split probabilities for all topological ESS measures and all 45 combinations of 9 datasets and 5  run lengths.
    Each point is the RMCE or ITMCE for a single split from one of the 45 simulations, colored by their estimated probabilities (see scale bar in top middle panel).
    Only splits with a posterior probability of at least 0.01 are shown.
    The two right panels are the same except for the scale of the $x$-axis.
    Whether a simulation appears in the left or right panels is determined by the estimated average ESS of each of the 45 simulations (the ``\textsf{ESS}'' in the column label), such that all splits from a simulation with average $\frechetCorrelationESS$ of 100 would appear in the left panel, while all splits from a simulation with an average Fr{\'e}chet correlation ESS of 600 would appear in the right two panels.
    As $\textsf{fixedN}$ always assumes $\ESS = 1000$, for this row we split by the number of MCMC iterations run, with the left panel including $10^3$ and $10^4$, and the right panel $10^5$, $10^6$, and $10^7$.
    The thinner light grey bar below the points shows the 80\% quantile range, the thicker dark grey bar the 50\% quantile range, and the vertical grey line is the median.
    Ideal performance is RMCE = 0 and ITMCE = 1 (perfect estimation of the Monte Carlo SE).
    As references we have plotted a solid black line for perfect performance, while the dashed (solid) red lines represent the 80\% quantile range (50\% quantile range) from the univariate $\operatorname{Normal}(0,1)$ experiment.
    The best performance that might reasonably be expected of a tree ESS measure would match the $\operatorname{Normal}(0,1)$ experiment, and thus have the grey line on the solid black line, the thinner light grey bar align with the dashed red lines, and the thicker dark grey bar align with the solid red lines.
    RMCE $<$ 0 (ITMCE $<$ 1) implies underestimating the ESS, while RMCE $>$ 0 (ITMCE $>$ 1) implies overestimating the ESS.
    The log-posterior ESS and assuming $\ESS = n$ tend to overestimate the ESS for splits, often substantially, while most tree ESS measures are much closer to the truth.
    }%
  \label{fig:benchmarks_splits}
\end{figure}

\begin{figure}
  \centering
  \begin{subfigure}[ht]{\textwidth}
    \centering
    \includegraphics[width=\textwidth]{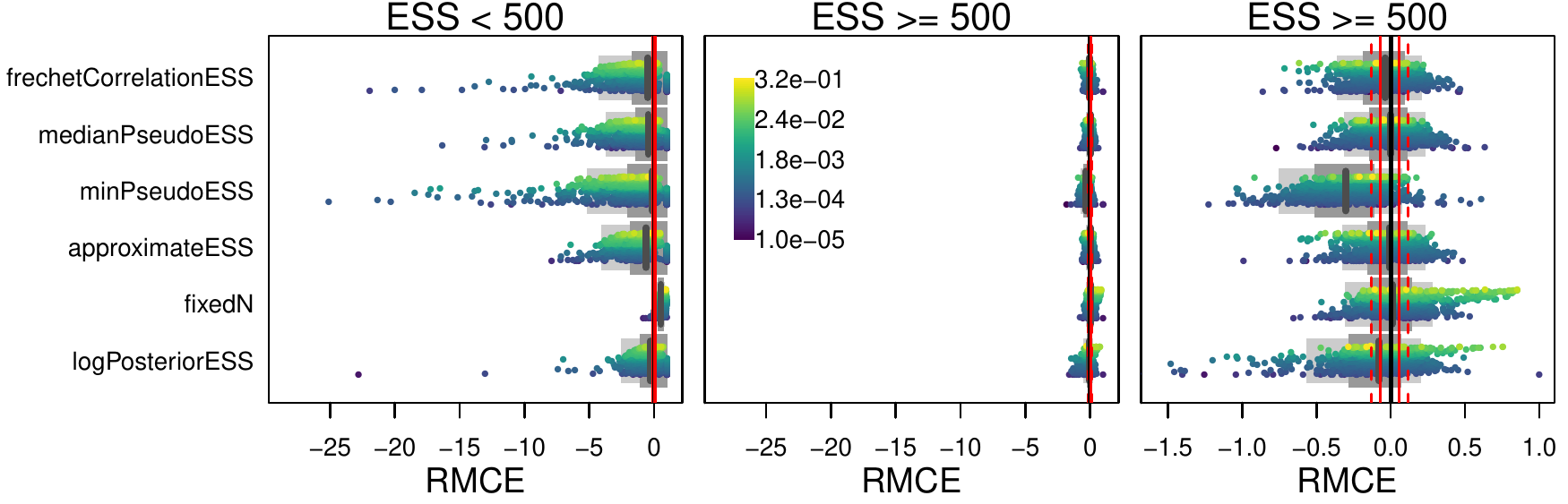}
  \end{subfigure}
  \begin{subfigure}[ht]{\textwidth}
    \centering
    \includegraphics[width=\textwidth]{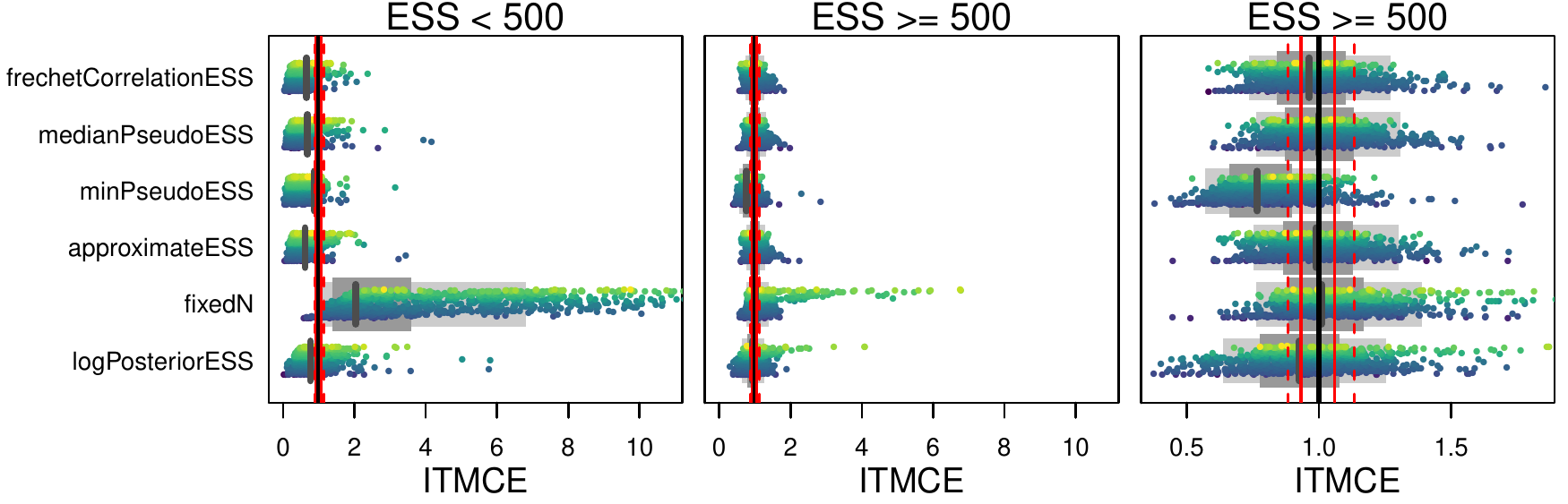}
  \end{subfigure}%
  \caption{
    The RMCE ($(\semcmc - \seess) / \semcmc$) and ITMCE ($\semcmc / \seess$) for topology probabilities for all topological ESS measures and all 45 dataset by run length combinations.
    Each point is the RMCE or ITMCE for a single tree topology from one of the 45 simulations, colored by their estimated probabilities (see scale bar in top middle panel).
    As there are too many distinct topology probabilities (nearly 100,000 across all 45 simulations), we plot only 1000 per row, preferentially keeping the highest-probability trees as these are the ones that contribute most to summary trees.
    For more explanation, see Figure \ref{fig:benchmarks_splits} caption.
    }%
  \label{fig:benchmarks_tree_probs}
\end{figure}

\begin{figure}
  \centering
  \begin{subfigure}[ht]{\textwidth}
    \centering
    \includegraphics[width=0.825\textwidth]{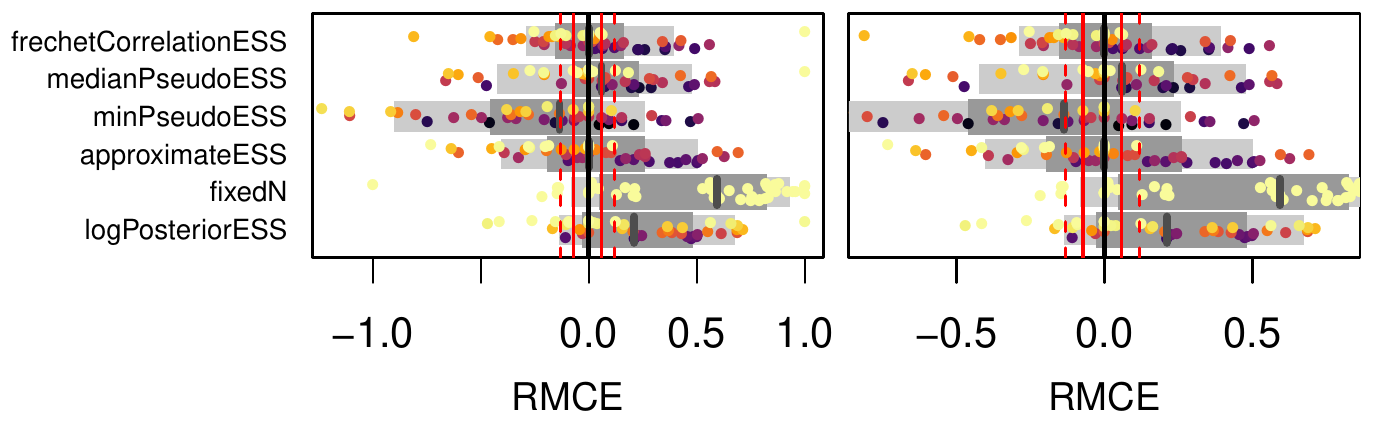}
  \end{subfigure}
  \begin{subfigure}[ht]{\textwidth}
    \centering
    \includegraphics[width=0.825\textwidth]{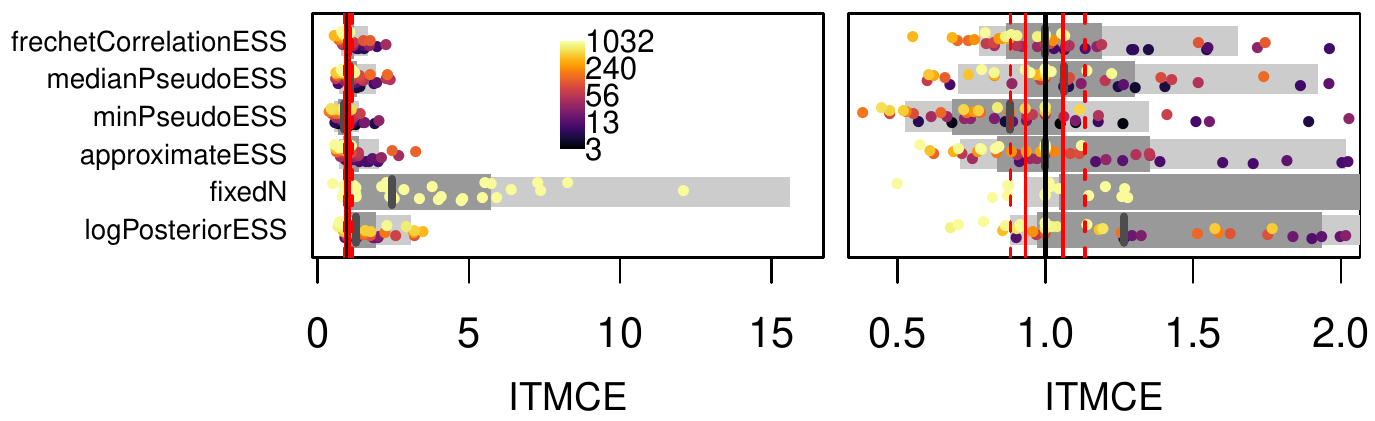}
  \end{subfigure}%
  \caption{
    The RMCE ($(\semcmc - \seess) / \semcmc$) and ITMCE ($\semcmc / \seess$) for the majority-rule consensus (MRC) tree for all topological ESS measures and all 45 dataset by run length combinations.
    Each point represents the RMCE or ITMCE for the summary tree from one of the 45 simulations.
    The standard error for the MRC tree is a Fr{\'e}chet Monte Carlo SE, rather than a classical Euclidean Monte Carlo SE\@.
    The right and left panels are the same except for the scale of the $x$-axis.
    As there are only 45 MRC trees, points are colored by the estimated average ESS (see scale bar in bottom left panel), and all 45 are plotted together rather than separated according to an ESS threshold as in Figures~\ref{fig:benchmarks_splits} and \ref{fig:benchmarks_tree_probs}.
    For more explanation, see Figure \ref{fig:benchmarks_splits} caption.
    }%
  \label{fig:benchmarks_MRC}
\end{figure}

Now that we have several possible summaries of how an ESS measure could work and a number of potential ESS measures, we can determine whether any of the potential tree ESS measures perform well with respect to any of the summaries.
We use a suite of common test datasets in phylogenetics \citep[\textit{e.g.}][]{lakner2008efficiency,hohna2012guided,larget2013estimation,whidden2015quantifying}, commonly referred to as DS1-DS8 and DS10.
We re-use previous MCMC analyses of these datasets consisting of 10 independent MCMC chains of 1 billion iterations, sampled every 100 iterations and pooled into a set of 100 million tree samples \citep{whidden2020systematic}.
As the full set of trees is too large to work with efficiently, we consider a subset of these trees which we term the ``best connected trees.''
To obtain this subset, we start with either the 95\% HPD set or the first 4096 trees in the 95\% HPD set, whichever is smaller.
Then, we ensure that every tree can be reached from every other tree using only NNI moves, and keep the largest connected subset.
In our setting this required us to drop at most 20\% of the trees and lose a maximum posterior probability mass of 0.065.
For each of these 9 real-data posterior distributions, we run simulated MCMC analyses (as described in the section ``Faking phylogenetic MCMC'') for run lengths of \{$10^3$, $10^4$, $10^5$, $10^6$, $10^7$\}, in each case thinning in order to retain 1000 tree samples.
Acceptance rates for the NNI moves are typical of those observed in real-data MCMC analyses, ranging from $\approx3\%$ (DS3) to $\approx11\%$ (DS6).
Since we do not need to integrate out any of the continuous parameters such as branch lengths, and since we start each chain in the stationary distribution (avoiding burn-in), these shorter run lengths should be equivalent to longer runs on real datasets.
For each dataset and run-length combination (of which there are 45), we run 100 replicate MCMC chains and use steps (1)-(5) to assess performance of the ESS measures for all summaries.

In the introduction, we mentioned that standard practice for phylogenetics often either ignores the mixing of the tree itself or uses the ESS of the log-posterior density as a proxy.
Based on these approaches, we define two baseline ESS measures as points of comparison to more sophisticated approaches.
One of these baselines, which we call $\textsf{fixedN}$, is to simply declare that the effective sample size is the sample size.
This is akin to ignoring the mixing of the tree, in which case one essentially assumes that the $n$ tree samples are worth $n$ independent samples.
Secondly, we consider using the univariate ESS (as implemented in \texttt{coda}) of the trace of the log-posterior-density, which we call the $\textsf{logPosteriorESS}$.

As an additional reference for our simulation results, we also run our testing setup in a non-phylogenetic context: estimating the mean of a $\operatorname{Normal}(0,1)$ variable.
We use 200 chain lengths between 1,000 and 100,000 and keep 1000 samples each time and use the \texttt{coda} univariate ESS.
The proposal is a $\operatorname{Normal}$ distribution, centered on the current value, with a standard deviation of 0.3.
This procedure produces ESS values from nearly 0 to nearly 1000, a range comparable to the observed tree ESS values computed.
The median RMCE is 0.01, with 50\% quantiles spanning [-0.073,0.057], and 80\% quantiles spanning [-0.13,0.12].
The median ITMCE is 1.01, with 50\% quantiles spanning [0.93,1.06] and 80\% quantiles spanning [0.88,1.13].
This establishes a reference for how well we might expect our tree ESS measures to work, and provides a simulation-based confirmation of our procedure for testing ESS measures.

Overall, performance of the tree ESS methods is variable across both simulated conditions and MCMCSE measures.
We account for variable performance across simulated conditions by binning our simulations based on the estimated ESS where possible.
For each of the 45 simulated data analyses, and for each ESS measure, we compute the average ESS across all 100 replicate MCMC runs, and then bin results for split and tree probabilities into the $\ESS < 500$ and $\ESS \geq 500$ regimes.
In the $\ESS < 500$ regime, the RMCE and ITMCE are quite variable, and often far from the optimal values of 0 and 1 (Figures~\ref{fig:benchmarks_splits},~\ref{fig:benchmarks_tree_probs}).
This is not ideal, as it means that the performance is quite variable across different splits, trees, and datasets.
In the $\ESS \geq 500$ regime, however, performance is much better for all methods.
For split probabilities, in both ESS regimes the tree ESS measures readily outperform the baseline measures ($\textsf{fixedN}$ and $\textsf{logPosteriorESS}$), and in the $\ESS \geq 500$ regime perform comparably to the univariate ESS of a Gaussian.
For the MRC tree (Figure~\ref{fig:benchmarks_MRC}), performance is not as good as the univariate ESS of a Gaussian, but the tree ESS measures readily outperform the baseline measures.
For tree probabilities (Figure~\ref{fig:benchmarks_tree_probs}), performance is somewhat worse.
In the $\ESS \geq 500$ regime, the tree ESS measures do not present as clear an improvement over the baselines as they do for split probabilities.
Further, none of them perform particularly well when compared to standard univariate ESS of a Gaussian (Figure~\ref{fig:benchmarks_tree_probs}).
However, in aggregate the tree ESS measures do capture the Monte Carlo error of the estimated tree probabilities, with the median RMCE $\approx$ 0 and the median ITMCE $\approx$ 1 in the $\ESS \geq 500$ regime for most measures.
It seems the tree ESS measures perform most usefully for split probabilities and the summary tree in the $\ESS \geq 500$ regime.
In the supplement, we present analogous figures for an ESS cutoff of 250, where the performance is poor both below and above the cutoff.

The $\frechetCorrelationESS$ and $\medianPseudoESS$ are the best-performing tree ESS measures.
Generally, the $\frechetCorrelationESS$ is a bit more conservative.
For both split and tree probabilities, it avoids some of the overestimation of the ESS evident in the $\medianPseudoESS$ in the $\ESS \geq 500$ regime.
For the MRC tree, the $\frechetCorrelationESS$ has the appropriate median RMCE and ITMCE, while the median of $\medianPseudoESS$ is slightly inflated (though there are fewer MRC trees to compare than splits).
The $\approximateESS$ is at best no better than the $\frechetCorrelationESS$ or the $\medianPseudoESS$, is in general more likely to overestimate the ESS, and for the MRC tree appears to overestimate the ESS when it is small and underestimate the ESS when it is large.
The $\minPseudoESS$ is overly conservative, and even in the $\ESS \geq 500$ regime almost exclusively underestimates the ESS.
In practice we recommend that practitioners consider multiple ESS measures, and note that any accurate quantification of Monte Carlo error requires an ESS of at least 500 for the ESS measure being used.
Using the $\minPseudoESS$ alone would be a conservative choice, though the cost is longer MCMC runs than strictly necessary to achieve a desired level of accuracy.
Using the $\minPseudoESS$ and either the $\frechetCorrelationESS$ or the $\medianPseudoESS$ (or both) allows for the best estimation of error in split probabilities, tree probabilities, and the MRC tree, while also providing an upper bound on the error.

\subsection*{Tree ESS measures reveal difficulties in MCMC sampling in real-world datasets}
Now that we have some understanding of how tree ESS measures work (and do not work), we turn our attention to applying them to real-world datasets.
Following \citet{lanfear2016estimating}, we apply our ESS measures to datasets from \citet{scantlebury2013diversification} spanning six taxonomic groups of Malagasy herpetofauna: \emph{Brookesia}, Cophylinae, \emph{Gephyromantis}, \emph{Heterixalus}, \emph{Paroedura}, \emph{Phelsuma}, and \emph{Uroplatus}.
These datasets are convenient because the same well-documented analysis methodology was applied to each, and because 4 independent replicate MCMC analyses for each dataset have been deposited at Dryad (\texttt{https://doi.org/10.5061/dryad.r1hk5}).

\begin{figure}[ht]
  \centering
  \includegraphics[width=\textwidth]{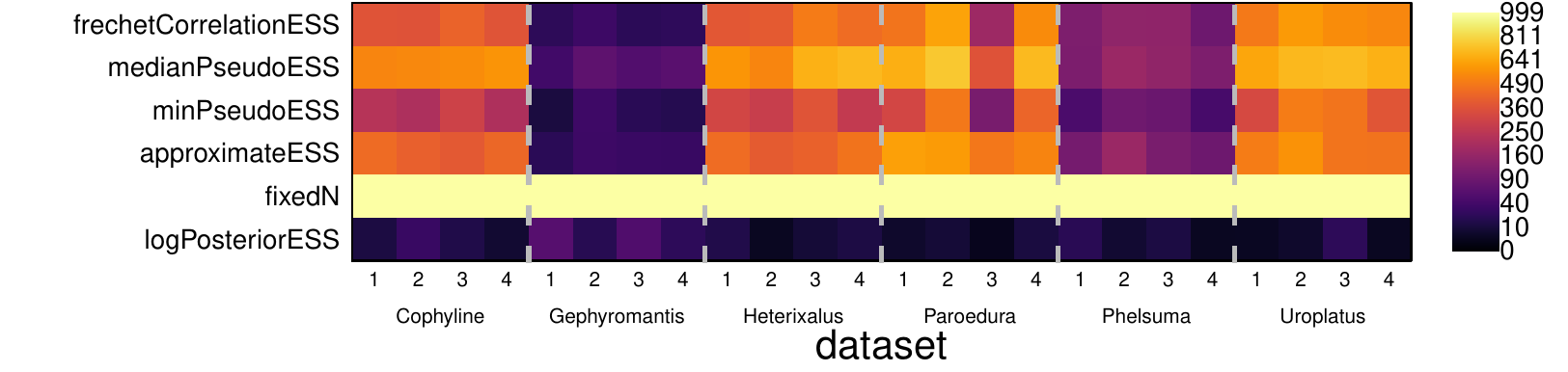}
  \caption{
    Tree ESS measures computed on 4 replicate chains for the 6 datasets from \citet{scantlebury2013diversification} as a heat map.
    To make differences clearer when ESS is low, the heatmap is spaced on the square-root scale.
    The ESS of the log-posterior and the $\textsf{fixedN}$ approaches are included as baselines, though neither captures the meaningful between-dataset differences in topological ESS.
  }%
  \label{fig:empirical_ess}
\end{figure}

In Figure~\ref{fig:empirical_ess}, we show ESS measures computed for 1001 samples from each of the 4 replicate chains for all 6 datasets.
While individual methods may often disagree, they all clearly agree that the \emph{Gephyromantis} and \emph{Phelsuma} datasets have low topological ESS across all chains, and other than the $\approximateESS$ they agree that the 3rd \emph{Paroedura} chain has a low ESS\@.
On these datasets, the performance of the baseline approaches ($\textsf{fixedN}$ and $\textsf{logPosteriorESS}$) is completely inadequate for assessing the mixing of the phylogeny.
Unlike in the simulations, though, here the $\textsf{logPosteriorESS}$ says the ESS is much smaller than any of the actual tree ESS measures.
That it is so low here, while the tree ESS measures are much larger, suggests that there were mixing problems with other model parameters than the tree, and highlights the need to address all parameters in the phylogenetic model, including the tree, on their own merits.
That is, the $\textsf{logPosteriorESS}$ is at best only loosely linked to the sampling of the tree topology, and its performance in our simulated scenarios is likely more closely linked to the actual Monte Carlo error in the tree than it ever will be in practice.

\begin{figure}[htp]
  \centering
  \includegraphics[width=0.7\textwidth]{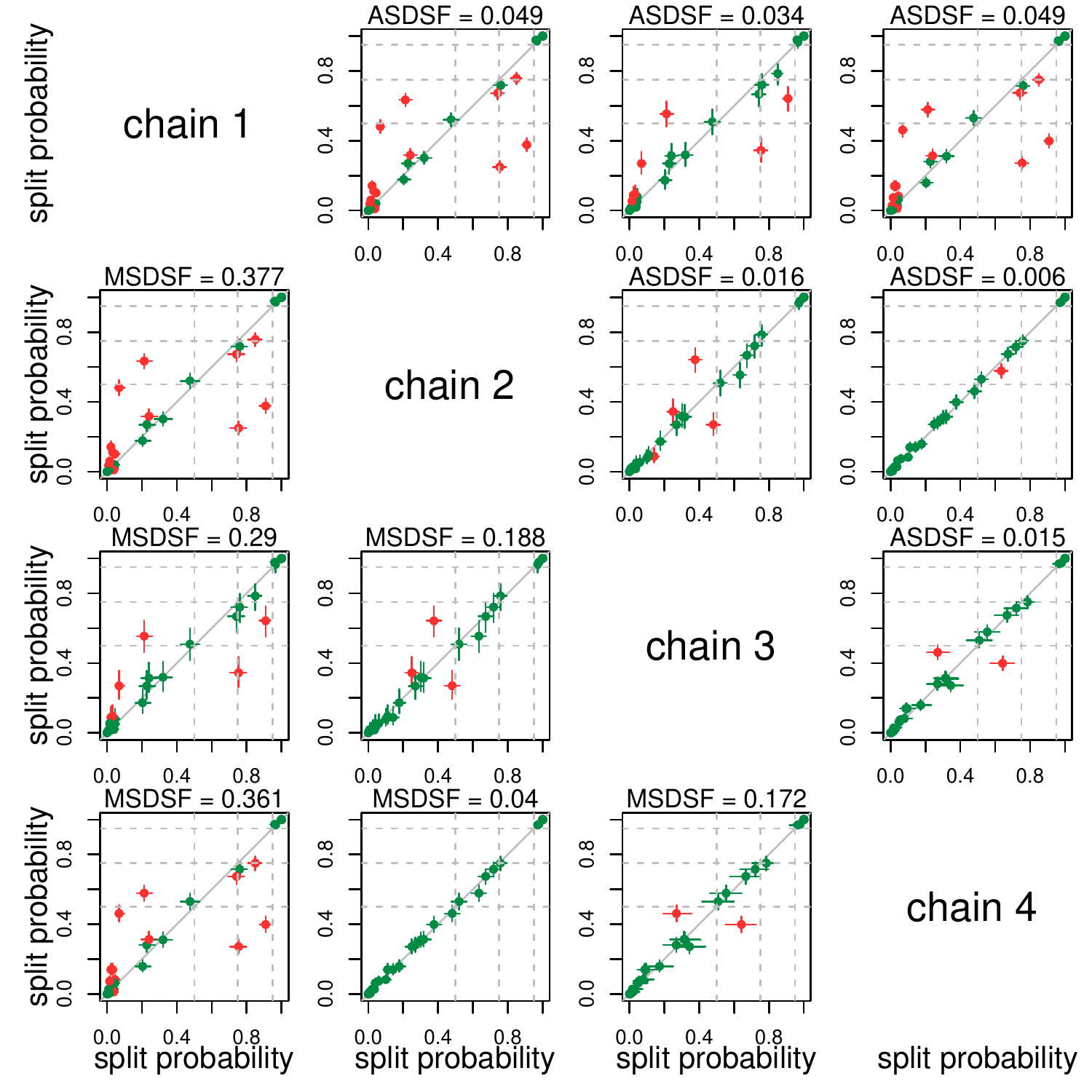}
  \caption{
    Split probabilities computed for all chains of the \emph{Paroedura} dataset of \citet{scantlebury2013diversification}, plotted against the probabilities computed for all other chains, with confidence intervals.
    Comparisons above the diagonal use the $\frechetCorrelationESS$ to compute confidence intervals, while comparisons below the diagonal use the $\minPseudoESS$, which is generally smaller and thus leads to larger confidence intervals.
    Each confidence interval is colored by whether or not the 95\% CI for the difference in split probability between chains $i$ and $j$ includes 0 (green for including 0, red for excluding 0).
    CIs for differences in probability that exclude 0 (or non-overlapping confidence intervals) are more likely to be indicative of convergence issues between chains.
    Narrower confidence intervals from larger tree ESS estimates will flag more splits as problematic (as in chains 1 and 4).
    Dashed grey lines indicate posterior probabilities of 0.5 (threshold for inclusion in the MRC tree), 0.75 (moderate support for a split), and 0.95 (strong support for a split).
    For comparison, we include the average standard deviation of split frequencies (ASDSF, above the diagonal) and maximum standard deviation of split frequencies (MSDSF, below the diagonal).
  }
  \label{fig:split_plots}
\end{figure}

In Figure~\ref{fig:split_plots}, we use the tree ESS to construct confidence intervals for split probabilities, and update a standard MCMC convergence plot, the plot of split frequencies for two chains.
Specifically, the confidence intervals are useful for determining whether the disagreement in split probabilities between two chains can be ascribed to a low sample size (CIs overlap), or whether it may be indicative of a convergence problem (CIs do not overlap).
By plotting thresholds, we can also examine whether a given split can be confidently assumed to be above a specific cutoff, such as 0.5 for inclusion in the MRC tree.
In this case we find that, even accounting for the topological ESS, only chains 2 and 4 for the \emph{Paroedura} dataset have complete agreement for all split probabilities.
This highlights the importance of performing multiple independent MCMC runs and using between-chain convergence diagnostics to assess between-chain differences, in addition to assessing the ESS of parameters.
On every dataset other than the \emph{Uroplatus} dataset, regardless of ESS, at least two chains disagree about the probability of at least one split (Supplemental Figures S8-S12).
Topological ESS measures flag the \emph{Gephyromantis} and \emph{Phelsuma} datasets as having relatively low ESS, but despite the low ESS, all four \emph{Gephyromantis} chains, and all \emph{Phelsuma} chains except one, show levels of conflict in line with the other datasets.
This further highlights the difference between low ESS (large uncertainty about parameter means) and between-chain convergence (two chains producing discordant estimates).

\subsection*{Data and code availability}
All necessary functions for computing the ESS of phylogenetic trees have been implemented in an \texttt{R} package named \texttt{treess} available at \texttt{bitbucket.com/afmagee/treess}.
This package also includes functions for creating our new split-split probability plots and running fake phylogenetic MCMC on arbitrary phylogenetic posterior distributions.
Code for the simulation study and real-data analyses is available at available at\\ \texttt{bitbucket.com/afmagee/tree\_convergence\_code}.

\section*{Discussion}

In this paper, we have investigated Monte Carlo error for phylogenetic trees.
We present three summaries of Monte Carlo error that a topological ESS measure might capture: the MCMCSEs of split probabilities, tree probabilities, and summary trees.
With simulations leveraging real-data posterior distributions, we use these MCMCSE summaries to assess four putative tree ESS measures (as well as two additional baselines based on standard practice).
We find that the performance of these measures varies across summary measures and within summaries based on the ESS\@.
At their best, the $\frechetCorrelationESS$ and $\medianPseudoESS$ can capture the Monte Carlo error inherent in MCMC estimates of split probabilities and MRC trees about as well as the univariate ESS captures the Monte Carlo error of the mean of a $\operatorname{Normal}(0,1)$ variable.
That is, these measures capture the Monte Carlo error of split probabilities and MRC trees about as well as could be hoped.
When the estimated ESS is less than 500, however, performance is notably less than ideal, severely overestimating the ESS for some split probabilities and underestimating it for others (similar patterns hold for tree probabilities and the summary tree).
Thus, if one wants to use ESS to accurately quantify the Monte Carlo error in their phylogeny, a larger cutoff, such as the 625 advocated for by \citet{fabreti2021convergence} would be more appropriate.
For split probabilities and MRC trees, the $\frechetCorrelationESS$ and $\medianPseudoESS$ clearly outperform the standard practices of ignoring Monte Carlo error or using the ESS of the log-posterior density as a proxy.
The tree ESS measures also capture Monte Carlo error in the estimated tree probabilities, though they do not clearly outperform the baselines.
When assessing the ESS of tree topologies, we recommend that practitioners consider the $\minPseudoESS$, $\frechetCorrelationESS$, and the $\medianPseudoESS$.
If the estimated ESS is at least 500, then this combination provides an upper bound on the error (the $\minPseudoESS$), and the best estimation of split probability error (the $\frechetCorrelationESS$) and MRC tree error (the $\medianPseudoESS$).
If the estimated ESS is less than 500, then all three approaches may overestimate the ESS quite severely.

The $\medianPseudoESS$ and the $\frechetCorrelationESS$ perform very similarly, which we found surprising given their quite distinct derivations.
The $\frechetCorrelationESS$ is derived as a generalization of the classical univariate ESS, and does not require a reference tree.
The $\medianPseudoESS$ is the ESS of a reduced-dimensional representation of the MCMC samples and requires fixing a reference tree.
We know of no obvious reason that they should work similarly.

While the simplicity of the purely topological measures we have considered are important as starting places for understanding MCMC for trees, extending this work to branch lengths via BHV space \citep{billera2001geometry} is an interesting possibility.
This would have the advantage of additionally assessing the mixing of the branch lengths, which is often unaddressed in phylogenetic workflows (we note, however, that MrBayes reports the potential scale reduction factors of branch lengths as a between-chain diagnostic).
By comparing purely topological and BHV-based measures, one could also understand how differences in branch lengths contribute to the tree-to-tree distances relative to the topological distances.
It is possible that most of the distance comes from differences in branch lengths, and thus the BHV-based tree ESS measures miss key differences in topology.
Alternately, BHV-based tree ESS measures could adequately capture both branch length and topological dynamics.
It is also possible that these two regimes co-exist in different regions of parameter space.
In either case, reducing the mixing of a very complex structure (a tree with branch lengths) to one or even two ESS measures would be a useful simplification.
In BHV space, it may also be possible to extend the work in \citet{willis2019confidence} on confidence sets for phylogenetic trees to better describe the uncertainty in summary trees.

Recently, \citet{harrington2020properties} performed an in-depth exploration of MCMC across thousands of empirical datasets.
They computed the approximate ESS of \citet{lanfear2016estimating} for all runs.
This represents the largest use to date of tree ESS measures, though until now the properties and performance of these ESS estimators have been somewhat ambiguous.
Our results show that effective sample sizes of at least 500 are generally capable of capturing the true mixing behavior of the tree, while smaller effective sample sizes are not.
The MCMC settings used by \citet{harrington2020properties} led to most topological ESS on the order of 750, meaning that in their usage the tree ESS is likely to adequately capture Monte Carlo error in the phylogeny.
We note, however, that a large within-chain ESS is not indicative of good between-chain convergence.
While 97\% of the analyses run by \citet{harrington2020properties} achieved a topological ESS of at least 200, only 37\% achieved an ASDSF below the standard cutoff of 0.01.

\citet{fabreti2021convergence} have recently published a paper on convergence diagnostics for phylogenetic MCMC, and an accompanying R package \texttt{convenience}.
We agree with \citet{fabreti2021convergence} (and \citet{rambaut2018posterior}) that $\ESS > 200$ is an arbitrary cutoff, and appreciate the approach \citet{fabreti2021convergence} take to deriving an ESS threshold which centers on Monte Carlo error.
Our work departs from theirs, though, in considering the entire tree, where the tree-based diagnostics in \texttt{convenience} consider each split separately.
As trees are inherently multivariate objects, we think it is important to attempt to consider them in their entirety.
Consider, for example, that the $\textsf{fixedN}$ approach estimates the Monte Carlo error for tree probabilities as well as the tree-based measures for longer runs (Figure~\ref{fig:benchmarks_tree_probs}), but severely underestimates the Monte Carlo error for split probabilities (Figure~\ref{fig:benchmarks_splits}).
In the supplement we also show that while the split probabilities may appear to have converged over the course of an MCMC run, there can still be notable uncertainty in the MRC tree (see section on ``Visualizing convergence of a single chain'' and Figure S1).

The tree ESS methods we have considered are applicable out-of-the-box to any data structure where one can compute a distance.
As distance measures exist for a wide range of objects, we hope that this work may prove useful for understanding Monte Carlo error in other non-Euclidean cases.
In phylogenetic applications, one might seek to understand the Monte Carlo error in estimates of phylogenetic networks, sequence alignments (for joint models of tree and alignment as in BAli-Phy \citep{redelings2021bali}) ancestral state estimates, or migration matrices.

We have identified tree ESS measures that can sufficiently describe the Monte Carlo error of both split probabilities and the summary tree.
Further, we have implemented these measures and approaches for constructing confidence intervals for split probabilities (as in Figure~\ref{fig:benchmarks_splits}) in the R package \texttt{treess}.
Focusing on Monte Carlo error as we have in this paper stands in contrast to the widespread use of ESS in phylogenetics, which is commonly treated as a ``box that should be checked'' by having ESS above a threshold before proceeding on to interpreting results.
We hope that this work motivates phylogenetic community to take more seriously the quality of MCMC estimation of its focal parameter, and more broadly that it helps de-mystify the matter of how long to run phylogenetic MCMC\@.
An analysis should be run long enough that the Monte Carlo error of the important quantities is small enough that conclusions are robust.

Most Bayesian phylogenetic MCMC run lengths are fixed in advance.
This can lead to runs which are too short to achieve good inference, or runs which are longer than needed and waste CPU time.
The problem of too-short runs is addressed by checkpointing, allowing post-hoc longer runs.
The problem of too-long runs is addressed by dynamic stopping criteria.
In non-phylogenetic contexts, dynamic stopping criteria have been considered based on between-chain and within-chain comparisons.
Among phylogenetic software, MrBayes enables auto-termination based on between-chain comparisons of split probabilities using the ASDSF.
While between-chain comparisons can capture Monte Carlo error, it appears that the number of chains required for accurate estimation of the Monte Carlo error is prohibitive (Figures~\ref{fig:benchmarks_splits_mrbayes},~\ref{fig:benchmarks_tree_probs_mrbayes}, and~\ref{fig:benchmarks_MRC_mrbayes}).
Future work may enable auto-termination when sufficient precision is achieved in split probabilities or the summary tree, leveraging the work done in this paper on phylogenetic effective sample sizes and methods developed for MCMC in Euclidean spaces \citep{gong2016practical,vats2019multivariate}.

\section*{Acknowledgements}
This work was supported by NSF grants CISE-1561334, CISE-1564137 and DGE-1762114, as well as NIH grant R01 AI162611.
Andrew Magee was supported by an ARCS Foundation Fellowship.
The research of Frederick Matsen was supported in part by a Faculty Scholar grant from the Howard Hughes Medical Institute and the Simons Foundation.
Dr.\ Matsen is an Investigator of the Howard Hughes Medical Institute.
Scientific Computing Infrastructure at Fred Hutch funded by ORIP grant S10OD028685.

\bibliographystyle{plainnat}
\bibliography{ms}

\beginsupplement

\clearpage
\setcounter{page}{1}
\begin{center}
\textbf{\large Supplementary Material for \\``How trustworthy is your tree? Bayesian phylogenetic effective sample size through the lens of Monte Carlo error''}
\end{center}

\section*{Visualizing convergence of a single chain}
To explore the uncertainty of estimates from a single MCMC chain through time, we employ a block-bootstrap approach in which we resample from the MCMC sample \citep{politis2003impact,suchard2003evolutionary}.
This approach requires a vector of subsample sizes, $n_1,\dots,n_s$.
For a given subsample length $n_i$, we define the batch size to be $b = \lfloor \sqrt{n_i} \rfloor$ and the number of batches $a = \lfloor n_i/b \rfloor$.
The summary tree is computed for the first $ab$ samples of the real chain, and then for some number of bootstrap replicates $r$ we re-estimate the summary tree.
We use block-bootstrapping to preserve autocorrelation in the samples.
Thus, for each of the $r$ bootstrap replicates (at a given $n_i$), we draw $a$ starting indices uniformly on $1,\dots,n-b+1$, and concatenate the resulting $a$ blocks of length $b$ into a bootstrap replicate chain.
Then we compute the median RF distance from the real-chain-subsample summary tree to the $r$ bootstrap replicates, as well as the 5th and 95th percentiles.
As the longer subsamples of the real chain include the shorter subsamples (all real-chain subsamples start at the first sample), this procedure allows us to track how the summary tree converges over the course of the MCMC run.
We can similarly track split or topology probabilities over the course of the run, in which case we use the average standard deviation of split frequencies (ASDSF) \citep{lemey2009phylogenetic} and the Euclidean distance, respectively, to compare the real chain estimates to the bootstrap estimates.

\begin{figure}[ht]
  \centering
  \includegraphics[width=0.8\textwidth]{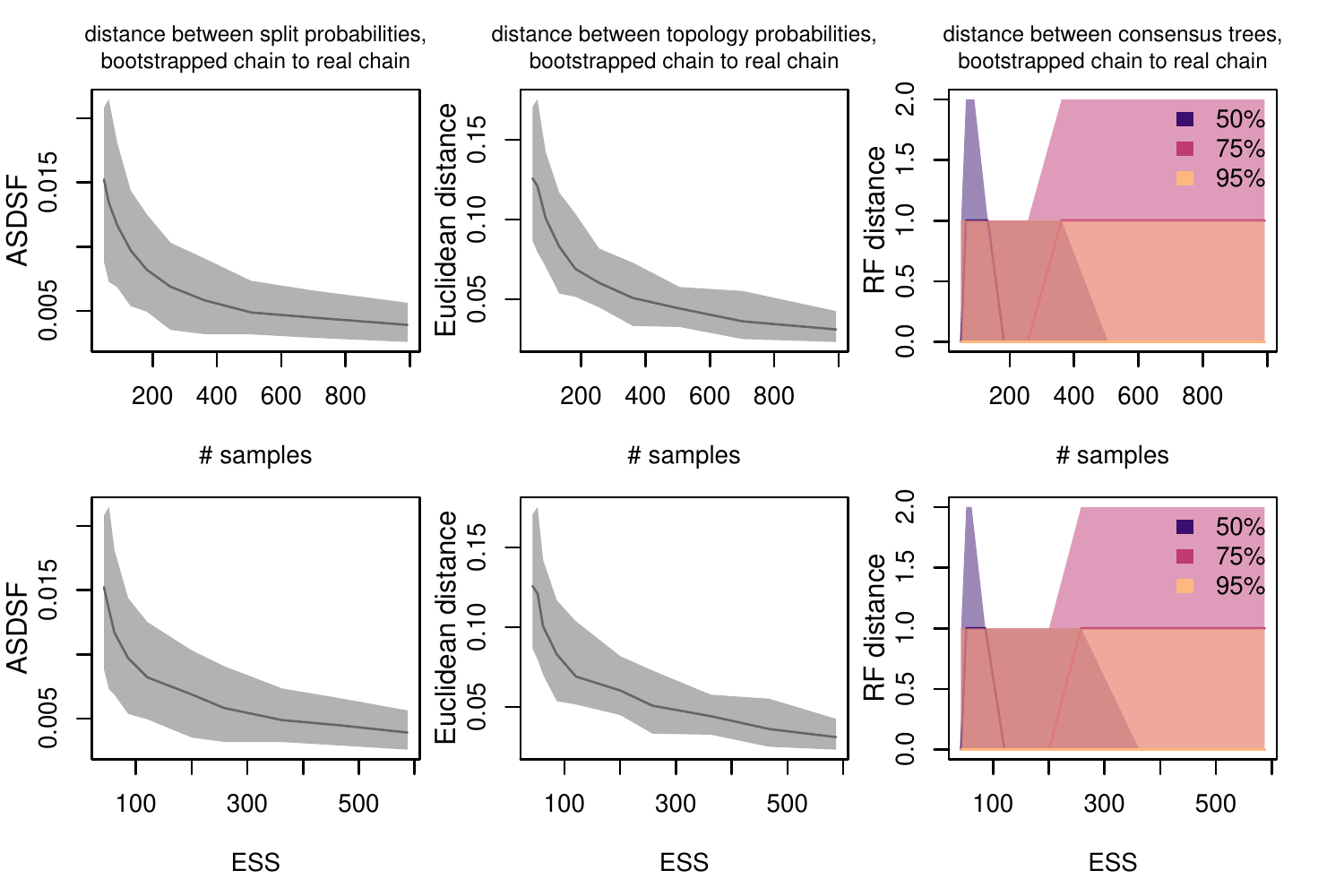}
  \caption{
    Monte Carlo error visualized over the length of one chain of the Paroedura dataset from \citet{scantlebury2013diversification}.
    The top and bottom rows are equivalent except that the $x$-axis is scaled to the absolute number of MCMC samples (top), and the split-frequency ESS (bottom).
    The left column plots the ASDSF between bootstrap replicate estimates of the split probabilities and the split probabilities estimated from the first $n_i$ samples of the chain.
    The central column plots the Euclidean distance between bootstrap replicate estimates of the (vector of) tree probabilities and the (vector of) tree probabilities estimated from the first $n_i$ samples of the chain.
    The right column plots the RF distance between bootstrap replicate estimates of consensus trees and the consensus trees estimated from the first $n_i$ samples of the chain.
    The different colors show consensus trees constructed with different minimum inclusion probabilities of splits, such that the purple curve shows the classical MRC tree, and the yellow curve shows a consensus tree containing only splits with 95\% probability.
    In all cases, the dark lines are the median and the shaded region is the central 90\% range.
  }%
  \label{fig:block_bootstrap}
\end{figure}

In Figure~\ref{fig:block_bootstrap}, we explore the convergence behavior of chain 1 of the Paroedura dataset using three summary measures.
These measures are the ASDSF between bootstrap and real-chain split probabilities, the Euclidean distance between bootstrap and real-chain estimates of the vector of split probabilities, and the RF distance between bootstrap and real-chain summary trees.
While both the split probabilities and tree probabilities appear to converge relatively well (the ASDSF quickly declines below the usual field-standard for good convergence of 0.01), there is still considerable Monte Carlo variability evident in summary trees.
This pattern holds across all datasets and almost all chains, indicating that classical ASDSF cutoffs for convergence of chains are not guarantees of the convergence of summary trees from those chains.
We note, however, that this can only ever help determine whether estimates from a single run have stabilized.
To diagnose issues such as convergence to a local mode, practitioners must run multiple chains.
We note that this is suggested as standard practice \citep{lemey2009phylogenetic} and is a widely available option, including in BEAST \citep{suchard2018bayesian} and RevBayes \citep{hohna2016revbayes}, and is the default in MrBayes \citep{ronquist2012mrbayes}.

\newpage

\section*{More efficient simulated phylogenetic MCMC}
Recall that our simulated phylogenetic MCMC is based on real-data phylogenetic posterior distributions, potentially truncated.
This consists of a vector of trees, $\btau$, and an associated probability mass function, $\phat(\btau)$ (we use the hat as a reminder that this target is based, indirectly, on real data).
We use NNIs to move between tree topologies, by uniformly drawing a tree from the set of neighbors, $N(\Psi)$.
Then we accept or reject according to the estimated topology probability $\phat(\Psi)$ (any tree not in the real-data posterior has probability 0).
If we redefine $N(\Psi)$ to instead be the NNI neighbors of $\Psi$ with positive probability (\textit{e.g.} $\Psi \in \btau$, which also requires far less storage), we can instead simulate the proposal in two steps.
First, draw $u \sim \text{Uniform}(0,1)$, and if $u < |N(\Psi)| / |N|$ (where $|N|$ is the number of NNI neighbors of any tree in the posterior), we draw our proposed tree $\Psi^*$ uniformly at random from $N(\Psi)$ and set $A = \min(1,\pr(\Psi^*)/\pr(\Psi))$.
Otherwise, we have drawn a tree outside the set of supported neighbors of $\Psi$ ($\Psi^* \not \in \btau$) and we do not need to specify which tree, as in this case it has probability 0 and so $A = 0$ and we will always reject the proposal.
Then we accept or reject the move with probability $A$ and proceed normally.
This approach requires us only to know what trees in the support of the posterior are neighbors, which for real phylogenetic posterior distributions is a much smaller set than the set of all NNI neighbors.

\section*{Explicit definitions and derivations of tree ESS measures}

In the following sections, we present more thorough derivations of the $\frechetCorrelationESS$ and $\approximateESS$ use in the main text, and derivations for 6 other potential tree ESS measures.
We have not presented these additional ESS measures in the main text as their performance is at best no better than the performance of the methods presented above, and in some cases is markedly worse (Figures S5-S7).
The ten total methods fall into the same three categories as the main text and are (using $*$ to denote those appearing in the main text):
\begin{itemize}
  \item ESS measures based on Fr{\'e}chet generalizations of Equation~\ref{eqn:ess_grand_definition} to trees
  \begin{itemize}
    \item[] $*$The Fr{\'e}chet Correlation ESS ($\frechetCorrelationESS$)
    \item[] The split frequency ESS ($\splitFrequencyESS$)
  \end{itemize}
  \item ESS measures based on projecting the tree to a single dimension and computing the ESS of that using standard univariate approaches
  \begin{itemize}
    \item[] The folded rank-medoid ESS ($\foldedRankmedoidESS$)
    \item[] $*$The median pseudo-ESS ($\medianPseudoESS$)
    \item[] $*$The minimum pseudo-ESS ($\minPseudoESS$)
    \item[] The total distance ESS ($\totalDistanceESS$)
    \item[] The classical multidimensional scaling ESS ($\CMDSESS$)
  \end{itemize}
  \item Ad-hoc ESS measures
  \begin{itemize}
    \item[] $*$The approximate ESS ($\splitFrequencyESS$)
    \item[] The unsmoothed bootstrap jump-distance ESS ($\jumpDistanceBootstrapUnsmoothedESS$)
    \item[] The (smoothed) bootstrap jump-distance ESS ($\jumpDistanceBootstrapESS$)
  \end{itemize}
\end{itemize}

\subsection*{Calculating the ESS by generalizing previous definitions}
In this section, we provide more in-depth derivations of our two ESS approaches that generalize Equation~\ref{eqn:ess_grand_definition} using concepts borrowed from the notions of Fr{\'e}chet mean and Fr{\'e}chet variance.
For a continuous random variable $X$, the sample mean minimizes the sum of squared deviations to all sampled points.
The Fr{\'e}chet mean generalizes this concept to other metric spaces and higher dimensions by keeping the idea of minimizing the sum of squared distances.
The Fr{\'e}chet mean of a set of samples is,
\[
  \bar{x} = \operatorname{argmin}_{y} \sum_{i=1}^n d(x_i,y)^2,
\]
where $d(\cdot,\cdot)$ is a distance metric.
Note that for the rest of this subsection on Fr{\'e}chet generalizations of univariate ESS approaches, we will use $\bar{x}$ to refer to the Fr{\'e}chet mean.
The Fr{\'e}chet mean may not be unique, in which case the collection of values that minimize the sum of squared distances are known as Karcher means.
Where the variance is the average squared deviation from the mean, the Fr{\'e}chet variance is the average squared distance from the Fr{\'e}chet mean.
In the case where $X$ is continuous and one-dimensional and $d(\cdot,\cdot)$ is the Euclidean distance, the Fr{\'e}chet mean is the mean and the Fr{\'e}chet variance is the variance.

These definitions take some adaptation to the setting considered here.
When using RF distances between trees, one can think of an ``RF space'' where topologies are encoded as a binary vector.
For a tree with $n_{\text{taxa}}$ tips, there are $2^{n_{\text{taxa}}} - n_{\text{taxa}}$ possible non-trivial splits.
Thus we can represent a tree as a vector of length $2^{n_{\text{taxa}}} - n_{\text{taxa}}$ which has a one entry exactly when the corresponding split is present in the tree.
There are $n_{\text{taxa}} - 3$ non-trivial splits in a fully resolved tree, thus the sum of entries in such a vector representation is $n_{\text{taxa}} - 3$.
The Hamming distance (or equivalently the Manhattan distance) between two trees represented as coordinate vectors in RF space is the classical RF distance.
This also means that we only need to consider coordinates in RF space which are non-zero in at least one tree in the set.
As we use RF distances in this paper, all this work can be seen to live in RF space.

\subsubsection*{The $\frechetCorrelationESS$}
In this section, we will explore how to generalize the sum-of-correlations ESS of Equation~\ref{eqn:ess_correlation_practical} to trees.
To do so, we first review several key identities, including relationships between pairwise distances and both covariance and variance.
For two real-valued variables, $X$ and $Y$, we can express the expected squared Euclidean distance as a function of the variances, the difference in means, and the covariance.
For convenience, we will write $\Delta^2 = (X - Y)^2$.
Then, taking advantage of the fact that $\cov(X,Y) = \E[XY] - \E[X]\E[Y]$, we get,
\begin{align*}
 \E[\Delta^2] &= \E[(X - Y)^2]\\
 &= \E[X^2] - 2\E[XY] + E[Y^2]\\
  &= \var(X) + E[X]^2 + \var(Y) + E[Y]^2 - 2(\cov(X,Y) + E[X]E[Y])\\
  &= \var(X) + \var(Y) - 2\cov(X,Y) + (\E[X] - \E[Y])^2\\
  &\geq \var(X) + \var(Y) - 2\cov(X,Y)
\end{align*}
Where the last line follows because $(\E[X] - \E[Y])^2 > 0$.
The last two lines of this equation block rearrange to:
\begin{align}
  \cov(X,Y) &= \frac{1}{2}(\var(X) + \var(Y) - \E[\Delta^2]  + (\E[X] - \E[Y])^2).
  \label{eqn:mvcor}
\end{align}
If $\E[X] \approx \E[Y]$, then we have the approximate equality,
\begin{align}
  \cov(X,Y) &\approx \frac{1}{2}(\var(X) + \var(Y) - \E[\Delta^2])
  \label{eqn:mvcor_lb}
\end{align}

It is worth noting that the sum of pairwise distances for a sample of a random variable can be used to estimate its variance.
\begin{align}
  \widehat{\var}(X) = \frac{1}{n-1} \sum_{i} (x_i - \bar{x})^2 = \frac{1}{n(n-1)} \sum_{j > i} (x_i - x_j)^2.
  \label{eqn:pairs_to_var}
\end{align}
To show this, first we need that,
\begin{align}
  \sum_i \sum_j (x_i - x_j)^2 = 2 n \sum_{i} (x_i - \bar{x})^2.
  \label{eqn:pairs_to_squares}
\end{align}
This can be shown as follows,
\begin{align*}
  \sum_i \sum_j (x_i - x_j)^2 &= \sum_i \sum_j ( (x_i-\bar{x}) - (x_j-\bar{x}) )^2\\
  &= \sum_i \sum_j (x_i - \bar{x})^2 - 2(x_i-\bar{x})(x_j-\bar{x}) + (x_j - \bar{x})^2\\
  &= n \sum_i (x_i - \bar{x})^2 + n \sum_j (x_j - \bar{x})^2 - 2 \sum_i \sum_j (x_i-\bar{x})(x_j-\bar{x})\\
  &= 2n \sum_i (x_i - \bar{x})^2 - 2 \sum_i \sum_j (x_i-\bar{x})(x_j-\bar{x})\\
  &= 2n \sum_i (x_i - \bar{x})^2 - 2 \sum_i \sum_j (x_i x_j - x_i \bar{x} - x_j \bar{x} + \bar{x}^2)\\
  &= 2n \sum_i (x_i - \bar{x})^2 - 2 (n^2 \bar{x}^2 - n^2 \bar{x}^2 - n^2 \bar{x}^2 + n^2 \bar{x}^2)\\
  &= 2n \sum_i (x_i - \bar{x})^2.
\end{align*}
Having shown that Equation~\ref{eqn:pairs_to_squares} is true, from it we can get,
\begin{align}
  \frac{1}{2n(n-1)} \sum_i \sum_j (x_i - x_j)^2 = \frac{1}{n-1} \sum_{i} (x_i - \bar{x})^2.
  \label{eqn:pairs_to_var2}
\end{align}
And Equation~\ref{eqn:pairs_to_var} results by noting that,
\[
  \sum_i \sum_j (x_i - x_j)^2 = 2 \sum_{j > i} (x_i - x_j)^2.
\]
(We note this relationship can also be derived analogously to Equation~\ref{eqn:mvcor} by letting X and Y be IID.)
Letting $d(\cdot,\cdot)$ be a distance measure, we can write a Fr{\'e}chet generalization of Equation~\ref{eqn:pairs_to_var} as,
\begin{align}
  \widehat{\var}(X) = \frac{1}{n(n-1)} \sum_{j > i} d(x_i,x_j)^2
  \label{eqn:variance_from_distances}
\end{align}

In the same way that mean and variance can be generalized using the Fr{\'e}chet mean and variance, Equation~\ref{eqn:mvcor} allows us to generalize covariance to a Fr{\'e}chet covariance.
This is accomplished by defining $\var(X)$ and $\var(Y)$ to be the Fr{\'e}chet variances, $\E[X]$ and $\E[Y]$ to be the Fr{\'e}chet means, and redefining $\Delta^2 = d(X,Y)^2$.
Note that $\E[\Delta^2]$ is simply the average distance between $X$ and $Y$,
\begin{align}
  \E[\Delta^2] = \frac{1}{n}  \sum_{i=1}^n d(x_i,y_i)^2.
  \label{eqn:estimating_delta_squared}
\end{align}
Thus, we can get a single-dimensional summary of the dependency of two random variables, and compute a single ESS measure for a high-dimensional object.
Equation~\ref{eqn:mvcor_lb} is particularly useful in this circumstance because it avoids the need to compute the topological mean of a set of trees, $\E[X]$ and $\E[Y]$, which may not be unique.
Equations~\ref{eqn:variance_from_distances}, and~\ref{eqn:estimating_delta_squared} are also useful, and they allow us to compute everything we need for Equation~\ref{eqn:mvcor_lb} from the sample distance matrix.

To compute the ESS for trees using Equations~\ref{eqn:ess_correlation_practical} and~\ref{eqn:mvcor_lb}, we specifically need to be able to compute the Fr{\'e}chet autocorrelation $\rho_s$ of the chain at time lag $t$, and thus $X$ and $Y$ are actually $X_t$ and $X_{t+s}$.
If the chain is stationary, then the mean does not change over time, and we should expect that $\E[X_t] \approx \E[X_{t+s}]$, and the use of Equation~\ref{eqn:mvcor_lb} rather than Equation~\ref{eqn:mvcor} is justified.
If instead of trees we had a time series of a Euclidean variable $\bX$, the estimated autocorrelation is the sample Pearson correlation coefficient between samples at the given time lag,
\begin{align}
  \hat{\rho}_s = \frac{\widehat{\cov}(\bX_t,\bX_{t+s})}{\sqrt{\widehat{\var}(\bX_t)\widehat{\var}(\bX_{t+s})}}.
  \label{eqn:rho_T}
\end{align}
For trees, instead we use Fr{\'e}chet variances and Equation~\ref{eqn:mvcor_lb} to get an approximation to the Fr{\'e}chet covariance, and plug this into Equation~\ref{eqn:rho_T},
\begin{align}
  \hat{\rho}_s = \frac{ \frac{1}{2} (\widehat{\var}(\btau_{t}) + \widehat{\var}(\btau_{t+s}) - \hat{\E}[\Delta^2]) }{ \sqrt{\widehat{\var}(\btau_{t}) \widehat{\var}(\btau_{t+s})} }
  \label{eqn:mvcor_practical}
\end{align}
Once we obtain our estimates $\hat{\rho}_s$, we use Equation~\ref{eqn:ess_correlation_practical} to estimate the ESS\@.
We call this approach the Fr{\'e}chet correlation ESS, or $\frechetCorrelationESS$.
In this set up, we must estimate $\var(\btau_{t})$, $\var(\btau_{t+s})$, and $\E[\Delta^2]$, which we compute using Equations~\ref{eqn:variance_from_distances} and \ref{eqn:estimating_delta_squared}.
Let $n$ be the total number of tree samples, $\btau$ be the vector of tree samples, and let us define $\btau_t,\btau_{t+s}$ to be the pair of vectors of trees separated by time lag $s$.
Then, we can compute the terms as,
\[
  \widehat{\var}(\btau_{t}) = \frac{1}{(n-s)(n-s-1)} \sum_{i=1}^{n-s-1} \sum_{j=i+1}^{n-s} d(\tau_i,\tau_j)^2,
\]
\[
  \widehat{\var}(\btau_{t+s}) = \frac{1}{(n-s)(n-s-1)} \sum_{i=s+1}^{n-1} \sum_{j=i+1}^{n} d(\tau_i,\tau_j)^2,
\]
\[
  \hat{\E}[\Delta^2] = \frac{1}{n-s} \sum_{i=1}^{n-s} d(\tau_i,\tau_{i+t})^2.
\]
Given that most distances will appear in several calculations, it is most efficient to pre-compute the sample distance matrix $\boldsymbol{D}$ with $D_{ij} = d(\tau_i,\tau_j)$.

In practice, while the above definition could theoretically permit $\ESS > n$, we enforce $\frechetCorrelationESS \leq n$.

\subsubsection*{The $\splitFrequencyESS$}

Our next generalization approach we term the split frequency ESS, or $\splitFrequencyESS$.
This is a generalization of the univariate \citet{vats2021revisiting} estimator of the effective sample size, which we will call the batch means ESS, which we will now review.
The batch means ESS is based on the relationship,
\begin{equation}
  \widehat{\ESS} = n \frac{\hat{\sigma}_{\pi}^2}{\hat{\lambda}_L^2},
  \label{eqn:ess_vats_1D_practical}
\end{equation}
where $\hat{\lambda}_L^2$ is an estimate of the limiting variance ($\limvar$) and $\hat{\sigma}_{\pi}^2$ is the estimate of the posterior variance computed from the samples.
Let $\bX$ be the vector of MCMC samples, $B$ be a batch size (define $a$ to be the according number of batches), and $\bY$ the vector of batch means, with $Y_i$ the mean in the $i$th batch (subset of the chain).
Then, \citet{vats2021revisiting} define,
\[
  \hat{\lambda}_B^2 = \frac{B}{a - 1} \sum_{i=1}^{a} (Y_i - \bar{X})^2.
\]
To use the batch means approach in practice, the batch size must scale with $n$.
Following \citet{vats2021revisiting}, we use a batch size $b = \lfloor n^{1/2} \rfloor$.
Then, the estimate of the limiting variance from the batch-means approach is given by,
\[
  \hat{\lambda}_L^2 = 2\hat{\lambda}_b^2 - \hat{\lambda}_{b/3}^2,
\]
where $\hat{\lambda}_{b/3}^2$ is computed using a batch size $\lfloor b/3 \rfloor$ \citep[Equation 5,][]{vats2021revisiting}.

To apply the batch means ESS to trees, we represent trees as vectors of splits.
We now walk through this generalization.
If the posterior distribution contains $S$ non-trivial splits, $s_1,\dots,s_S$, then we transform the vector of trees into a matrix, where in each row we represent that tree as its vector of coordinates in RF-space $\bX$.
Namely,
\begin{equation*}
  X_{ij} =
  \begin{cases}
    1 & \text{if } s_j \in \tau_i, \\
    0 & \text{otherwise},
  \end{cases}
\end{equation*}

As our distance metric we take the Euclidean distance, so the Fr{\'e}chet mean is the arithmetic mean, $\bar{\bX}$.
We choose a batch size $b$ that scales with $n$ (we use $\lfloor n^{1/2} \rfloor$).
Again, $a = \lfloor n/b \rfloor$ is the number of batches, and $\bY$ the matrix of batch means, with $\bY_i$ the vector of means in the $i$th batch (subset of the chain).
For a fixed split $j$,
\[
  Y_{ij} = \frac{1}{b} \sum_{k=(i-1)b+1}^{ib} X_{kj}.
\]
We use a Fr{\'e}chet-based generalization for $\hat{\lambda}_b^2$, namely,
\[
  \hat{\lambda}_b^2 = \frac{b}{a-1} \sum_{i=1}^{a} d(\bY_i,\bar{\bX})^2.
\]
Similarly, We use a Fr{\'e}chet-based generalization for $\hat{\sigma}_{\pi}^2$,
\[
  \hat{\sigma}_{\pi}^2 = \frac{1}{n-1} \sum_{i=1}^{n} d(\bX_i,\bar{\bX})^2.
\]
Given these modified $\hat{\lambda}_b^2$ and $\hat{\sigma}_{\pi}^2$, we use Equation~\ref{eqn:ess_vats_1D_practical} to compute the ESS\@.
We note that all the batch means,$\bY_i$, are in fact the split frequencies (or estimates split probabilities) in those batches, and the global mean, $\bar{\bX}$ are the marginal split frequencies across the entire posterior distribution.
We thus call this the split frequency ESS, or $\splitFrequencyESS$.

\subsubsection*{Approaches to calculating the ESS by projecting the tree to a single dimension}

All dimension-reduction approaches entail first transforming the trees into a 1-D representation, then taking the ESS of that.
We use the The R package \texttt{coda} \citep{plummer2006coda} implementation of the ESS, and before we discuss our approaches we first outline how it works.

\subsubsection*{The ESS computation in \texttt{coda}}
The R package \texttt{coda} \citep{plummer2006coda}, commonly used for MCMC diagnostics, fits an autoregressive model to the MCMC samples to estimate the ESS\@.
Specifically, the \texttt{coda} estimate of the ESS, which we will call the power spectrum ESS, is,
\begin{equation}
  \ESS = n \frac{\hat{\sigma}_{\pi}^2}{\widehat{\Gamma(0)}},
  \label{eqn:ess_coda}
\end{equation}
where $\widehat{\Gamma(0)}$ is an estimate of the power spectrum at frequency 0 \citep[see ][ for details]{heidelberger1981spectral}, and $\hat{\sigma}_{\pi}^2$ is the estimate of the posterior variance computed from the samples.
This follows from Equation~\ref{eqn:ess_grand_definition} and the fact that the standard error of the mean of a covariance-stationary process is $\Gamma(0)/n$ \citep{heidelberger1981spectral}.
The power spectrum at 0, $\Gamma(0)$, can be linked to the autoregressive parameters by,
\[
  \Gamma(0) = \frac{\sigma^2_e}{{(1 - \sum_{i=1}^{p} \phi_i)}^2},
\]
where $\sigma^2_e$ is the error variance (the variance unexplained by the autoregressive model, also called the noise variance), also called the noise variance \citep{vonstorch2001statistical}.
In practice, \texttt{coda} estimates $\widehat{\Gamma(0)}$ using an autoregressive process of unknown order $p$.
With an estimated order, $\hat{p}$, a fitted set of autoregression coefficients $\phi_1,\dots,\phi_{\hat{p}}$, and an estimated error variance $\hat{\sigma}^2_e$, the estimate is,
\[
  \widehat{\Gamma(0)} = \frac{\hat{\sigma}^2_e}{{(1 - \sum_{i=1}^{\hat{p}} \hat{\phi}_i)}^2}.
\]

\subsubsection*{The $\foldedRankmedoidESS$}
Vehtari \textit{et al.} introduce two new approaches for computing ESS measures, one of which, the folded rank-transformed ESS, can be co-opted for phylogenies relatively painlessly \citep{vehtari2021rank}.
For a real-valued parameter $x$, this ESS is computed for the transformed variable $z$, where there are a few layers of transformations:
\begin{align*}
  \zeta &= |x - \text{median}(x)|,\\
  r &= \text{rank}(\zeta),\\
  z &= \Phi^{-1}\left( \frac{r - 3/8}{n - 1/4} \right).
\end{align*}
The first step is to ``fold'' the variable, and track the absolute deviations from the median.
Then a rank transformation is applied, which stabilizes for any extreme deviations.
Lastly, a Normal inverse-CDF is applied (with an offset).
\citet{vehtari2021rank} then take the folded rank-transformed ESS to be the ESS of $z$ using Equation~\ref{eqn:ess_correlation_practical}.
In the case there is not a unique medoid tree, we compute the ESS using all possible reference trees and take the minimum.

To use this approach for trees, we make a few generalizations, and we call the resulting ESS the $\foldedRankmedoidESS$.
First, we replace the sample median with the medoid, which is a generalization of the median to higher dimensions.
Specifically, the medoid tree is the (sampled) tree with the minimum sum of distances to all other sampled trees, $\text{medoid}(\btau) = \underset{\Psi \in \btau}{\text{argmin}}\ \sum_i d(\Psi,\tau_i)$.
Then, we replace the absolute divergence with the distance (in one dimension, these are equivalent).
The $\foldedRankmedoidESS$ is computed for the transformed variable $z$, where we obtain $z$ through the following transformations:
\begin{align*}
  \zeta &= d(\tau,\text{medoid}(\tau)),\\
  r &= \text{rank}(\zeta),\\
  z &= \Phi^{-1}\left( \frac{r - 3/8}{n - 1/4} \right).
\end{align*}

\subsubsection*{The $\totalDistanceESS$}
As an alternative to picking a specific reference tree, as in the $\foldedRankmedoidESS$, $\medianPseudoESS$, or $\minPseudoESS$, we also consider an ESS based on the sum of distances between each tree and all the other trees.
In this setup, we compute the ESS of the transformed variable $y$, defined by $y_i = \sum_{j=1}^{n} d(\tau_i,\tau_j)$.
We call this the total distance ESS, or $\totalDistanceESS$.

\subsubsection*{The $\CMDSESS$}
We also consider multidimensional scaling of the (squared) distance matrix $\boldsymbol{D}^{2}$ to compute an ESS\@.
Specifically, we use classical multidimensional scaling.
This approach seeks to find a matrix $\boldsymbol{Y}$ which minimizes a loss function called the strain between $\boldsymbol{Y}$ and the $\boldsymbol{B}$, a doubly centered version of $\boldsymbol{D}$.
As our new variable we take the first column of the new matrix, $\boldsymbol{Y}_{\cdot 1}$
We call this the classical multidimensional scaling ESS, or $\CMDSESS$.

\subsection*{Ad-hoc approaches to computing the ESS}
If we define $s_0$ to be the time lag at which samples from our MCMC become independent of each other, then we could somewhat conservatively estimate the ESS as $n/s_0$.
This approach can be seen as a naive implementation of the idea that the effective sample size is the hypothetical number of independent samples contained within the $n$ MCMC samples.
This is not tied to any mathematical definition of the ESS, and is not without problems.
For one, the approach is expected to be overly conservative, as it effectively discards all samples in between an estimated autocorrelation time, whereas classical ESS approaches keep fractions of all samples.
Additionally, in this approach ESS can take on only $n$ distinct values because it is guaranteed that $s_0$ is an integer between 1 and $n$ (inclusive).
The approximate ESS of \citet{lanfear2016estimating} can be seen as one approach to overcoming these limitations, as it requires estimating $s_0$ then uses identities about expected distances.
In the following sections, we consider methods for estimating $s_0$ and simply using this to estimate the ESS directly, $\widehat{\ESS} = n/\hat{s}_0$.

\subsubsection*{The $\jumpDistanceBootstrapUnsmoothedESS$ and the $\jumpDistanceBootstrapESS$}

We now define two new approaches to computing the ESS based on estimating $s_0$.
In both approaches, we start with a similarity or dissimilarity measure for trees at time lag $s$, $g(s)$, which we then smooth into a monotonically increasing function $G(s)$.
We do this by defining $G(s) = \max(g(s),g(t-1))$ for dissimilarity measures and $G(s) = \max(-g(s),-g(t-1))$ for similarity measures.
In essence, regardless of $g(s)$, $G(s)$ is a distance or dissimilarity measure.
We also consider a smoother version of $G(s)$, which we call $G^*(s)$, which we define below.
We do not search for an asymptote of either curve directly, as \citet{lanfear2016estimating} do for the approximate ESS.
Rather, we seek the point at which the dissimilarity of trees at time lag $s$ is indistinguishable from the dissimilarity of a pair of trees drawn independently from the posterior distribution.

Let $\pr(G(1) \mid \text{iid})$ be the distribution of $G(1)$ given a set of iid samples from the posterior.
Given a probability $\alpha$, we define a threshold $\epsilon$ to be the $(1 - \alpha)$th percentile of $\pr(G(1) \mid \text{iid})$.
We estimate $\hat{\epsilon}$ using bootstrap resampling of the posterior samples, which breaks the autocorrelation but preserves the fact that the samples are from the posterior distribution.
Given a choice of $\alpha$ and an estimate $\hat{\epsilon}$, $s_0$ is the first time lag $s$ for which $G(s) > \epsilon$.
In this paper, we define $G(s)$ to be the median RF distance between trees at time lag $s$, and call the resulting estimate the unsmoothed jump-distance bootstrap ESS, $\jumpDistanceBootstrapUnsmoothedESS$, though we note that any choice of $G(s)$ could rightfully be called a bootstrap ESS\@.
In practice, we set $\alpha = 0.05$, such that $s_0$ is the time lag at which the tree-to-tree dissimilarity is at least as big as the 5th percentile of the tree-to-tree dissimilarity for trees drawn identically and independently from the posterior distribution.

To circumvent the fact that the $\jumpDistanceBootstrapUnsmoothedESS$ can only take on values in $n/1,n/2,n/3,\dots,n/(n-1),1$, we also consider using smoothing.
Specifically, we use linear interpolation to smooth $G(s)$ into $G^*(s)$.
If $\boldsymbol{s}^{\text{step}}$ is the vector of times at which $G(s)$ changes, we can define a piecewise linear function $G^*(s)$ as,
\begin{equation}
G^*(s) = G(s_i^{\text{step}}) + \frac{s - s_{i}^{\text{step}}}{s_{i+1}^{\text{step}} - s_{i}^{\text{step}}} (G(s_{i+1}^{\text{step}}) - G(s_i^{\text{step}})),
\end{equation}
where $s$ is in the $i$th interval ($s_i^{\text{step}} \leq s < s_{i+1}^{\text{step}}$).
Defining $s_0$ to be the time $s$ such that $G^*(s) >= \epsilon$ allows us to assign fractional $s_0$, and have a continuous estimator.
We call the resulting estimator the (smoothed) jump distance bootstrap ESS, $\jumpDistanceBootstrapESS$.

There are a few constraints that must be imposed to complete this approach.
In the pathological case where all trees are the same tree, we set $\widehat{\ESS} = 1$, as clearly if we have only sampled one topology we have an effective sample size of 1.
The unsmoothed approach would not have a defined answer, as there is $s$ for which $G(s) > \epsilon = 0$, and the smoothed approach would yield $\widehat{\ESS} = n$ as $G(1) = \epsilon = 0$.
It is also possible that there is no observed $s$ for which $G^*(s) = \hat{\epsilon}$ and that $G^*(s) < \hat{\epsilon}$ for all $s$, in which case we enforce a minimum ESS of 1.
Further, while $g(0)$ is defined, and we could infer $s_0 < 1$, we enforce a maximum ESS of $n$.

There is evidence that the curve-fitting approach of \citet{lanfear2016estimating} for underestimates $s_0$.
It is not completely clear how well these jump-distance bootstrap approaches work to estimate $s_0$, but it is possible that they may be useful in combination with the approximate ESS, which requires estimating $s_0$.
We leave this to future work.

\section*{Performance of the ESS measures below $\ESS = 500$}
\begin{figure}
  \centering
  \begin{subfigure}[ht]{\textwidth}
    \centering
    \includegraphics[width=\textwidth]{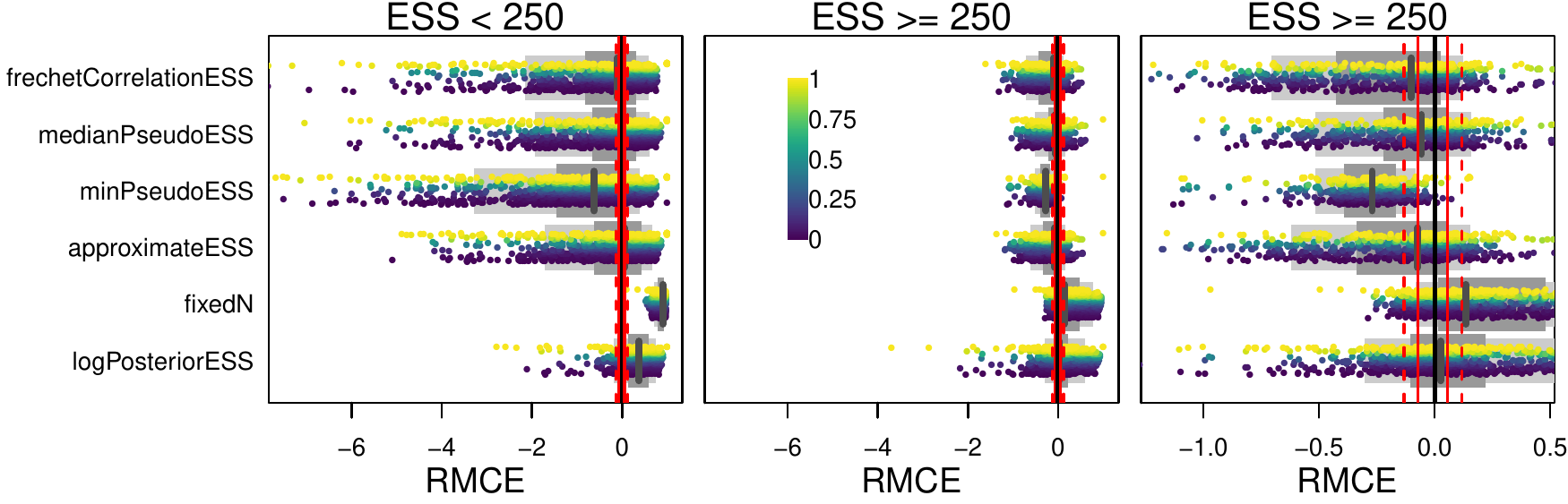}
  \end{subfigure}
  \begin{subfigure}[ht]{\textwidth}
    \centering
    \includegraphics[width=\textwidth]{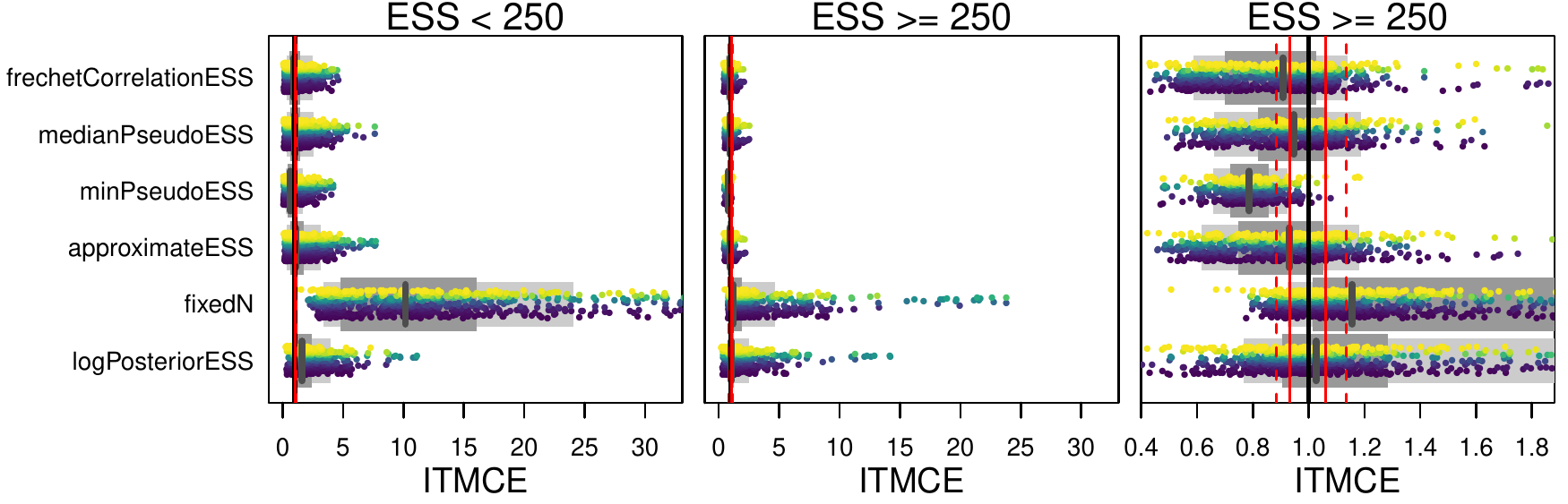}
  \end{subfigure}%
  \caption{
    The RMCE ($(\semcmc - \seess) / \semcmc$) and ITMCE ($\semcmc / \seess$) for split probabilities for all topological ESS measures and all 45 combinations of 9 datasets and 5  run lengths.
    This figure uses an ESS cutoff of 250 instead of 500, but is otherwise the same as Figure 2.
    }%
  \label{fig:benchmarks_splits_250}
\end{figure}

As an alternative to binning ESS performance using a cutoff of 500, we also consider a laxer cutoff of 250.
From Figures~\ref{fig:benchmarks_splits_250} and \ref{fig:benchmarks_tree_probs_250} it is evident a cutoff of 250 is not sufficient.
In the $250 \leq \ESS < 500$ regime, it would appear most methods are generally conservative and underestimate the ESS.
However, there are splits and tree topologies where the error in the estimated probability can be quite large compared to the $\ESS \geq 500$ regime.

\begin{figure}
  \centering
  \begin{subfigure}[ht]{\textwidth}
    \centering
    \includegraphics[width=\textwidth]{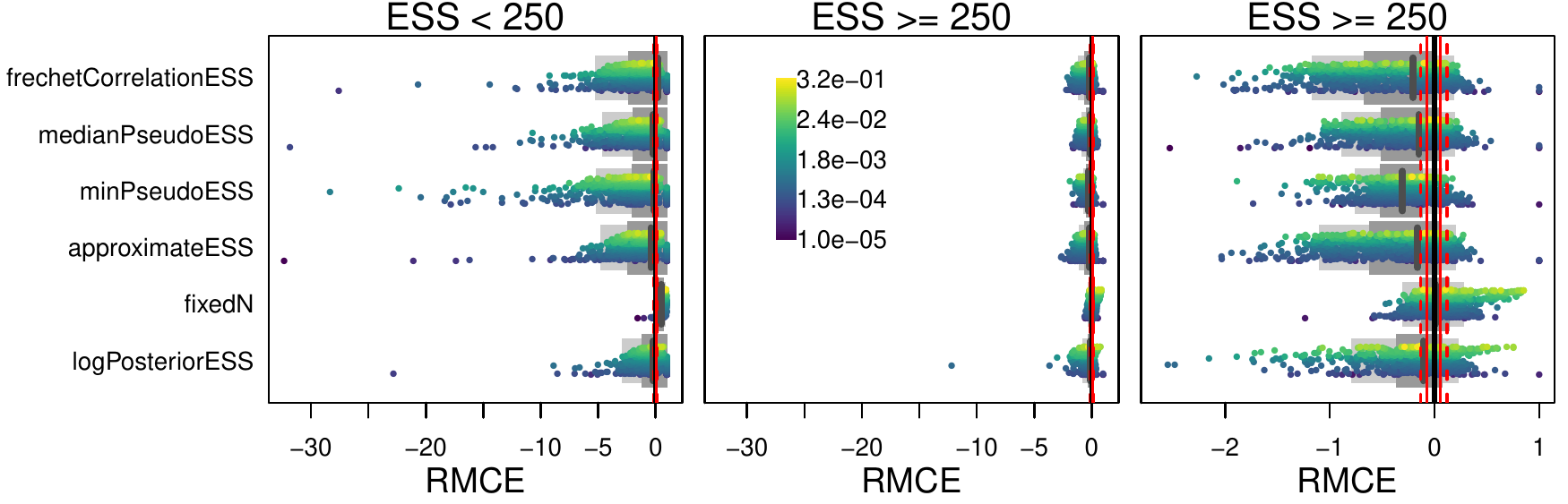}
  \end{subfigure}
  \begin{subfigure}[ht]{\textwidth}
    \centering
    \includegraphics[width=\textwidth]{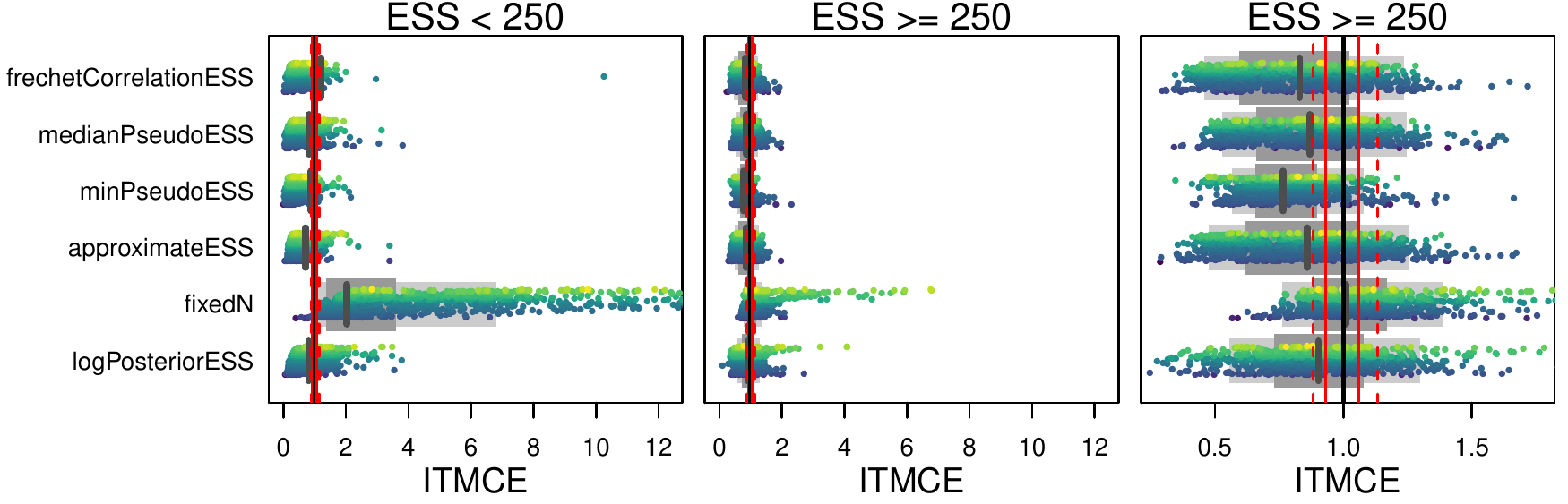}
  \end{subfigure}%
  \caption{
    The RMCE ($(\semcmc - \seess) / \semcmc$) and ITMCE ($\semcmc / \seess$) for topology probabilities for all topological ESS measures and all 45 dataset by run length combinations.
    This figure uses an ESS cutoff of 250 instead of 500, but is otherwise the same as Figure 3.
    }%
  \label{fig:benchmarks_tree_probs_250}
\end{figure}

\pagebreak

\section*{Comparing the estimates of the effective sample size}
In the main text, we focused on the performance of each ESS measure separately.
Here we examine how similar their estimates are, using the 4500 simulated analyses (9 datasets, 5 chain lengths, 100 replicates each).
There is large-scale agreement, but there are also clearly effects of different datasets and run lengths.

\begin{figure}[htpb]
  \centering
  \includegraphics[width=0.9\textwidth]{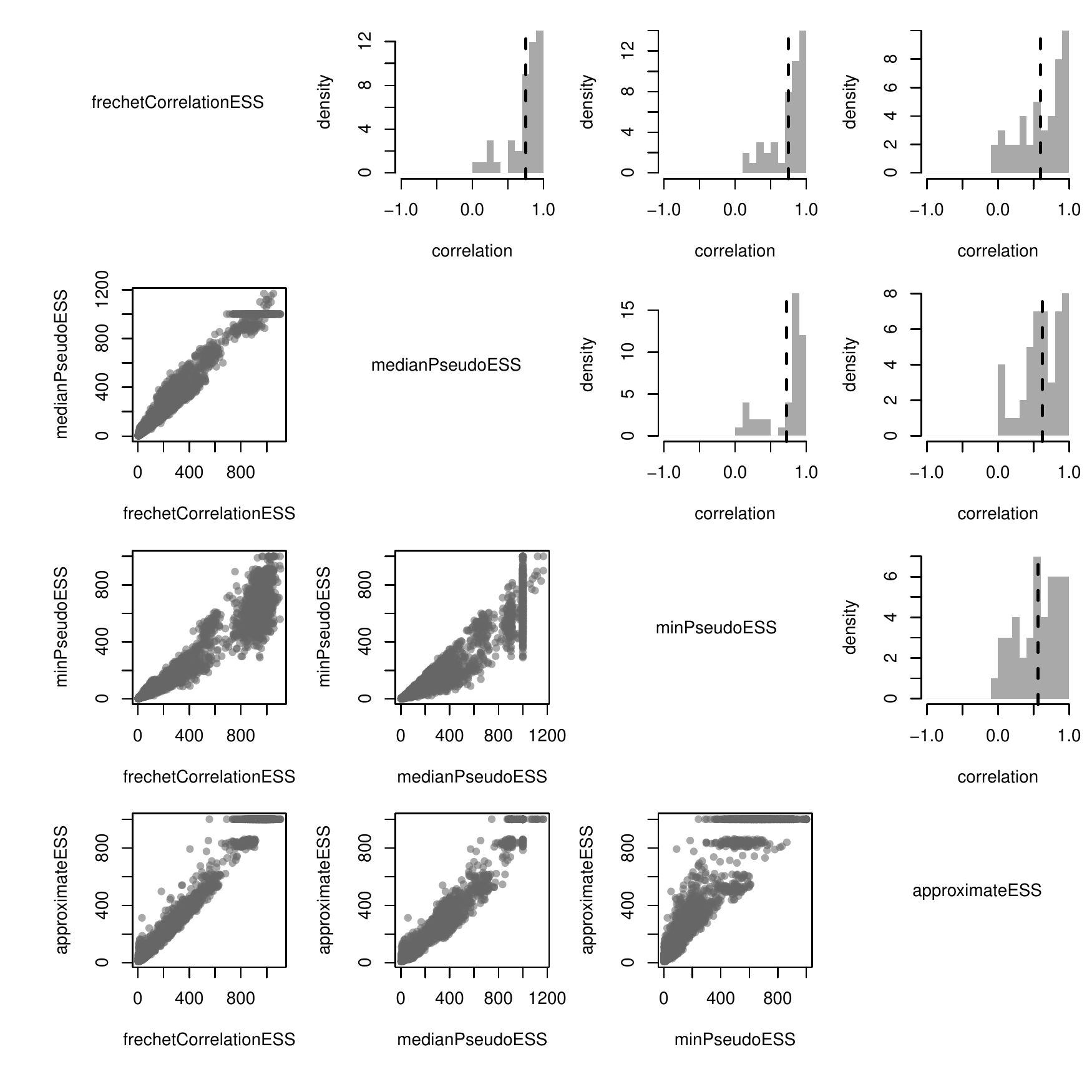}
  \caption{
      Comparison of the estimated ESS for the main-text tree ESS measures on all 4500 simulated analyses (9 datasets, 5 chain lengths, 100 replicates each).
      Below the diagonal, we simply plot the estimated ESS for each pair of methods.
      However, there is clearly variability across the different datasets and run lengths in concordance.
      We summarize this variability in the histograms above the diagonal.
      These summarize the 45 correlation coefficients computed for all 100 replicates for each dataset and run length combination.
      The dashed vertical line is the mean.
    }%
  \label{fig:benchmarks_tree_probs_250}
\end{figure}

\section*{Performance of all additional tree ESS measures}

Across all 10 ESS methods (4 main text ESS methods and the 6 introduced in the supplement), performance is mostly similar to the main-text results.
The two \textit{ad-hoc} ``jump distance'' approaches that only use $s_0$ to estimate the ESS drastically underestimate the ESS.
The $\splitFrequencyESS$ performs about as well as the $\frechetCorrelationESS$, with slightly worse performance in the $\ESS < 500$ regime and slightly better performance in the $\ESS \geq 500$ regime.
The dimension-reduction approaches all perform relatively similarly.
The $\foldedRankmedoidESS$ generally performs equivalently to $\medianPseudoESS$, the $\CMDSESS$ is slightly more conservative, and the performance of the $\totalDistanceESS$ is a bit more variable.
In Figures S5-S7, we plot all 10 methods for all 3 MCMCSE measures.
For simplicity, and since the results are similar, we present only the RMCE.
Overall, a combination of the $\minPseudoESS$ and either the $\frechetCorrelationESS$, $\splitFrequencyESS$, $\foldedRankmedoidESS$, or $\medianPseudoESS$ should cover both the $\ESS < 500$ and $\ESS \geq 500$ regimes in practice.

\begin{figure}[ht]
  \centering
  \includegraphics[width=\textwidth]{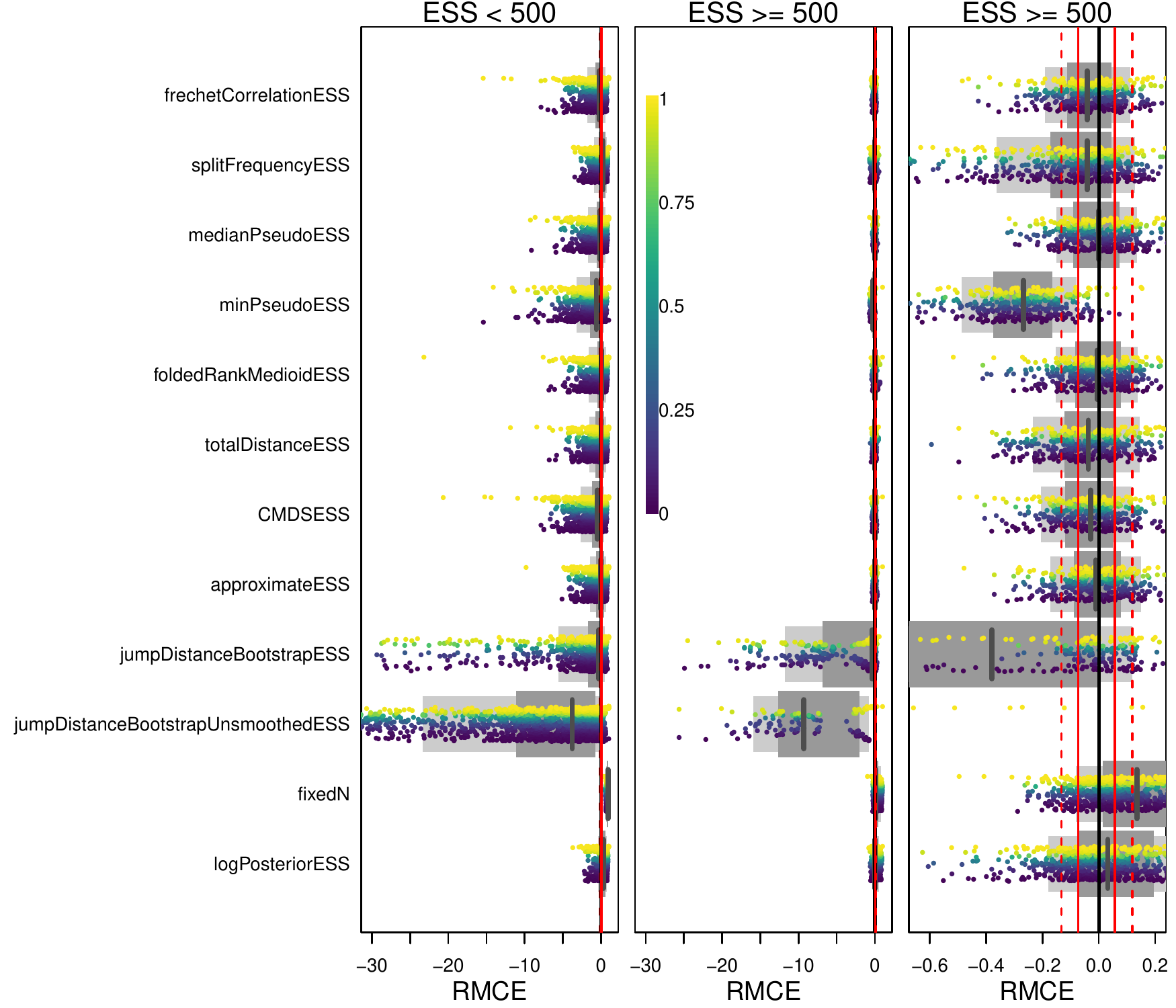}
  \caption{
    The RMCE ($(\semcmc - \seess) / \semcmc$) for split probabilities for all topological ESS measures and all 45 combinations of 9 datasets and 5  run lengths.
    This figure is a more comprehensive version of Figure 2 including all ESS measures considered in the paper.
    }%
  \label{fig:benchmarks_splits_supp}
\end{figure}

\begin{figure}[ht]
  \centering
  \includegraphics[width=\textwidth]{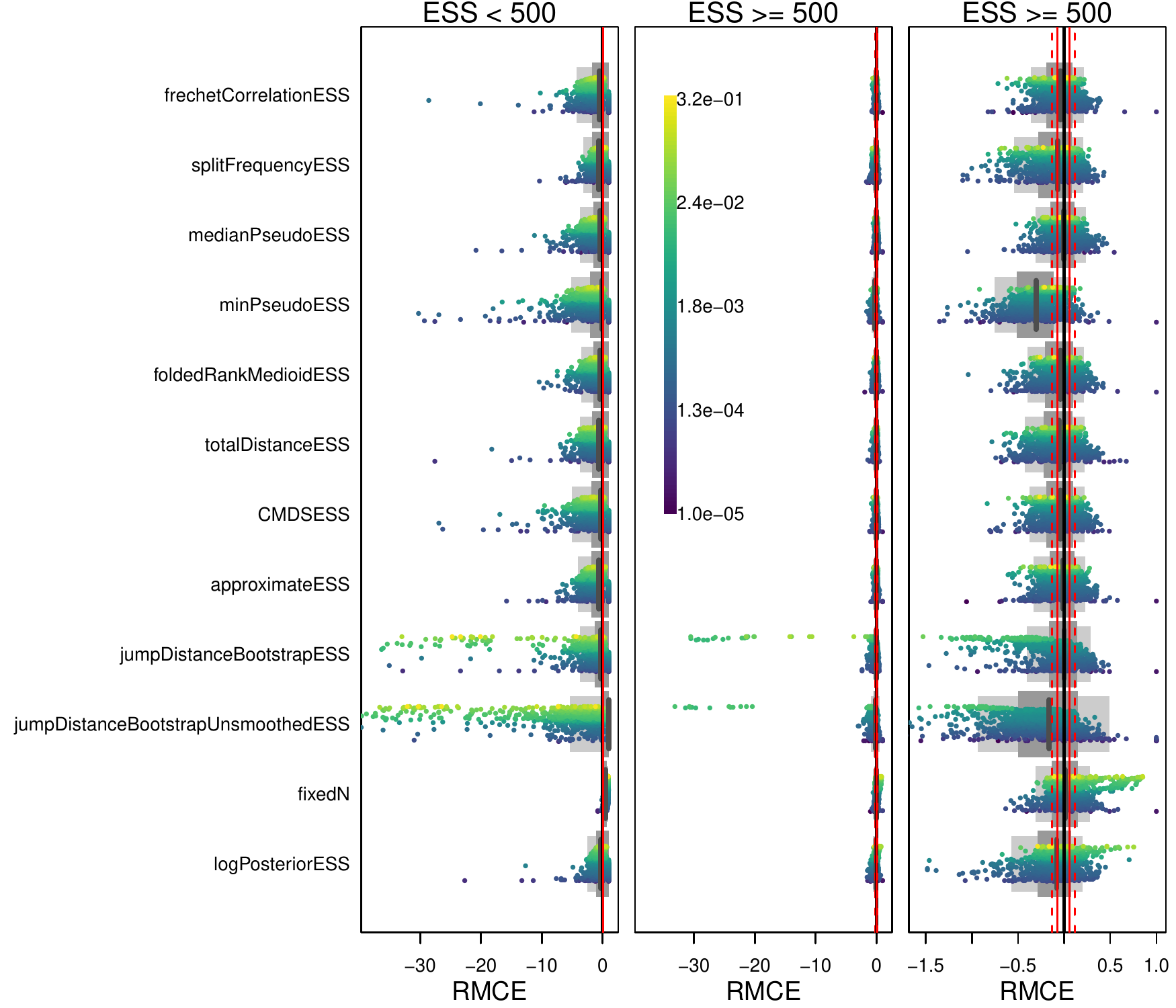}
  \caption{
    The RMCE ($(\semcmc - \seess) / \semcmc$) for topology probabilities for all topological ESS measures and all 45 dataset by run length combinations.
    This figure is a more comprehensive version of Figure 3 including all ESS measures considered in the paper.
    }%
  \label{fig:benchmarks_tree_probs_supp}
\end{figure}

\begin{figure}[ht]
  \centering
  \includegraphics[width=\textwidth]{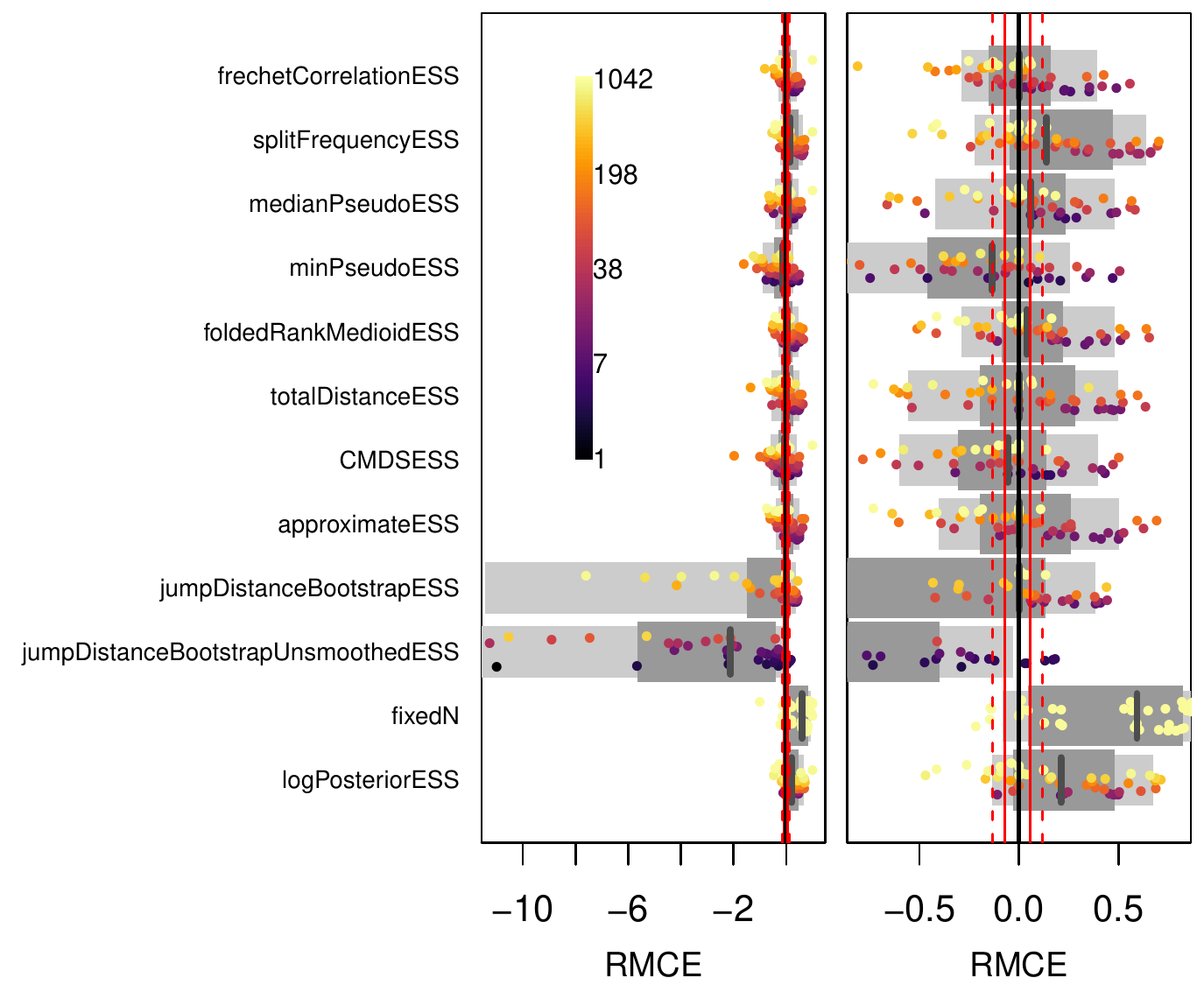}
  \caption{
    The RMCE ($(\semcmc - \seess) / \semcmc$) for the majority-rule consensus (MRC) tree for all topological ESS measures and all 45 dataset by run length combinations.
    This figure is a more comprehensive version of Figure 4 including all ESS measures considered in the paper.
    }%
  \label{fig:benchmarks_MRC_supp}
\end{figure}

\subsection*{Scalability}
Computing most of the tree ESS measures described requires computing the entire $n \times n$ distance matrix, which is computationally costly and scales with the square of the number of trees.
Many of the described methods can be altered to accommodate subsampling, and the RWTY implementation implements this for both the approximate ESS and $\medianPseudoESS$ \citep{warren2017rwty}.
Future work will be needed to determine whether any methods perform adequately with subsampling, and which methods provide an adequate runtime for either very large samples of trees or samples of very large trees.

\clearpage

\section*{Additional empirical results}

For completeness, we now present split-split plots with confidence intervals for the other 5 datasets of \citet{scantlebury2013diversification}.
The confidence intervals for split probabilities here are computed using the Jeffreys interval \citep{brown2001interval}, while the confidence intervals for the difference in split probabilities are computed here using the approach of Agresti and Caffo \citep{agresti2000simple}.
These approaches appear to work well in practice, though in \texttt{treess} we implement alternatives to both.
We also present an aggregation across all these plots comparing the confidence interval approach to more standard approaches based on the average and maximum standard deviations of split frequencies (ASDSF and MSDSF).
This aggregation highlights the fact that similar ASDSF or MSDSF can correspond to a range of numbers of splits whose probabilities disagree across runs (and vice-versa).

\begin{figure}[htbp]
  \centering
  \includegraphics[width=0.7\textwidth]{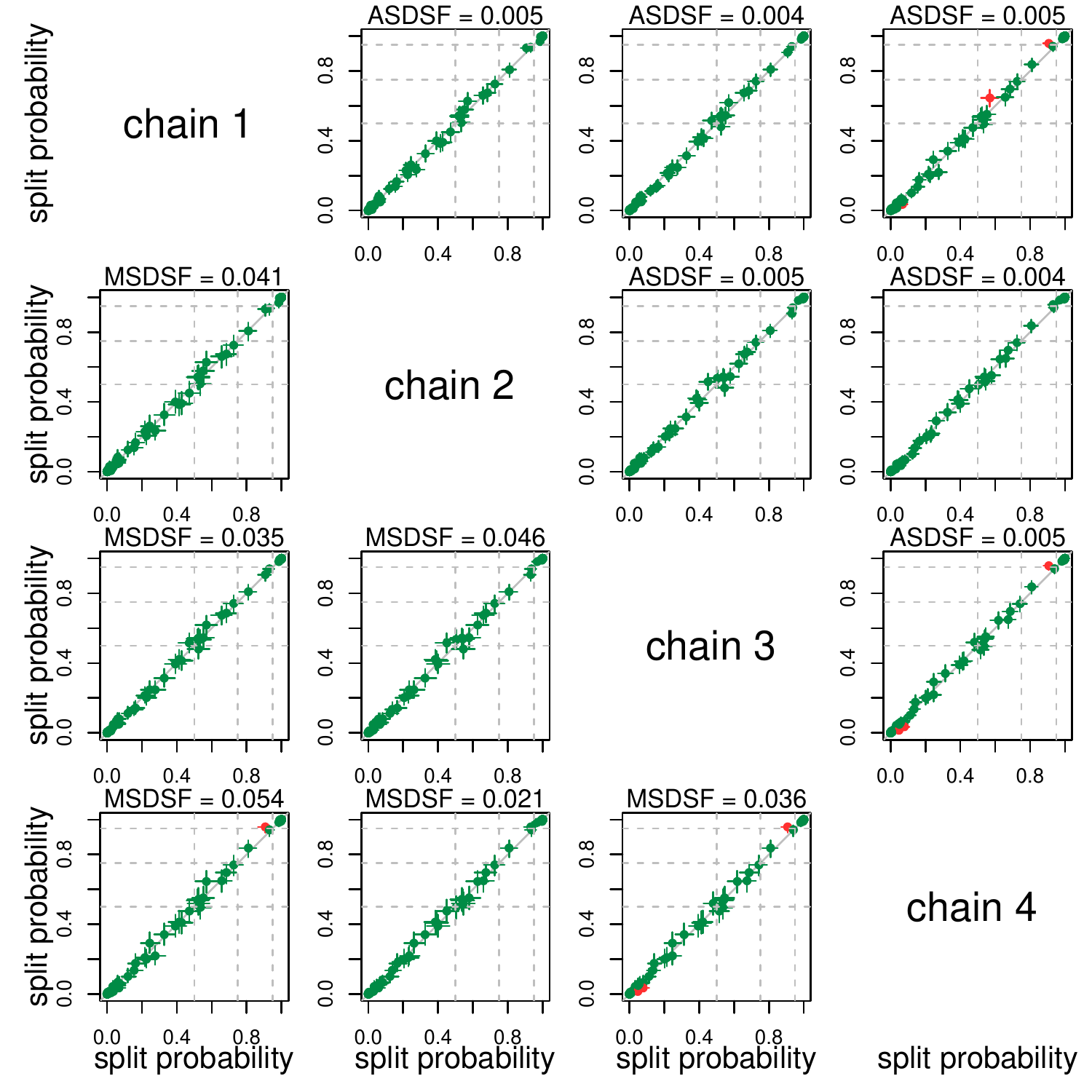}
  \caption{
    Split probabilities computed for all chains of the \emph{Cophyline} dataset of \citet{scantlebury2013diversification}, plotted against the probabilities computed for all other chains, with confidence intervals.
    Comparisons above the diagonal use the $\frechetCorrelationESS$ to compute confidence intervals, while comparisons below the diagonal use the $\minPseudoESS$, which is generally smaller and thus leads to larger confidence intervals.
    Each confidence interval is colored by whether or not the 95\% CI for the difference in split probability between chains $i$ and $j$ includes 0 (green for including 0, red for excluding 0).
    CIs for differences in probability that exclude 0 (or non-overlapping confidence intervals) are more likely to be indicative of convergence issues between chains.
    Narrower confidence intervals from larger tree ESS estimates will flag more splits as problematic (as in chains 1 and 4).
    Dashed grey lines indicate posterior probabilities of 0.5 (threshold for inclusion in the MRC tree), 0.75 (moderate support for a split), and 0.95 (strong support for a split).
    For comparison, we include the average standard deviation of split frequencies (ASDSF, above the diagonal) and maximum standard deviation of split frequencies (MSDSF, below the diagonal).
  }
  \label{fig:split_plots_Cophyline}
\end{figure}

\begin{figure}[htbp]
  \centering
  \includegraphics[width=0.7\textwidth]{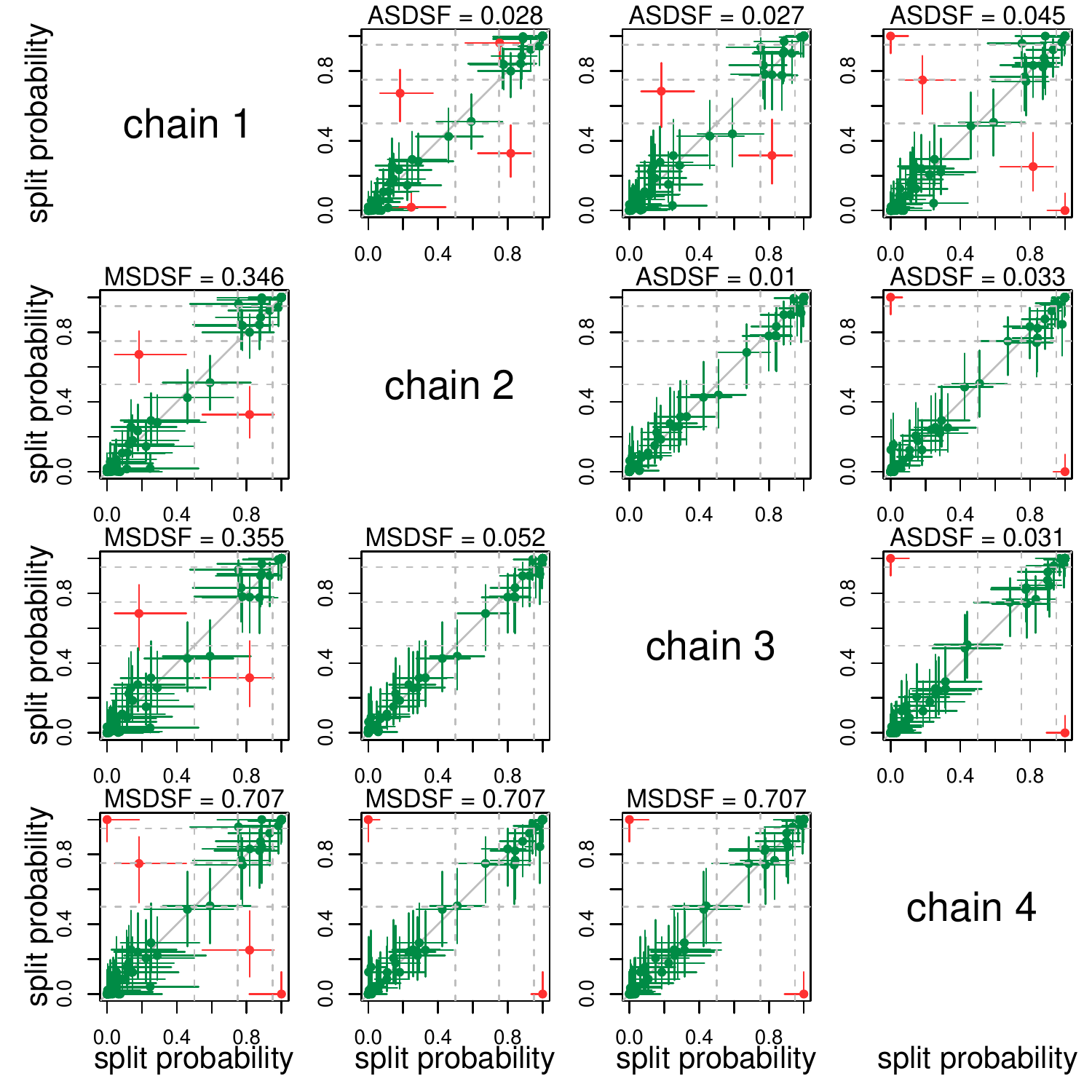}
  \caption{
    Split probabilities computed for all chains of the \emph{Gephyromantis} dataset of \citet{scantlebury2013diversification}, plotted against the probabilities computed for all other chains, with confidence intervals.
    For more explanation, see Figure \ref{fig:split_plots_Cophyline} caption.
  }
  \label{fig:split_plots_Gephyromantis}
\end{figure}

\begin{figure}[htbp]
  \centering
  \includegraphics[width=0.7\textwidth]{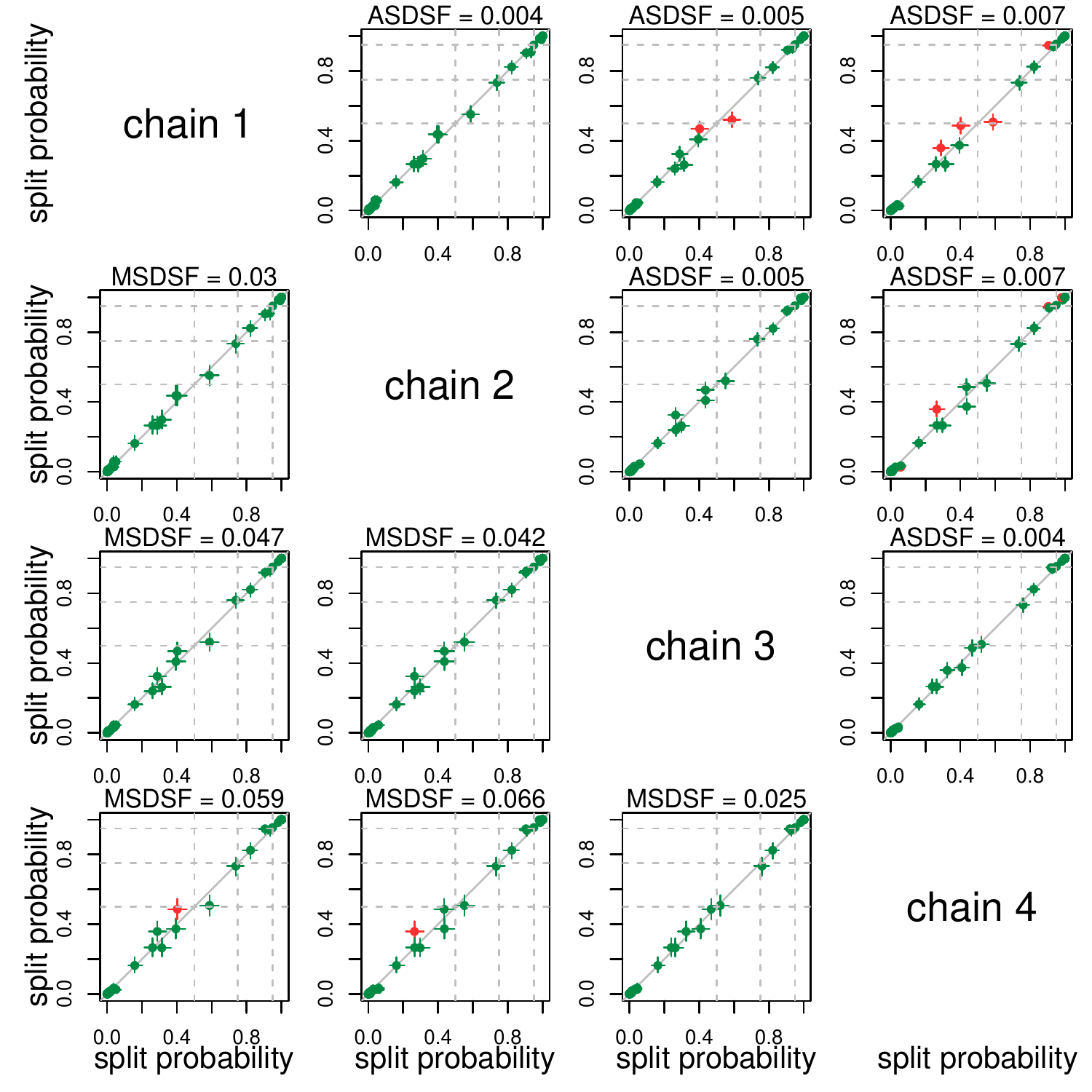}
  \caption{
    Split probabilities computed for all chains of the \emph{Heterixalus} dataset of \citet{scantlebury2013diversification}, plotted against the probabilities computed for all other chains, with confidence intervals.
    For more explanation, see Figure \ref{fig:split_plots_Cophyline} caption.
  }
  \label{fig:split_plots_Heterixalus}
\end{figure}

\begin{figure}[htbp]
  \centering
  \includegraphics[width=0.7\textwidth]{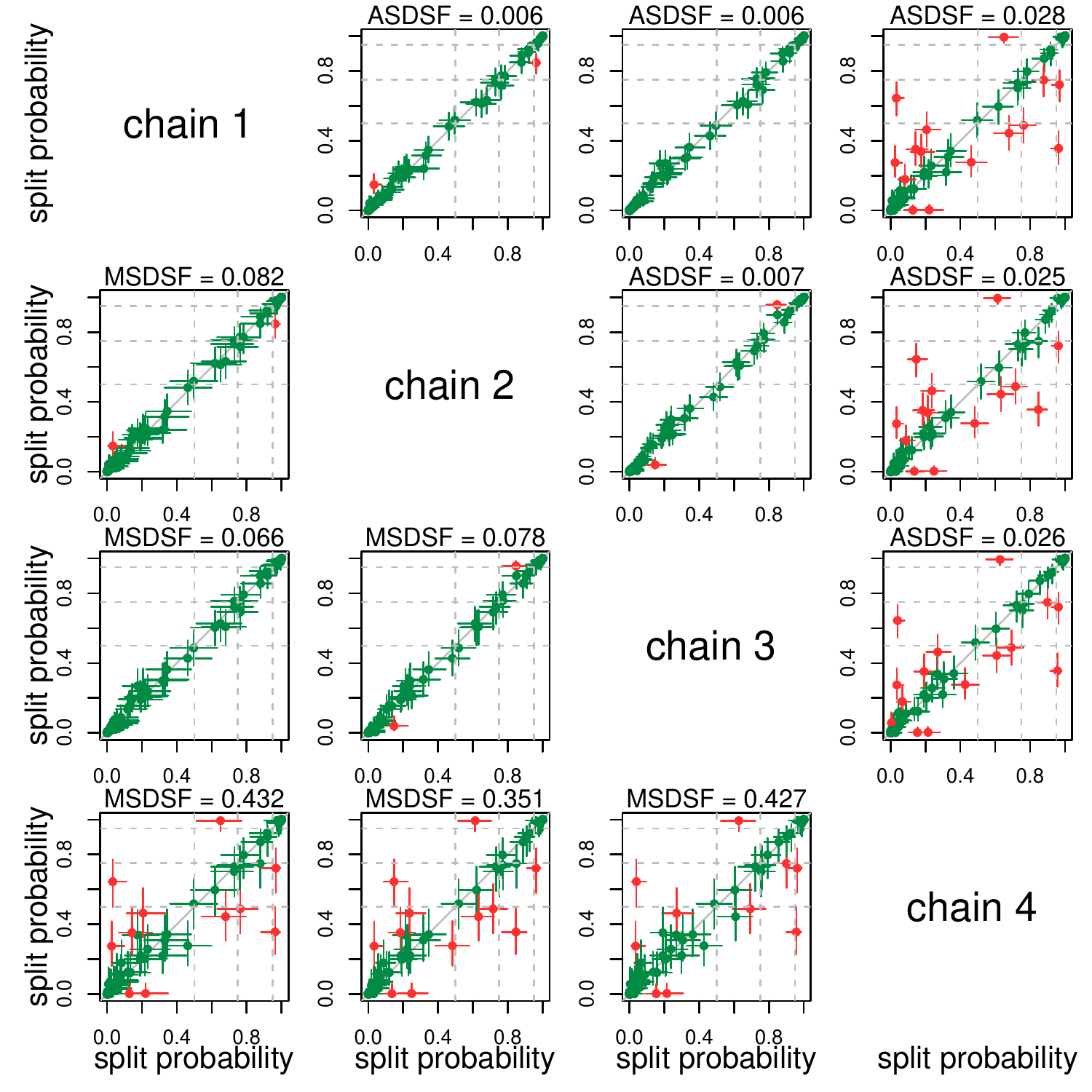}
  \caption{
    Split probabilities computed for all chains of the \emph{Phelsuma} dataset of \citet{scantlebury2013diversification}, plotted against the probabilities computed for all other chains, with confidence intervals.
    For more explanation, see Figure \ref{fig:split_plots_Cophyline} caption.
  }
  \label{fig:split_plots_Phelsuma}
\end{figure}

\begin{figure}[htbp]
  \centering
  \includegraphics[width=0.7\textwidth]{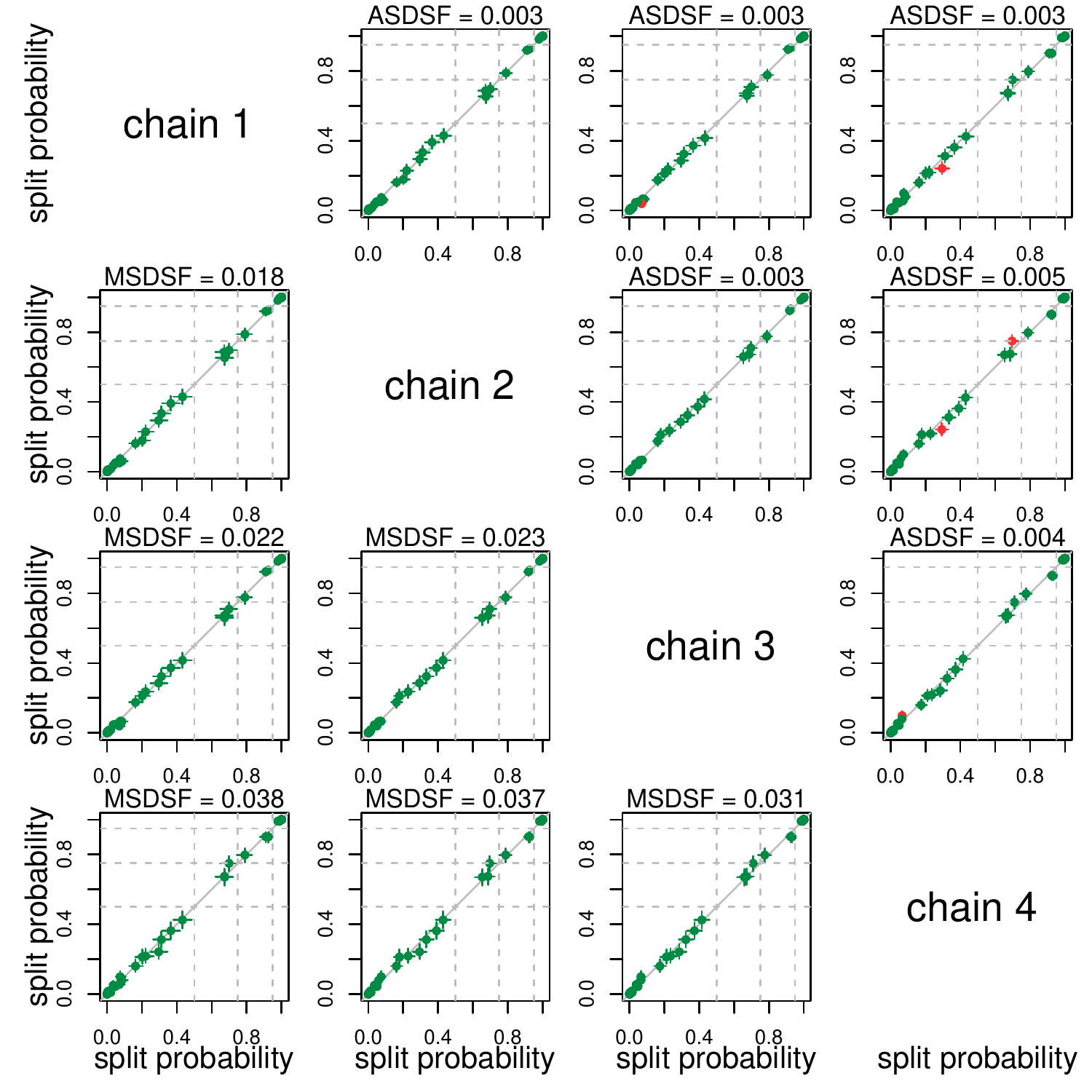}
  \caption{
    Split probabilities computed for all chains of the \emph{Uroplatus} dataset of \citet{scantlebury2013diversification}, plotted against the probabilities computed for all other chains, with confidence intervals.
    For more explanation, see Figure \ref{fig:split_plots_Cophyline} caption.
  }
  \label{fig:split_plots_Uroplatus}
\end{figure}

\newpage

\section*{Comparing split confidence intervals and the ASDSF}
Here we consider how the ASDSF (and MSDSF) compare to our confidence-interval-based approach to comparing splits.

Let us first take the \textit{Gephyromantis} and \textit{Phelsuma} datasets together as a case study.
The ESS is low for both (Figure~\ref{fig:empirical_ess}), with an average $\frechetCorrelationESS$ of 26 for the \textit{Gephyromantis} dataset and 129 for the \textit{Phelsuma} dataset.
The MSDSF is large for many pairwise chain comparisons, and visual inspection of split-split probability plots shows notable discrepancies.
Clearly the analyses of these datasets encountered MCMC difficulties.
What do we learn from the various comparisons available to us?
The MSDSF clearly indicates that there are between-chain convergence problems in both datasets, at which point we might plot the split probabilities to see what is going on.
We would clearly see that chain 4 is distinct in both datasets, and that chain 1 also appears discordant in the \textit{Gephyromantis} dataset.
The confidence intervals provide additional information and suggest different failure modes between the chains.
That many splits (roughly a dozen) disagree for the \textit{Phelsuma} dataset suggests the possibility that the fourth chain has converged to a different local mode than the others.
In this case, running the chains longer should solve the problem.
Without accounting for the effective sample size, we might think that the \textit{Gephyromantis} dataset experienced similar problems.
But accounting for the low ESS, we see that the pattern is being driven by only 2-4 splits.
When considered with the fact that there appear to be 3 clusters of chains, 1, 2+3, and 4, we may begin to suspect that a peculiarity of the treespace which is causing difficulty mixing.
In this case, longer runs alone may not easily solve the problem and we may wish to consider alternatives like Metropolis-coupling.

\begin{figure}[htpb]
  \centering
  \includegraphics[width=0.7\textwidth]{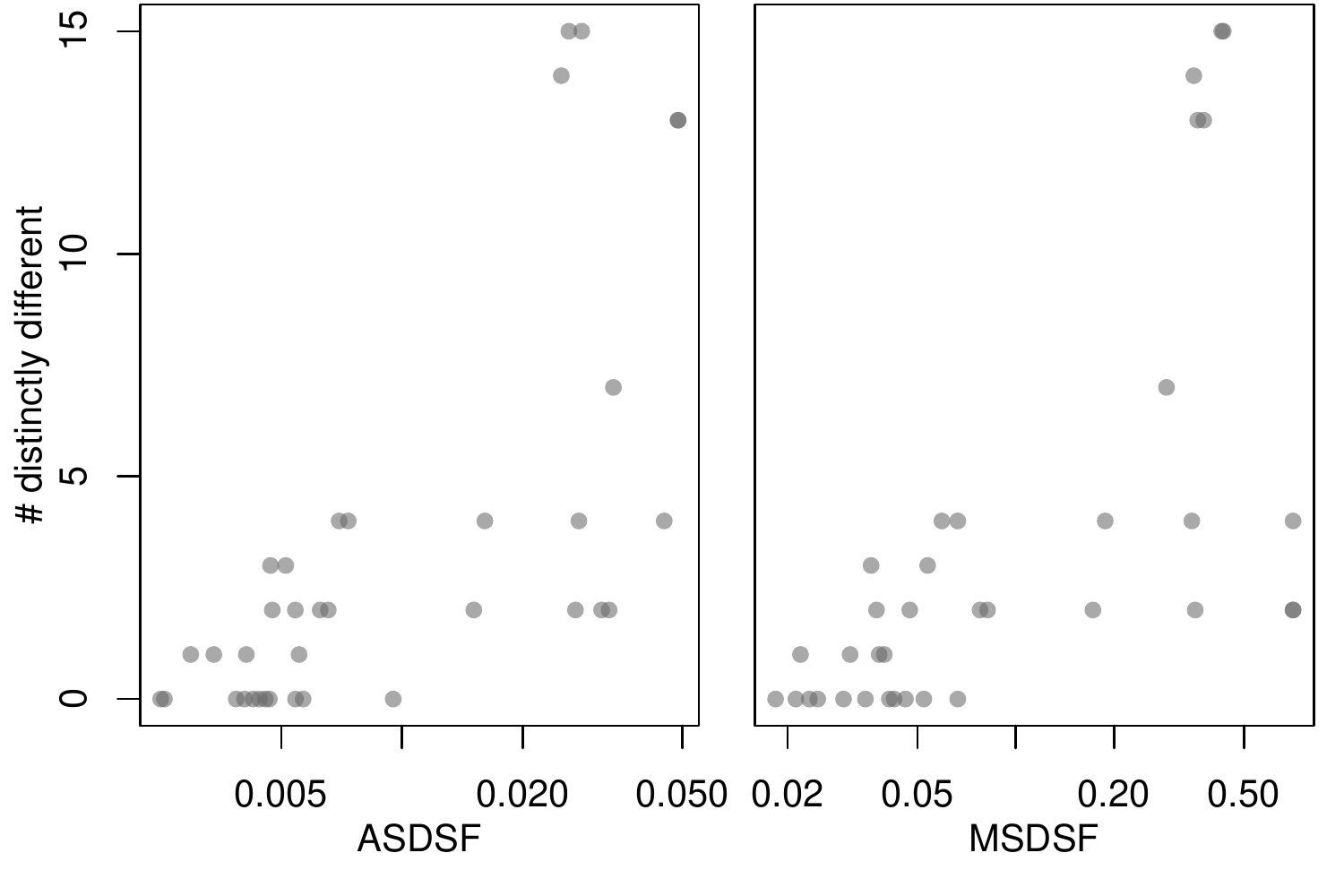}
  \caption{
    An aggregated comparison of our confidence interval approach to comparing split probabilities and previously-existing approaches.
    For all 24 pairwise comparisons of chains in the Malagasy analyses, we compute the average standard deviation of split frequencies (ASDSF) and maximum standard deviation of split frequencies (MSDSF).
    We also count the number of splits for which the 95\% CI for the difference in split probability between chains excludes 0, which we term the number of distinctly different splits.
    We us the $\frechetCorrelationESS$ to compute the CIs.
    While this number is correlated with the ASDSF and MSDSF, there is notable variation.
    Similar ASDSF and MSDSF can correspond to a range of numbers of failing splits, and vice-versa.
  }
  \label{fig:CI_vs_ASDSF_and_MSDSF}
\end{figure}

We also consider a coarser comparison of the confidence interval approach with the ASDSF and MSDSF.
In Figure~\ref{fig:CI_vs_ASDSF_and_MSDSF}, we plot the number of splits which appear to be distinctly different using the confidence interval approach (those colored red) against the ASDSF and MSDSF.
While the measures are correlated, we can see that for a given ASDSF or MSDSF, there is a notable range of numbers of splits which differ, and vice-versa.

\newpage

\section*{Comparing ESS-based measures of Monte Carlo error to multiple-chain-based measures}
\begin{figure}[ht]
  \centering
  \includegraphics[width=\textwidth]{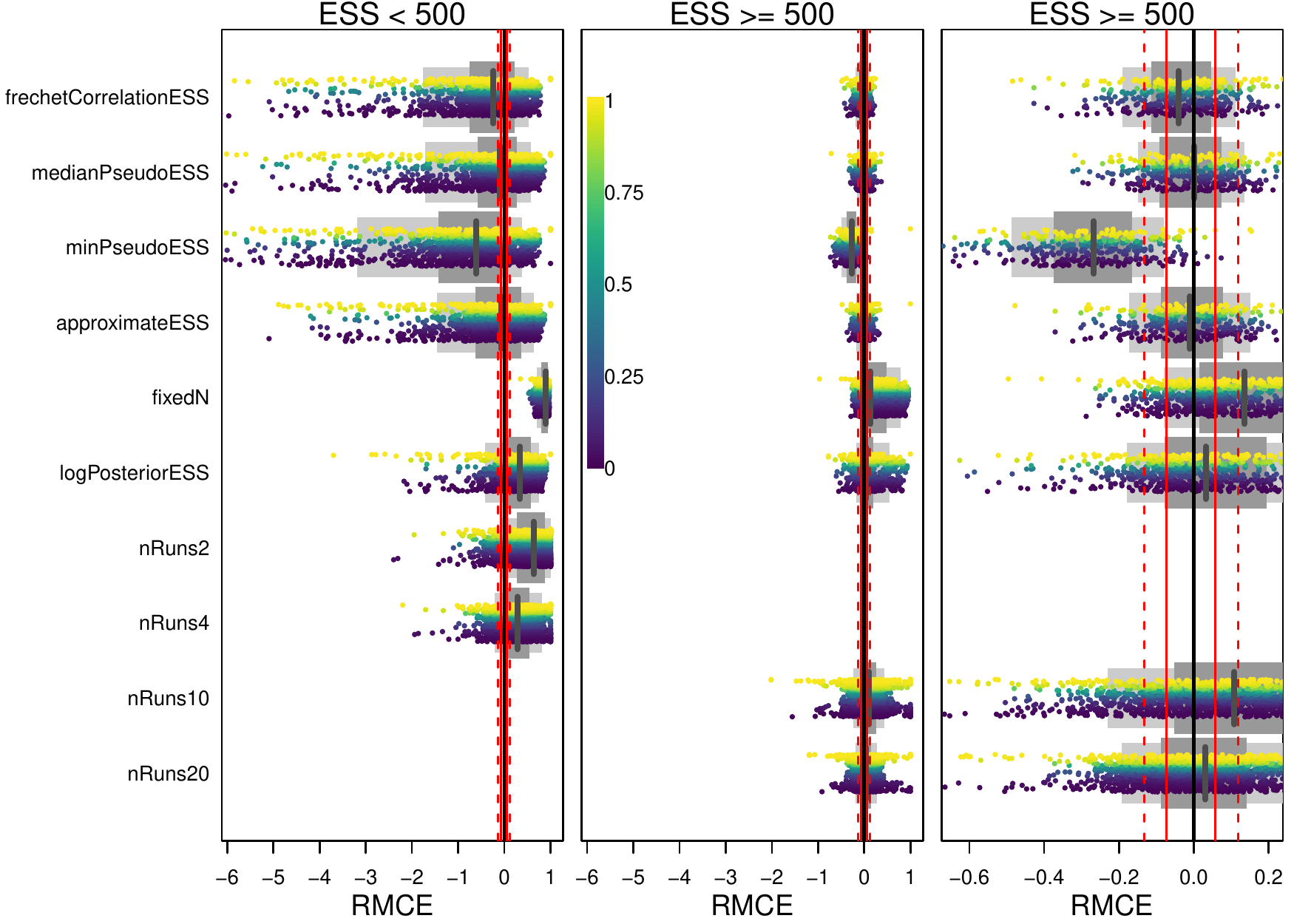}
  \caption{
    The RMCE ($(\semcmc - \seess) / \semcmc$) for split probabilities for all topological ESS measures and all 45 combinations of 9 datasets and 5  run lengths.
    This figure reproduces Figure 3 and adds four approaches to directly estimating the Monte Carlo error.
    We use 2, 4, 10, or 20 independent MCMC runs (nRuns2 to nRuns20) and the same brute-force approach employed to obtain $\semcmc$.
    These brute-force approaches do not use an ESS to estimate the Monte Carlo error.
    Given the performance differential between $< 10$ and $\geq 10$ runs, we place the results from 2 and 4 chains in the $\ESS < 500$ column and the results from 10 and 20 chains in the $\ESS \geq 500$ columns.
    }%
  \label{fig:benchmarks_splits_mrbayes}
\end{figure}

\begin{figure}[ht]
  \centering
  \includegraphics[width=\textwidth]{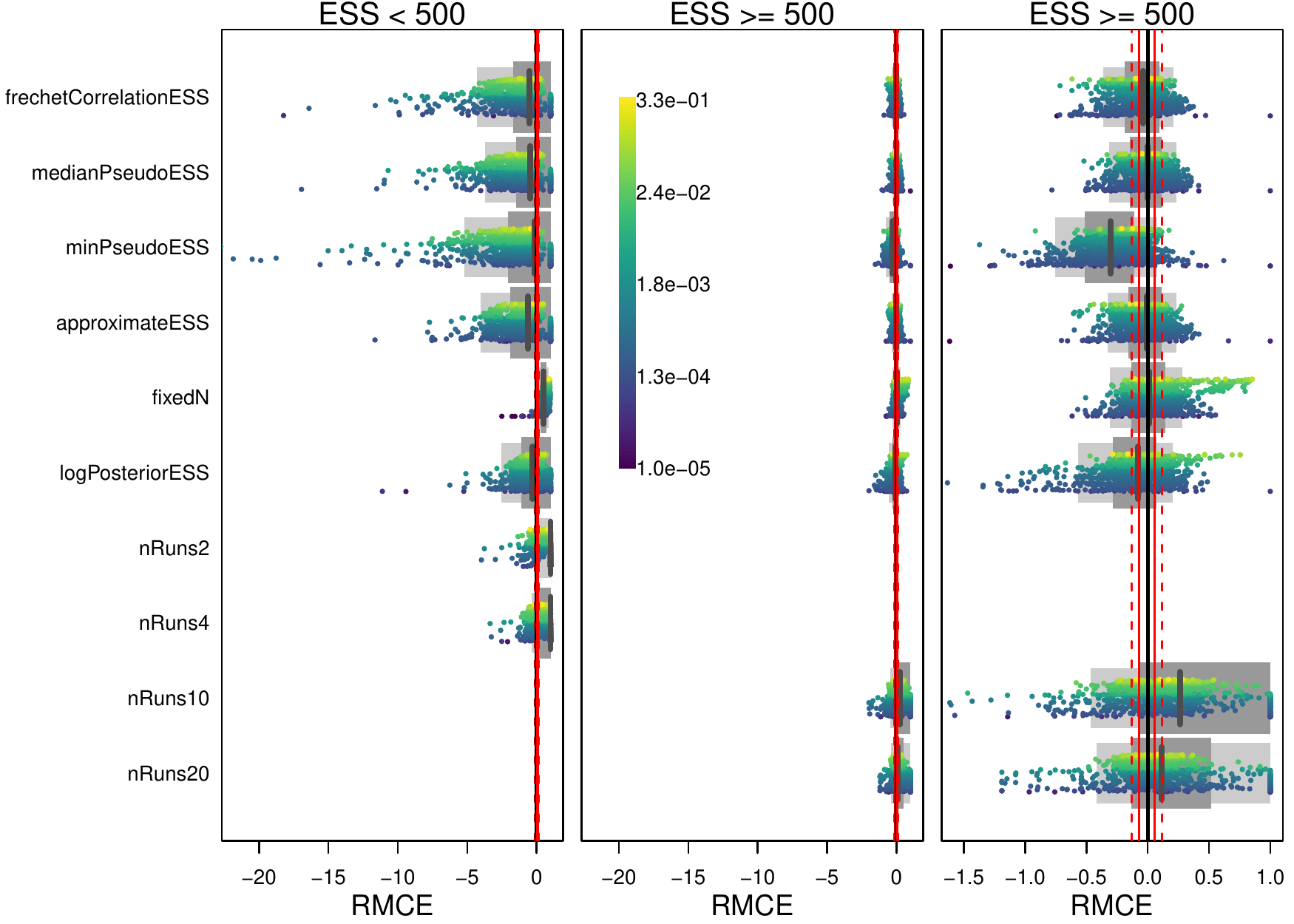}
  \caption{
    The RMCE ($(\semcmc - \seess) / \semcmc$) for topology probabilities for all topological ESS measures and all 45 dataset by run length combinations.
    This figure reproduces Figure 4 and adds four approaches to directly estimating the Monte Carlo error as in Figure~\ref{fig:benchmarks_splits_mrbayes}.
    }%
  \label{fig:benchmarks_tree_probs_mrbayes}
\end{figure}

\begin{figure}[ht]
  \centering
  \includegraphics[width=\textwidth]{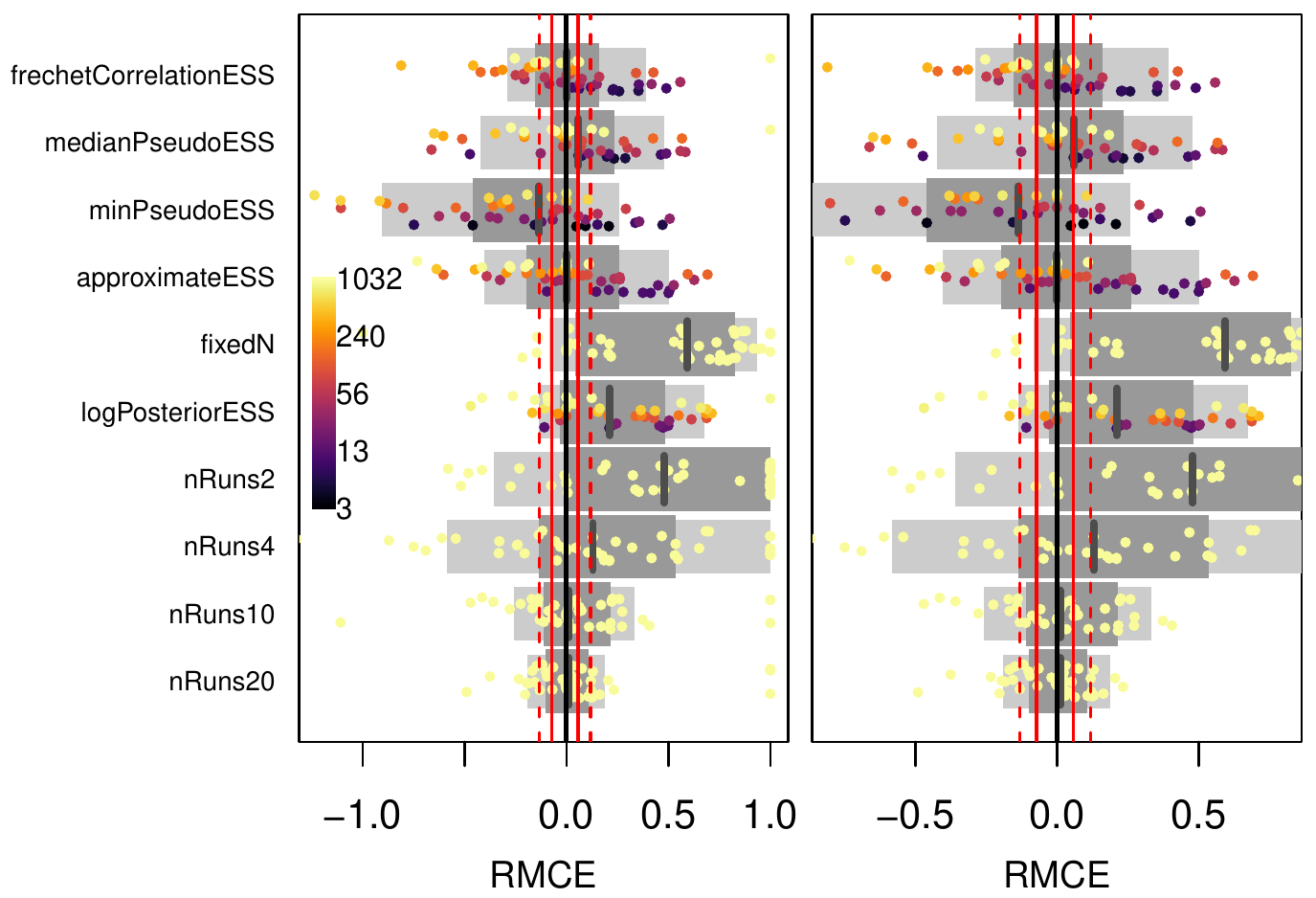}
  \caption{
    The RMCE ($(\semcmc - \seess) / \semcmc$) for the majority-rule consensus (MRC) tree for all topological ESS measures and all 45 dataset by run length combinations.
    This figure reproduces Figure 5 and adds four approaches to directly estimating the Monte Carlo error as in Figure~\ref{fig:benchmarks_splits_mrbayes}.
    As these brute-force approaches do not use an ESS to estimate the Monte Carlo error, we arbitrarily color the points for the nRuns approaches as if they had ESS 1000.
    }%
  \label{fig:benchmarks_MRC_mrbayes}
\end{figure}

For splits in the consensus tree, MrBayes reports the standard deviation (across runs) of the split freqencies (the SDSF).
This is a direct approach to quantifying the Monte Carlo error in split probabilities.
To determine how well this approach can capture Monte Carlo error, we perform an additional experiment.
For each of our 45 dataset x run length combinations, we take a small number of the independent MCMC runs (2, 4, 10, and 20) and use them to estimate the Monte Carlo error.
These runs are a subset of the 100 runs used to compute the $\semcmc$.
We compare the performance to the main-text ESS measures for split probabilities (Figure~\ref{fig:benchmarks_splits_mrbayes}), tree probabilities (Figure~\ref{fig:benchmarks_tree_probs_mrbayes}), and the MRC tree (Figure~\ref{fig:benchmarks_MRC_mrbayes}).
We find that using 2 or 4 chains fails to capture the Monte Carlo error well for any of these three quantities, performing notably worse than the $\textsf{logPosteriorESS}$ and $\textsf{fixedN}$ approaches.
Using 10 chains can capture the Monte Carlo error for the MRC tree adequately, but not for split or tree probabilities.
Using 20 chains can capture the Monte Carlo error in the MRC tree about as well as the tree ESS measures.
For split and tree probabilities, the performance using 20 chains falls between the $\textsf{logPosteriorESS}$ and $\textsf{fixedN}$ approaches and using the tree ESS measures considered in this paper.

\end{document}